\documentclass[a4paper,12pt]{article}

\pdfoutput=1

\usepackage[T1]{fontenc}
\usepackage{jheppub}
\usepackage[font=footnotesize]{caption}
\usepackage{multirow}
\usepackage{cleveref}

\newcommand{\CL}{\mathrm{CL}}
\newcommand{\eps}{\varepsilon}
\newcommand{\ii}{\mathrm{i}}
\newcommand{\real}{\mathrm{Re}}
\newcommand{\imag}{\mathrm{Im}}
\newcommand{\obs}{\mathcal{O}}

\graphicspath{{./figures}}

\title{Correctness criteria for complex Langevin} 

\author{Michael Mandl}
\affiliation{Institute of Physics, NAWI Graz, University of Graz, Universit{\"a}tsplatz 5, 8010 Graz, Austria}
\emailAdd{michael.mandl@uni-graz.at}

\abstract{The complex Langevin approach is a promising method for the numerical treatment of systems with a sign problem, for which conventional lattice field theory techniques based on importance sampling cannot be applied. However, complex Langevin dynamics may fail to converge in some cases and converge to a wrong limit in others, motivating the development of various diagnostic tools over the years to assess the correctness of given simulation results. This work aims at providing a systematic comparison between the most prominent such correctness criteria. In particular, the main goal is to contrast their applicability, ease of use, and -- most importantly -- their predictive power. To this end, four simple but nontrivial models are considered and the criteria applied to each of them. It is found that no single correctness criterion is entirely foolproof on its own and that it is advisable in practice to employ multiple different criteria in combination. These conclusions are expected to carry over to more realistic theories as well.}

\keywords{Lattice Quantum Field Theory, Algorithms and Theoretical Developments,  Other Lattice Field Theories}

\arxivnumber{2604.12388}

\begin{document}

	\maketitle
	\flushbottom
	
	\section{Introduction}
	\label{sec:introduction}In lattice quantum field theory the nonperturbative computation of physical observables from first principles involves an integration over high-dimensional spaces of weight functions $\rho$; for instance, $\rho\propto e^{-S_E}$ if $S_E$ is the action of a theory in Euclidean space. The most effective and scalable general-purpose way of computing such integrals is by means of Monte-Carlo methods, among which the importance sampling technique stands out as the workhorse. The central idea is to generate ensembles of field configurations distributed according to $\rho$ and estimate expectation values by averaging over these ensembles. A necessary prerequisite for this approach to be applicable is that $\rho$ can be interpreted as a probability density, i.e., that it is real and nonnegative. 

In many cases of physical interest, however, this prerequisite does not hold. Such cases include, but are not restricted to, quantum field theories in Minkowski space time, most Euclidean quantum field theories with fermions in the presence of a nonzero chemical potential, and various systems in condensed-matter physics. In the case that the probabilistic interpretation breaks down, conventional lattice techniques relying on importance sampling cannot be applied and one is confronted with the infamous sign problem (see, e.g., the recent review \cite{AS26r}), for which no general solution exists. The underlying difficulty is the fact that one integrates over contributions with wildly fluctuating phases, leading to strong cancellations and an exponential increase of statistical errors with the system's volume \cite{TW05}.

One particular family of methods aimed at bypassing the sign problem makes use of the complexification of the dynamical degrees of freedom based on insights from complex analysis. For instance, using Cauchy's theorem one may deform the original integration contour into the complex plane, which leaves expectation values of holomorphic observables unaltered, but might come with a weaker sign problem \cite{BK18,BK21,BM22,BM24}. Particularly appealing integration manifolds, at least from a theoretical point of view, are the theory's Lefschetz thimbles \cite{CDS12}, on each of which the imaginary part of the action is constant. Note, however, that the Lefschetz thimbles are not necessarily optimal in the sense of the sign problem \cite{PP25}. Another interesting approach for sampling oscillatory distributions, using so-called physics-informed kernels, was put forward recently in \cite{IKP26}.

A different class of methods, also based on complexification, attempts to replace the complex distribution $\rho$ on the original (real) integration manifold $\mathcal{M}_r$ by a genuine probability distribution $P$ on the complex extension of $\mathcal{M}_r$. If $P$ is equivalent to $\rho$ in the sense that it gives rise to identical expectation values on the space of holomorphic observables, this idea constitutes a solution to the sign problem. One possible way to obtain such a probability distribution and the main topic of this work is the complex Langevin approach \cite{Par83,Kla83}, which is based on the stochastic evolution of the complexified field variables in an auxiliary time dimension. This method, however, is plagued by the notorious wrong-convergence problem, causing the evolution to give rise to incorrect results in various cases. Needless to say, in such a scenario it is essential to have criteria at one's disposal that allow one to discern correct results from incorrect ones without having to compare to exact solutions, which are unavailable in all but the most simple models. 

For the complex Langevin method, various such correctness criteria have emerged over the years, some of which are well motivated on theoretical grounds, while some are more heuristic in nature. Moreover, some of the criteria are believed (or proven) to be equivalent to one another -- if not on paper then at the very least in most practical applications --, whereas some should be applied only in combination with others. It is the purpose of this work to systematically compare different correctness criteria for complex Langevin with regard to their accuracy and predictive power as well as their practical applicability. To the best of the author's knowledge, no large-scale systematic analysis of this sort exists in the literature to date and any type of guideline would be serviceable, perhaps even more so for new practitioners of the method. This is the aim that shall be pursued here. 

This work is structured as follows: The complex Langevin approach is introduced and its drawbacks and failure modes discussed in \cref{sec:cle}. The most commonly used correctness criteria are then introduced in \cref{sec:criteria}. Next, after a brief discussion on the details of the simulations employed in this work (\cref{sec:setup}), the criteria are put to the test in the context of model theories in \cref{sec:models}. Conclusions and main takeaways are presented in \cref{sec:conclusions}, while useful auxiliary derivations are given in the appendices.	
	
	\section{The complex Langevin equation}
	\label{sec:cle}The complex Langevin approach \cite{Par83,Kla83} (see, e.g., \cite{OSZ93p,BRL21} for reviews) is, in essence, a generalization of Parisi--Wu stochastic quantization \cite{PW81} to complex actions. Similar in spirit, it introduces an auxiliary (dimensionless) time dimension $\tau$, the so-called Langevin time, in which the dynamical degrees of freedom are evolved stochastically. However, contrary to conventional stochastic quantization, the complex action requires an extension of field space into the complex domain. The underlying evolution is governed by the complex Langevin equation. For a theory of $N$ real scalar variables\footnote{The scalar field notation is used throughout as it is sufficient to cover all models discussed in this work. It should be emphasized, however, that the formalism is by no means restricted to scalar theories, but like stochastic quantization (see, e.g., \cite{BKK85} for a lattice field theory) can be generalized to include gauge fields as well as fermions.} $\Phi_{R,i}$ that is described by a complex Euclidean\footnote{The subscript `$E$' on the Euclidean action is omitted throughout for simplicity.} action $S[\Phi_R]$, it reads, after complexification\footnote{Like $\Phi_R$, $\Phi_I$ is real-valued. Note that the formalism can also be applied to theories with complex scalar fields after complexifying their real and imaginary parts individually.} to $\Phi=\Phi_R+\ii\Phi_I$, 
\begin{equation}\label{eq:cle}
	\frac{\partial\Phi_i}{\partial\tau} = 
		-H_{ij}[\Phi]H_{kj}[\Phi]\frac{\partial S[\Phi]}{\partial \Phi_k} + \frac{\partial H_{ij}[\Phi]H_{kj}[\Phi]}{\partial\Phi_k} 
		+ H_{ij}[\Phi]\eta_j(\tau)\;.
\end{equation}
Here, $\eta_i(\tau)$ is a real-valued $N$-component noise vector whose components follow independent Gaussian distributions with mean $0$ and variance $2$ for each $\tau$. Furthermore, the complex $N\times N$ matrix $H$, which can depend on the fields in general, is referred to as the kernel. $H$ (or, to be more precise, the product $HH^T$ in \eqref{eq:cle}) plays the role of a complex diffusion matrix and can be used to change the convergence properties of the complex Langevin evolution. Other than having to be holomorphic in $\Phi$, there are no general restrictions on the kernel, which means it provides a large amount of freedom for possible improvement \cite{Sod88,OOS89,OSZ91,BHM23,BHM24}. Such improvement has, for instance, been found by applying machine-learning techniques to obtain optimal kernels \cite{ALR23,LS23,ARS24}. 

Note that the notation in \eqref{eq:cle} implicitly assumes that the theory in question is defined on a lattice. To go to the continuum, some of the sums in \eqref{eq:cle} have to be replaced by integrals. Moreover, from \eqref{eq:cle}, a more common version of the complex Langevin equation results after choosing a trivial kernel, $H_{ij}[\Phi]=\delta_{ij}$. In the following, the first two terms on the right-hand side of \eqref{eq:cle}, i.e., the deterministic contributions to the complex Langevin equation, are together referred to as the drift term, while the last expression, i.e., the stochastic contribution, is called the noise term. Note that a nontrivial kernel modifies both of them. It should be mentioned that a more rigorous definition of the Langevin equation as well as its complex extension can be written down in terms of a Wiener process, which underlies the noise term; see, e.g. \cite{DH87r}. 

The stochastic evolution of the complex field $\Phi$ in the Langevin time $\tau$ given by \eqref{eq:cle} gives rise to a $\tau$-dependent probability distribution $P[\Phi_R,\Phi_I;\tau]$ in the space spanned by the real and imaginary parts $\Phi_R$ and $\Phi_I$ of $\Phi=\Phi_R+\ii\Phi_I$ \cite{Par83}. Its $\tau$-dependence is governed by an associated Fokker--Planck equation,
\begin{equation}\label{eq:fpe}
	\frac{\partial}{\partial \tau}P[\Phi_R,\Phi_I;\tau] = L^TP[\Phi_R,\Phi_I;\tau]\;,
\end{equation}
where the (real) Fokker--Planck operator $L^T$ is a rather complicated expression depending on derivatives of the drift term whose precise form is not important for this work. 

The complex Langevin approach is applicable only if the equilibrium limit
\begin{equation}
	P[\Phi_R,\Phi_I] = \lim_{\tau\to\infty}P[\Phi_R,\Phi_I;\tau]
\end{equation}
exists. In that case, one may use the method to compute expectation values with respect to the probability distribution $P[\Phi_R,\Phi_I]$:
\begin{equation}\label{eq:cle_expectation-values}
	\langle\obs[\Phi]\rangle_\CL = \int\mathcal{D}\Phi_R\int\mathcal{D}\Phi_I\obs[\Phi_R+\ii\Phi_I]P[\Phi_R,\Phi_I]\;.
\end{equation}
This should be contrasted with the usual path-integral definition of observables in Euclidean space,
\begin{equation}\label{eq:expectation_values}
	\langle\obs[\Phi_R]\rangle = \int\mathcal{D}\Phi_R\obs[\Phi_R]\rho[\Phi_R]\;,
\end{equation}
where
\begin{equation}
	\rho[\Phi_R] = \frac{e^{-S[\Phi_R]}}{Z}\;, 
	\quad
	Z = \int\mathcal{D}\Phi_R e^{-S[\Phi_R]}\;,
\end{equation}
and the notation indicates that the integrals should be taken over the fields' original domain before complexification. Thus, it is sensible to say that the complex Langevin approach gives correct results, i.e., solves the sign problem, if and only if
\begin{equation}\label{eq:correctness}
	\langle\obs[\Phi]\rangle_\CL = \langle\obs[\Phi_R]\rangle
\end{equation}
for all holomorphic observables $\obs$. The validity of \eqref{eq:correctness} is by no means guaranteed, as there are various potential failure modes that can cause the complex Langevin evolution to produce incorrect results. This is the infamous wrong-convergence problem of the complex Langevin method. It is therefore sensible to dissect \eqref{eq:correctness} in some more detail. 

As \eqref{eq:correctness} suggests, the underlying idea of the approach is to represent the complex density $\rho$, which is defined on a real manifold, by a probability density $P$ on a complex manifold in such a way that both give rise to the same expectation values on the space of holomorphic observables. An immediate question that arises is whether or not a given $\rho$ admits such a $P$ in the first place. As it turns out, however, this appears to be true quite generally \cite{Sal97,Wei02,Sal07}. Moreover, in general, multiple $P$ could represent one and the same $\rho$, but this nonuniqueness is of little consequence for the purposes at hand. After all, if one is capable of finding any such $P$ for a given $\rho$, one has solved the sign problem for that theory. 

The remaining question is then how to find a probability distribution satisfying the requirement \eqref{eq:correctness} and, in particular, whether the complex Langevin approach can provide such a $P$. There is no theorem that guarantees this for general theories and, in fact, there are numerous counterexamples in which the complex Langevin evolution converges, but to the wrong limit \cite{AY85,AFP86,AJ10,AJP13,PZ13,PZ13_2,MS13,MST15,KS19}. The main failure modes of the complex Langevin method can be summarized as follows (see also \cite{MSS25} and references therein for some more details):
\begin{itemize}
	\item {\bf Runaway trajectories:} These are trajectories within the complexified field space along which the classical evolution (i.e., the one without the noise term in \eqref{eq:cle}) diverges. In a simulation, they can lead to discretization artifacts or even divergences. Their effect can often be dealt with entirely by using adaptive step-size algorithms \cite{AJS10} or implicit solvers \cite{ALR21}.
	\item {\bf Slowly-decaying distributions or boundary terms:} If the probability distribution $P[\Phi_R,\Phi_I]$ decays slowly towards infinity or near poles of the drift term, an assumption underlying the formal proof of correctness of the method in \cite{ASS10,AJS11} is violated due to the appearance of boundary terms, which prevent the use of partial integration necessary for the proof. A different point of view on this problem was provided in \cite{NNS16,NNS18_2}, which investigated the probability distribution of the drift term itself and found that correct convergence requires a sufficiently fast decay of said distribution. While the issues of slow decay of $P$ (towards infinity or near drift-term poles) and that of the drift-term distribution (towards infinity) are likely related to one another, it should be mentioned that there exist different diagnostic tools for their detection, as outlined in \cref{sec:criteria}.
	\item {\bf Unwanted integration cycles:} The complex Langevin evolution may receive contributions \cite{Sal93,GP09,GP09_2,Sal16} from unwanted integration cycles (using the terminology of \cite{Wit10,Wit11}), which constitute unphysical solutions to the theory's Dyson--Schwinger equations \cite{GGG96,GG10}. Indeed, it has been shown in \cite{SS19} for theories with a single degree of freedom that, if the complex-Langevin expectation values \eqref{eq:cle_expectation-values} obey the Dyson--Schwinger equations (see \cref{sec:dse}), they are given by a linear combination of the form 
	\begin{equation}\label{eq:sase_theorem}
		\langle\obs[\Phi]\rangle_\CL = \sum_{i=1}^{n_\gamma}a_i\langle\obs[\Phi]\rangle_{\gamma_i}\;, 
	\end{equation}
	with
	\begin{equation}\label{eq:cycle_integral}
		\langle\obs[\Phi]\rangle_{\gamma_i} = \int_{\gamma_i}\mathcal{D}\Phi\obs[\Phi]\rho[\Phi]
	\end{equation}
and $a_i\in\mathbb{C}$. Here, the $\gamma_i$ denote a set of linearly independent integration cycles of the theory, which are either contours that connect two distinct zeros of $\rho[\Phi]$ or closed noncontractible loops. The total number of independent cycles is denoted as $n_\gamma$. Evidence for the validity of \eqref{eq:sase_theorem} for theories with more than one variable was found in \cite{HMS25}.
\end{itemize}

The existence of the wrong-convergence problem motivates the search for practical tools that can reliably detect incorrect results. The discussion and comparison of different correctness criteria for complex Langevin shall thus be the main goal of this work. Needless to say, discovering incorrect convergence is only one part of the story, while the other is finding ways to avoid it. Indeed, there exist countless attempts for fixing wrongly-converging complex Langevin simulations, including various reweighting attempts \cite{BS08,HHT16,Blo17}, gauge cooling \cite{SSS13,NNS16_2}, dynamical stabilization \cite{AJ19,CKT22,AJZ22u,HS25}, weight regularization \cite{BHM25}, and the use of an appropriate kernel \cite{OOS89,OSZ91,BHM23,BHM24,ALR23,LS23,ARS24}. Here, however, the main focus shall not be on how to fix wrong convergence, but on how to detect it.
		
	\section{Correctness criteria for complex Langevin}
	\label{sec:criteria}The previous section has made clear the necessity for reliable diagnostic tools for complex Langevin simulations. In particular, in practice one would like to perform measurements on some quantity, the outcome of which provides information about whether the obtained simulation results are correct or not. Over the years, a plethora of possible candidates for such a correctness criterion has emerged. Early attempts not discussed here can be found in \cite{GL93,Lee94,Gau94,Gau98}. In the following, the most prominent criteria are summarized and contrasted with one another. It is important to stress that, while some of these criteria were believed to be both necessary and sufficient to guarantee correctness at the time of their development, the true situation is different. Indeed, while most criteria really are necessary conditions for correct convergence, only one of them turns out to be sufficient in the general case as well. 

The main aim of this work is to compare different correctness criteria, to point out their insufficiencies if possible, and to contrast their practical applicability. Only such criteria are considered that can be evaluated reasonably straightforwardly, i.e., without too much numerical effort, in view of real-world applications. For instance, criteria based on the spectral properties of the associated Fokker--Planck operators are not practical for realistic systems \cite{SSS24}. It should be mentioned, however, that it was pointed out in \cite{NS15} that the usual spectral condition (i.e., that the nonzero eigenvalues of the Fokker--Planck Hamiltonian have a positive real part) is fulfilled automatically once the complex Langevin evolution converges to a unique equilibrium distribution, which was confirmed by a direct computation of the spectrum within a simple model. Hence, while this spectral condition does not guarantee correct convergence on its own, it should not be viewed as an independent practical obstacle.

Note that, while some of the correctness criteria investigated in this work are provably necessary for correct convergence, others are more heuristic in nature. It should be noted that the described criteria can be extended to gauge theories, but in the context of this work only scalar theories will be considered. Finally, the criteria will only be stated but not derived here; for derivations, the reader is referred to the original literature.

\subsection{Dyson--Schwinger equations}\label{sec:dse}
The Dyson--Schwinger equations are exact relations between the correlation functions of a theory. Their validity for a given observable $\obs$ is expressed in the context of this work as the relations \cite{GP09}
\begin{equation}\label{eq:dse}
	\langle A_i\obs[\Phi]\rangle_\CL = 0 
	\quad \textnormal{for} \quad
	 1\leq i\leq N
\end{equation}
with the Dyson--Schwinger operators
\begin{equation}\label{eq:ds_operators}
	A_i = \frac{\partial}{\partial\Phi_i} + \frac{1}{\rho[\Phi]}\frac{\partial\rho[\Phi]}{\partial\Phi_i}\;.
\end{equation}
The condition \eqref{eq:dse} is thus a trivial necessary criterion for correctness. However, as was mentioned in the previous section, it is not sufficient on its own. In particular, the Dyson--Schwinger equations are -- by definition -- insensitive to contributions from unwanted integration cycles.

\subsection{Histograms}\label{sec:histograms}
A milestone in the development of the complex Langevin approach was the realization \cite{ASS10,AJS11,NS15,ASS17} that, formally, the validity of \eqref{eq:correctness} can be proven under the assumption of a sufficiently fast decay of the probability distribution $P$ towards infinity as well as in the vicinity of poles of the drift term. If these requirements are not fulfilled, however, one is confronted with the appearance of boundary terms, spoiling the integration by parts the formal proof relies on. To be more precise, it is the fast decay of the quantity $P[\Phi_R,\Phi_I;\tau-\tau']\obs[\Phi_R+\ii\Phi_I;\tau']$ and its derivatives for all $\tau$ and $\tau'\leq\tau$ that is necessary for the proof. Here, however, $\obs[\Phi_R+\ii\Phi_I;\tau']$ denotes the Langevin-time-evolved observable $\obs$, which one usually does not have access to. A possible way out is the observation that the limiting case $\tau\to\infty$ at $\tau'=0$, which is easier to access in practice, tends to be especially problematic \cite{AJS11,SSS19}.

Thus, a possible requirement for the absence of boundary terms for a given observable $\obs$ is the sufficiently fast decay of the product $P[\Phi_R,\Phi_I]\obs[\Phi_R+\ii\Phi_I]$ and its derivatives at infinity and near poles of the drift term. Note, however, that the decay behavior will depend on the chosen observable. This means that, even if $P$ were to decrease exponentially towards infinity, its decay might be counteracted by an exponential increase of the observable in question. Hence, one should expect boundary terms to be nonzero for some observables even if they vanish for others. 

An advantage of such an analysis is that -- in principle -- each observable may be checked for correctness individually. Alternatively, in order to make more general statements, one may study the decay behavior of the equilibrium distribution $P$ on its own. An exponential decay of $P$ can be interpreted as a sign of correctness at least for polynomial observables, which shall be the main concern in the context of this work. Finally, it should be noted that it is generally expected that the decay properties towards infinity are most problematic in the imaginary directions of the complexified field space.

\subsection{Boundary terms}\label{sec:boundary_terms}
Concerning the boundary terms at infinity mentioned in the previous subsection, a practical method for their direct estimation has emerged in recent years \cite{SSS20}. It involves the measurement of additional observables after the introduction of a cutoff $Y$ on a suitably chosen norm of the fields, $\mathcal{N}[\Phi]$. Concretely, one measures the quantities
\begin{equation}\label{eq:boundary_terms}
	\mathcal{B}_{\obs[\Phi]}(Y) = \langle\Theta(Y-\mathcal{N}[\Phi])L_c\obs[\Phi]\rangle_\CL\;,
\end{equation}
with
\begin{equation}\label{eq:L_c}
	L_c = \left(\frac{\partial}{\partial\Phi_i} - \frac{\partial S[\Phi]}{\partial\phi_i}\right)H_{ij}[\Phi]H_{kj}[\Phi]\frac{\partial}{\partial\Phi_k}\;,
\end{equation}	
as a function of $Y$ and searches for plateaus that can be extrapolated to $Y\to\infty$. A plateau at a non-vanishing value of $\mathcal{B}_\obs$ for a given observable $\obs$ indicates the presence of boundary terms at infinity and thus that the complex Langevin results for that observable are incorrect. Furthermore, the absence of any such plateau is a signal for incorrect convergence as well. Regarding the norm in \eqref{eq:boundary_terms}, a safe choice is
\begin{equation}
	\mathcal{N}[\Phi] = \max_i\vert\Phi_i\vert\;,
\end{equation}
since it takes into account both the imaginary and the real directions of the fields. Indeed, this is the norm employed in this work, although other choices are possible.

The study of boundary terms at poles of the drift term is more involved. In particular, it has been argued that such boundary terms should not be visible in the equilibrium limit $\tau\to\infty$, but only at finite Langevin times \cite{ASS18e,Sei20}. Whether a situation in which boundary terms at poles are nonzero for finite Langevin times but vanish in the equilibrium limit implies wrong convergence is not entirely clear, even though it seems likely. Either way, the vanishing of all types of boundary terms is again only a necessary condition for correctness, but, as it turns out, not a sufficient one \cite{ALR23,HMS25}.

\subsection{Convergence conditions}\label{sec:convergence_conditions}
Upon taking the limit $Y\to\infty$ in \eqref{eq:boundary_terms} directly, one arrives at the so-called convergence (or consistency) conditions \cite{AJS11}
\begin{equation}\label{eq:convergence_conditions}
	\langle L_c\obs[\Phi]\rangle_\CL = 0\;,
\end{equation}
the validity of which could, again, be checked separately for every observable of interest. While \eqref{eq:convergence_conditions} is another necessary condition for correct convergence, it is automatically fulfilled once the stochastic process has equilibrated, assuming that there are no boundary terms in equilibrium \cite{AJS11}. Thus, the convergence conditions cannot be expected to be a sufficient criterion for correctness either. Moreover, they are commonly plagued by large fluctuations and the ensuing signal-to-noise issues.

\subsection{Drift criterion}\label{sec:drift_histograms}
An alternative way of looking at the proof of correctness in \cite{ASS10,AJS11} was presented in \cite{NNS16,NNS18_2}. There, by considering the complex Langevin equation with a discrete Langevin-time step-size $\eps$, it was realized that the failure of the integration by parts discussed in \cite{ASS10,AJS11} can be understood as the breakdown of the expansion in $\eps$ that describes the Langevin-time evolution due to the appearance of a large drift term. Consequently, in \cite{NNS16} a correctness condition was formulated that only depends on the decay properties of the drift term itself. In particular, one approximates the probability distribution of 
\begin{equation}\label{eq:max_drift}
	u[\Phi] = \max_i\left\vert D_i[\Phi]\right\vert\;, 
\end{equation}
where $D_i$ is the drift term defined in \eqref{eq:cle},
\begin{equation}\label{eq:drift}
	D_i[\Phi] = -H_{ij}[\Phi]H_{kj}[\Phi]\frac{\partial S[\Phi]}{\partial \Phi_k} + \frac{\partial H_{ij}[\Phi]H_{kj}[\Phi]}{\partial\Phi_k}\;,
\end{equation}
via a histogram, and studies its decay properties towards infinity. If this decay is at least exponential, one would like to conclude that the complex Langevin results are correct. Indeed, this criterion, too, is necessary for correct convergence. However, as will be demonstrated explicitly in this work, it is again not sufficient in general. 

From \cite{NNS16}, the drift criterion is expected to be essentially equivalent to the boundary-term analysis and perhaps even slightly stronger. This is because it requires the drift to decay at least exponentially, whereas the absence of boundary terms follows already from a decrease faster than any power law. It should be noted that one assumption made by both correctness criteria may be violated in practical applications, namely that the complex density $\rho$ is the unique stationary distribution of the Fokker--Planck evolution governed by $L_c^T$, with $L_c$ defined in \eqref{eq:L_c}. This assumption can be violated if $L_c^T$ has degenerate zero modes.

Note that the applicability of the two criteria is rather different: The application of the drift criterion is straightforward, only requiring the computation of histograms of a real and nonnegative quantity. In return, it allows one to make statements about the correctness of the complex Langevin evolution as a whole, i.e., it is agnostic to differences between observables. This implies that in a scenario in which certain observables come out correct while others do not, this criterion would indicate wrong convergence. Moreover, a practical disadvantage is that the drift distributions' tails typically probe low-statistics regions of the field space, which can impede one's ability to draw definite conclusions about correctness from the drift criterion alone.

The boundary-term criterion, on the other hand, is sensitive to different observables. One of its advantages is that one could test correctness for each observable separately, allowing one to discard certain observables as incorrect, while others may still be correct -- at least within the statistical uncertainties. It is, however, unlikely that such a scenario would survive the limit of infinite sample size. The downside of the boundary-term criterion is that with it making hard statements about overall correctness is not as straightforward, since that would require the measurement of boundary terms for (at least) all observables of interest, which is often cumbersome in practise.

\subsection{Observable bounds}\label{sec:observable_bounds}
To the best of current knowledge, there is only one condition for correctness for complex Langevin that has been proven to be both necessary and sufficient \cite{MSS25}. In order to introduce it, one first needs to establish some notation and make a few mild assumptions that are fulfilled in most physical systems of interest. Let $\mathcal{M}_r$ denote the field space before complexification and $\mathcal{M}_c$ its complex extension. Assume that the path integral density $\rho$ in \eqref{eq:expectation_values} can be extended to $\mathcal{M}_c$ and that it has no zeros\footnote{This is not really an assumption since one may always deform $\mathcal{M}_r$ via Cauchy's theorem in such a way that it avoids the zeros of $\rho$ \cite{MSS25}.} in $\mathcal{M}_r$. Furthermore, assume that the partial derivatives of $\log\rho$ grow at most polynomially on $\mathcal{M}_r$. 

Now, consider the space of observables $\mathfrak{A}$ whose elements are polynomials in the field components $\Phi_i$ as well as the enlarged space $\mathcal{H}=\mathfrak{A}+\sum_{i}A_i\mathfrak{A}$, with $A_i$ defined in \eqref{eq:ds_operators}. Notice that $\mathcal{H}=\mathfrak{A}$ is possible only if the complex extension of $\rho$ has no zeros in $\mathcal{M}_c$. For the definition of the theorem, one decomposes $\rho[\Phi]$ as $\rho[\Phi]=w[\Phi]\rho_r[\Phi]$, where $w$ and $\rho_r$ are assumed to be continuous, but may be otherwise chosen essentially arbitrarily apart from certain decay properties \cite{MSS25}. Then, if one defines the $p$-norms ($p\geq1$) of functions $f$ on $\mathcal{M}_r$ as
\begin{equation}\label{eq:p_norms}
	\Vert f\Vert_p = \left(\int_{\mathcal{M}_r}\mathcal{D}\Phi\vert f[\Phi]\vert ^p\right)^{1/p}\;, \qquad
	\Vert f\Vert_\infty = \sup_{\Phi\in\mathcal{M}_r}\vert f[\Phi]\vert\;,
\end{equation}
the correctness condition reads as follows: If and only if for all observables $\obs\in\mathcal{H}$
\begin{enumerate}
	\item the Dyson--Schwinger equations \eqref{eq:dse} hold and
	\item the bounds 
	\begin{equation}\label{eq:observable_bounds}
		\left\vert\langle\obs[\Phi]\rangle_\CL\right\vert\leq C^{(p)}\Vert\obs\Vert_w^{(p)}
	\end{equation}
	with 
	\begin{equation}\label{eq:observable_bounds_aux}
	C^{(p)}=\Vert\rho_r\Vert_p 
	\quad \textnormal{and} \quad 
	\Vert f\Vert_w^{(p)}=\Vert w[\Phi]f[\Phi]\Vert_{\frac{p}{p-1}}
	\end{equation}
	are satisfied,
\end{enumerate}
then the complex Langevin expectation results $\langle\obs[\Phi]\rangle_\CL$ are correct \cite{MSS25,MSS25p}. 

While the necessity and sufficiency of this condition have indeed been proven, its downside is apparent in its limited practical applicability. Needless to say, it is impossible to measure all polynomial observables, let alone all observables in the extended space $\mathcal{H}$ in practice. However, the criterion has proven successful in ruling out incorrect results \cite{MSS25,MSS25p}. After all, if the bounds in \eqref{eq:observable_bounds} are violated for even a single choice of $\obs$, $w$ and $\rho_r$, $p$, or even $\mathcal{M}_r$, the obtained complex Langevin results must necessarily be incorrect. This property is explored below. Note that the criterion, in contrast to the others, is sensitive to contributions from unwanted integration cycles. Also note that a similar (albeit different in detail) criterion has already been proposed in \cite{AJS11}.

\subsection{Unitarity norm}\label{sec:unitarity_norm}
Like all solutions to the sign problem based on the complexification of the underlying field space, the complex Langevin method defines its dynamical degrees of freedom in an enlarged space $\mathcal{M}_c$. In a simulation, one may measure the distance between the simulation trajectory and the original field manifold $\mathcal{M}_r$ via quantities usually referred to as unitarity norms, a terminology that was coined (and makes most sense) in the context of gauge theories; see, e.g., \cite{SSS13}. A common practice is to monitor the Langevin-time dependence of the unitarity norm $\mathcal{N}_U$: If $\mathcal{N}_U$ becomes `too large', the expectation is that the simulation will become unstable and, correspondingly, the complex Langevin results will be incorrect. One thus usually discards configurations whose unitarity norm becomes too large -- by hand if necessary. 

Quite obviously, one cannot force $\mathcal{N}_U$ to be too small either, as a vanishing unitarity norm would amount to a phase-quenched simulation, which is ultimately not what one is after. There is thus a certain tradeoff when it comes to the unitarity norm. This is used, for instance, in methods such as gauge cooling \cite{SSS13} or dynamical stabilization \cite{AJ19}, where one attempts to keep the unitarity norm reasonably small in order to stabilize simulations. Moreover, in recent work the unitarity norm was used as a loss function in machine-learning approaches to finding optimal stabilizing kernels for complex Langevin simulations \cite{LS23,ARS24,MSS25p}. 

While gauge theories are not discussed in this work, it would nonetheless be interesting to see if a suitably defined `unitarity norm' could allow one to draw any conclusions regarding the (in)correct convergence of simulations of scalar theories. For this work, the unitarity norm shall be defined as
\begin{equation}\label{eq:unitarity_norm}
	\mathcal{N}_U = \sum_{i=1}^N\imag(\Phi_i)^2\;.
\end{equation}

\subsection{Configurational temperature}\label{sec:configurational_temperature}
The final correctness criterion employed in this work has recently emerged and is a configuration-based thermodynamic consistency check \cite{JK25_1,JK25_2} based on the work of \cite{Rug97}. Termed the configurational thermometer, the underlying idea is to test whether or not the field configurations in a simulation are being generated with the correct statistical weight $e^{-S}$. In practice, this is done by comparing a geometric definition of temperature, which can be measured configuration-by-configuration, with an `input' temperature. If those two notions of temperature agree with one another, one concludes correct convergence. Indeed, test studies of the criterion are promising and suggest that it may also be capable of detecting algorithmic deficiencies as well as thermalization properties \cite{JK25_1,JK25_2}. Moreover, it is not restricted to complex-Langevin simulations but also finds use in conventional approaches based on importance sampling \cite{DJL25}.

In the context of lattice quantum field theory, the role of the aforementioned input temperature is not played by the physical temperature of the system (which, for instance, in the Euclidean formalism would correspond to the inverse system size in the temporal direction), but the temperature corresponding to the formal analogy between the Euclidean path integral and the canonical ensemble in statistical mechanics. Concretely, if the action $S$ is interpreted as the Hamiltonian $H$ of a statistical system, the formal equivalence between $e^{-S}$ in the Euclidean path integral and $e^{-\beta H}$ in the canonical ensemble suggests that $\beta=1$, implying that the input temperature $T=1/\beta$ one compares the configurational temperature with is unity.

In the microcanonical ensemble, the configurational temperature is defined as $\tilde{T}=1/\tilde{\beta}$, where $\tilde{\beta}$ can be written as \cite{JK25_1}
\begin{equation}\label{eq:configurational_temperature}
	\tilde{\beta} = \left\langle\frac{\partial}{\partial\Phi_i}\frac{\frac{\partial S[\Phi]}{\partial\Phi_i}}{\frac{\partial S[\Phi]}{\partial\Phi_j}\frac{\partial S[\Phi]}{\partial\Phi_j}}\right\rangle = \Bigg\langle \frac{\mathbb{H}_{ii}}{g_jg_j} - 2\frac{g_i\mathbb{H}_{ij}g_j}{(g_kg_k)^2}\Bigg\rangle\;,
\end{equation}
and where the gradient and Hessian of the action are defined as
\begin{equation}
	g_i = \frac{\partial S}{\partial\Phi_i}\;,
	\quad \textnormal{and} \quad
	\mathbb{H}_{ij} = \frac{\partial^2 S}{\partial\Phi_i\partial\Phi_j}\;,
\end{equation}
respectively. By ensemble equivalence, the expression \eqref{eq:configurational_temperature} is valid in the canonical ensemble as well, albeit only in the thermodynamic limit. However, the models considered in this work have very limited numbers of degrees of freedom and are thus far away from the thermodynamic limit. Hence, within those systems it is questionable whether the study of \eqref{eq:configurational_temperature} would allow one to make any conclusive statements about the correct convergence of complex Langevin simulations at all. Nevertheless, since the physical intuition behind this criterion makes it particularly attractive, it shall be investigated here as well.
		
	\section{Simulation setup}
	\label{sec:setup}Details regarding the simulations employed in this work can be found in \cite{HMS25}. The most important points are nonetheless summarized in the following. In accordance with the FAIR guiding principles \cite{FAIR16}, the raw data underlying the results of this work \cite{data} as well as the employed simulation code base \verb|CLaSS| (\textbf{C}omplex \textbf{La}ngevin for \textbf{S}imple \textbf{S}ystems) \cite{code} and analysis scripts \cite{scripts} are available online.

\subsection{Discrete Langevin evolution}
For this work, the complex Langevin equation \eqref{eq:cle} is solved using one of two different discrete update schemes for each theory under consideration. The simpler of the two is a straightforward Euler--Maruyama update. Denoting the field and noise variables at the $n$-th discrete Langevin-time-step as $\Phi_i^{(n)}$ and $\eta_i^{(n)}$, respectively, and the step-size as $\eps$, this update prescription reads
\begin{equation}\label{eq:euler_maruyama}
	\Phi_i^{(n+1)} = \Phi_i^{(n)} + \eps D_i[\Phi^{(n)}] + \sqrt{\eps}H_{ij}[\Phi]\eta_j^{(n)}\;,
\end{equation}
where the drift term $D_i$ was defined in \eqref{eq:drift}. The other employed update scheme is an improved one \cite{UF85}, which reads
\begin{align}\label{eq:improved_scheme}
	\begin{aligned}
		\tilde{\Phi}_i^{(n)} &= \Phi_i^{(n)} + \eps D_i[\Phi^{(n)}] + \sqrt{\eps}H_{ij}\eta_j^{(n)}\;,\\
		\tilde{D}_i[\Phi^{(n)}] &= \frac{1}{2}\left(D_i[\Phi^{(n)}]+D_i[\tilde{\Phi}^{(n)}]\right)\;, \\
		\Phi_i^{(n+1)} &= \Phi_i^{(n)} + \eps\tilde{D}_i[\Phi^{(n)}] + \sqrt{\eps}H_{ij}\eta_j^{(n)}\;,
	\end{aligned}
\end{align}
where the kernel matrix $H_{ij}$ was assumed to be independent of $\Phi$.

In the simulations using the scheme \eqref{eq:euler_maruyama}, the step-size $\eps$ is determined adaptively in such a way as to keep the product $\eps\max_i\vert D_i\vert$ within the interval
\begin{equation}\label{eq:adaptive_step_size}
	\frac{\mathcal{D}}{2}\leq\eps\max_i\vert D_i\vert\leq2\mathcal{D}\;,
\end{equation}
where $\mathcal{D}$ is some reference value one may choose at will. If this procedure were to  result in an $\eps$ larger than some predefined maximum value $\eps_{\max}$, this maximum value is chosen as the new $\eps$ instead. For the improved scheme \eqref{eq:improved_scheme} the procedure is analogous, but with $D_i$ replaced by $\tilde{D}_i$ in \eqref{eq:adaptive_step_size}. In practice, this is realized by first computing the full drift $\tilde{D}_i$, checking whether it satisfies \eqref{eq:adaptive_step_size}, and if that is not the case repeating the drift computation with an adapted step-size until convergence. Such convergence is, however, not guaranteed, which is the main reason why the scheme \eqref{eq:improved_scheme} is not employed for all simulations in the first place. In those cases where the adaptive step-size computation does not converge sufficiently fast in the improved scheme, the standard scheme \eqref{eq:euler_maruyama} is used instead. 

\subsection{Sample generation}
The numerical studies performed in this work consist of a number $N_{\mathrm{sim}}$ of simulations for every data point, which are run in parallel on a GPU using the \verb|CUDA| application programming interface \cite{cuda}. In particular, each \verb|CUDA| thread is responsible for its own simulation with randomly distributed initial values. Due to the massive parallelization capabilities of modern GPUs, $N_{\mathrm{sim}}$ can (and will) be chosen rather high, such that an attempt to store each individual simulation trajectory in its entirety would produce excessive amounts of data. This is why what is ultimately stored is only the averages of observables over all these trajectories, taken after every measurement step $\tau_{\mathrm{meas}}$ up to a maximum Langevin time $\tau_{\max}$. In the end, a Langevin-time-average over these thread-averages is performed and this entire procedure is repeated $N_{\mathrm{runs}}$ times. This gives rise to $N_{\mathrm{runs}}$ completely independent averages for each observable that one may then conveniently extract estimators from. 

The procedure just described thus includes the full generated ensemble via averages. This, however, has the disadvantage that retroactive measurements are essentially impossible. To allow for some more flexibility, a handful of simulation trajectories (written after every $\tau_{\mathrm{meas}}$) are stored as well in order to be able to perform offline measurements. This is referred to as the restricted ensemble. While it certainly comes with larger statistical uncertainties, the estimators of the full and restricted ensembles will agree for sufficiently large sample sizes. For more details on the sampling process, as well as on the computation of observables, including histograms and boundary terms, see \cite{HMS25}. 

The following simulation parameters are employed in all simulations discussed in this work: $N_{\mathrm{sim}}=2^{13}$, $N_{\mathrm{runs}}=100$, $\tau_{\max}=1000$, $\tau_{\mathrm{meas}}=0.1$, $\mathcal{D}=2\times10^{-4}$. The chosen update scheme as well as the maximum step-size $\eps_{\max}$ may differ between simulations.

	\section{Comparing correctness criteria for different models}
	\label{sec:models}In this section, the various correctness criteria for complex Langevin discussed in \cref{sec:criteria} shall be put to the test. In the following, four different models with known solutions are considered and complex Langevin simulations thereof are analyzed via each correctness criterion separately. It should be emphasized that some of the presented results have been discussed in the literature before, but are nonetheless compiled here in order to give as complete a picture as possible. The reasoning behind choosing this particular set of models is twofold: On the one hand, their simplicity makes them particularly attractive for such a detailed study. In fact, out of the four theories considered, only one describes more than a single degree of freedom. On the other hand, they already incorporate some of the major difficulties one is confronted with in simulations of more realistic theories, such as poles in the drift term, encountered, e.g., in lattice quantum chromodynamics (QCD). However, no models with compact degrees of freedom are investigated in this work, which, importantly, also excludes gauge theories from consideration. Nonetheless, the conclusions drawn here likely also apply to such theories, as well as higher-dimensional ones, without major modifications.

\subsection{One-dimensional quartic model}\label{sec:quartic_1D}
A particularly instructive toy model from the point of view of complex Langevin simulations is given by
\begin{equation}\label{eq:quartic_1D}
	S(z) = \frac{\lambda}{4}z^4\;,
\end{equation}
where $\lambda$ is a complex coupling constant with unit modulus and $z$ denotes a single complex variable. Despite (or perhaps because of) its simplicity, the model has received quite a substantial amount of attention in the complex Langevin literature \cite{OOS89,MHS24p,HMS25,MSS25,MSS25p}. Indeed, many of the difficulties of the approach can be studied (and fully cured) in this model, without the complication of having to deal with multiple degrees of freedom or gauge symmetry. In particular, it is well known that a suitably chosen constant kernel $H$ in \eqref{eq:cle} (with $N=1$) can change the convergence behavior for a given $\lambda$ from incorrect to correct \cite{OOS89}. In fact, a simple $z$-independent phase,
\begin{equation}\label{eq:quartic_1D_kernel}
	H = e^{-\ii\phi}\;, \quad \phi\in[0,\dots2\pi)\;,
\end{equation}
is sufficient for that purpose. Moreover, this choice of kernel has been shown to affect which integration cycles of \eqref{eq:quartic_1D} are sampled in a simulation \cite{HMS25}. Some insights on this matter are summarized in \cref{app:kernel_1D}. 

The following analysis considers the choice $\lambda=e^{5\ii\pi/12}$ without loss of generality. Notice that $\mathrm{Re}\,\lambda>0$, such that the integral of $\rho=e^{-S}$ over the real axis is well defined, enabling the choice $\gamma_1=\mathbb{R}$ in \eqref{eq:cycle_integral} as the integration cycle of interest. If that were not the case, one could simply redefine $\gamma_1$ via an appropriate rotation of the real axis \cite{HMS25} without affecting any of the results presented below. The phase $\phi$ in \eqref{eq:quartic_1D_kernel} is parametrized as $\phi=\frac{m}{48}\pi$ with an integer $m$. In the following, the choices $m\in\{0,1,2,4,5,17,28,29\}$ are studied in detail. Since the present setup can be straightforwardly translated to the one considered in \cite{HMS25} via the replacement $m\to m+5$ derived in \cref{app:kernel_1D}, the results of \cite{HMS25} apply here as well. From this, one expects to find incorrect convergence due to the appearance of boundary terms for $m=0$ and $m=1$, correct convergence for $m=2$, $m=4$, and $m=5$, and incorrect convergence due to unwanted integration cycles for $m=17$, $m=28$, and $m=29$.

The model \eqref{eq:quartic_1D} (with $\real\,\lambda>0$) allows for the analytical computation of monomial expectation values along the real integration cycle, giving \cite{OOS89}
\begin{align}\label{eq:quartic_1D_exact}
	\langle z^n\rangle_{\gamma_1}= 
	\begin{cases}
		\left(\frac{4}{\lambda}\right)^{\frac{n}{4}}\frac{\Gamma\left(\frac{n+1}{4}\right)}{\Gamma\left(\frac{1}{4}\right)} \quad &\textnormal{for even $n$}\\
		0 \quad &\textnormal{for odd $n$}
	\end{cases}\;.
\end{align}
In the following, these shall be referred to as the exact (or correct) solutions. Moreover, it is not difficult to see that the expectation values of the same observables, but computed along the imaginary cycle  $\gamma_2=\ii\mathbb{R}$ are related to \eqref{eq:quartic_1D_exact} via
\begin{equation}
	\langle z^n\rangle_{\gamma_2} = \ii^n\langle z^n\rangle_{\gamma_1}\;.
\end{equation}

\begin{table*}[t]
    \centering
    \renewcommand{\arraystretch}{1.2}
	\resizebox{\textwidth}{!}{
    \begin{tabular}{|c|cccc|}
    \hline
    $m$ & $\langle z\rangle_\CL$ & $\langle z^2\rangle_\CL$ & $\langle z^3\rangle_\CL$ & $\langle z^4\rangle_\CL$ \\
    \hline
	$0$ & $1(3)\times10^{-5}-2(2)\times10^{-5}\ii$ & 
		  $0.53445(1)-0.41343(2)\ii$ & 
		  $0.0005(4)-0.0003(3)\ii$ & 
		  $0.29(3)-0.91(7)\ii$ \\
	$1$ & $0(3)\times10^{-5}+0(2)\times10^{-5}\ii$ & 
		  $0.53628(1)-0.41150(1)\ii$ & 
		  $1(2)\times10^{-5}-1(5)\times10^{-5}\ii$ & 
		  $0.25881(2)-0.96589(4)\ii$ \\
	$2$ & $-1(3)\times10^{-5}+0(2)\times10^{-5}\ii$ & 
		  $0.53629(1)-0.41151(1)\ii$ & 
		  $-1(2)\times10^{-5}+3(5)\times10^{-5}\ii$ & 
		  $0.25882(1)-0.96591(4)\ii$ \\
	$4$ & $-9(3)\times10^{-5}+4(1)\times10^{-5}\ii$ & 
		  $0.53630(1)-0.41152(1)\ii$ & 
		  $-7(2)\times10^{-5}+0.00012(4)\ii$ & 
		  $0.258826(9)-0.96597(5)\ii$ \\
	$5$ & $0(3)\times10^{-5}-0(1)\times10^{-5}\ii$ &
		  $0.53628(1)-0.41150(1)\ii$ & 
		  $-1(3)\times10^{-5}+2(4)\times10^{-5}\ii$ & 
		  $0.25881(1)-0.96590(4)\ii$ \\
	$17$ & $-0.001(1)-0.0004(5)\ii$ &
		   $0.41153(3)+0.53627(2)\ii$ & 
		   $-2(5)\times10^{-5}+0(1)\times10^{-5}\ii$ & 
		   $0.25887(5)-0.96592(4)\ii$ \\
	$28$ & $-1(1)\times10^{-5}-3(3)\times10^{-5}\ii$ &
		   $-0.53631(1)+0.41153(1)\ii$ & 
		   $4(4)\times10^{-5}+2(3)\times10^{-5}\ii$ & 
		   $0.258833(9)-0.96600(4)\ii$ \\
	$29$ & $1(1)\times10^{-5}+3(4)\times10^{-5}\ii$ &
		   $-0.53630(1)+0.41152(1)\ii$ & 
		   $-4(5)\times10^{-5}-3(3)\times10^{-5}\ii$ & 
		   $0.25883(1)-0.96596(4)\ii$ \\
   	\hline
    correct & $0$ &
    		$0.536290 - 0.411509\ii$ &
    		$0$ & 
    		$0.258819 - 0.965926\ii$ \\
    \hline
    \end{tabular}}
    \caption{Comparison of expectation values $\langle z^n\rangle$ between complex-Langevin simulations of the model \eqref{eq:quartic_1D} with different kernels (given by \eqref{eq:quartic_1D_kernel} with $\phi=\frac{m}{48}\pi$) and exact results for small $n$. The parentheses indicate the statistical uncertainties rounded to their respective first significant digits.}
    \label{tab:quartic_1D_low}
\end{table*}

The simulations of \eqref{eq:quartic_1D} employ the improved update scheme \eqref{eq:improved_scheme} and a maximum discrete Langevin-time step-size of $\eps_{\max}=10^{-5}$. A comparison between complex-Langevin expectation values for different kernels and exact results for small $n$ is shown in \cref{tab:quartic_1D_low}. One finds overall agreement for the kernels with $m=0$, $1$, $2$, $4$, and $5$. For the other kernels, there is a severe discrepancy: As expected from \cite{HMS25}, for $m=28$ and $m=29$ the numerical results rather agree with $\langle z^n\rangle_{\gamma_2}$, while those for $m=17$ are consistent with a non-trivial linear combination of $\langle z^n\rangle_{\gamma_1}$ and $\langle z^n\rangle_{\gamma_2}$ given by $a_{1,2}=\frac{1}{2}(1\pm\ii)$ and $a_3=0$ in \eqref{eq:sase_theorem}. Curiously, for $m=4$ one observes statistically significant deviations from $\langle z^n\rangle_{\gamma_1}$, but only for (small) odd $n$. This, however, is a statistical outlier, as discussed in \cref{app:resolution}.

The slight deviations that remain in the results become visible only because of the immense statistics that went into their computation. In a realistic simulation, generating such a large ensemble is rarely possible and the statistical uncertainties one is confronted with are usually much larger. In other words, while the results in \cref{tab:quartic_1D_low} allow one to exclude $m=0$ as a valid kernel upon closer inspection, the same would not be possible if the error estimates were somewhat larger.

Experience has shown that an apparent agreement for small $n$ can be misleading as discrepancies may arise only for sufficiently large $n$. Therefore, another comparison, but for larger values of $n$, is shown in \cref{tab:quartic_1D_high}. For these high powers, it becomes clear that the choices $m=0$ and $m=1$ lead to incorrect results. In particular, they cause large fluctuations, making it impossible to obtain a usable signal, an effect that is enhanced by larger powers. For all other choices of kernel, however, the complex-Langevin evolution converges and the expectation values are well defined and finite. As before, however, only $m=2$, $4$, and $5$ give correct results, while for the other kernels the imaginary integration cycle makes an unwanted contribution. The discrepancies for $m=4$ observed in \cref{tab:quartic_1D_low} are absent in \cref{tab:quartic_1D_high}. Note, however, that the statistical uncertainties in the latter are much larger in comparison.

\begin{table*}[t]
    \centering
    \renewcommand{\arraystretch}{1.2}
	\resizebox{\textwidth}{!}{
    \begin{tabular}{|c|cccc|}
    \hline
    $m$ & $\langle z^{13}\rangle_\CL$ & $\langle z^{14}\rangle_\CL$ & $\langle z^{15}\rangle_\CL$ & $\langle z^{16}\rangle_\CL$ \\
    \hline
	$0$ & 
		  $3(3)\times10^{18}-3(3)\times10^{18}\ii$ & 
		  $4(3)\times10^{20}+6(6)\times10^{20}\ii$ & 
		  $-1(1)\times10^{23}+3(3)\times10^{22}\ii$ & 
		  $1(1)\times10^{24}-2(2)\times10^{25}\ii$ \\
	$1$ & 
		  $1(4)\times10^{5}+1(4)\times10^{5}\ii$ & 
		  $-6(8)\times10^{6}-8(5)\times10^{6}\ii$ & 
		  $1.2(8)\times10^{8}-1(1)\times10^{8}\ii$ & 
		  $2(2)\times10^{9}+0(1)\times10^{9}\ii$ \\
	$2$ & 
		  $-0.005(8)-0.02(1)\ii$ & 
		  $-20.38(1)+154.81(3)\ii$ & 
		  $-0.09(4)-0.04(7)\ii$ & 
		  $292.5(1)+506.6(1)\ii$ \\
	$4$ & 
		  $0.008(5)-0.01(1)\ii$ & 
		  $-20.386(4)+154.85(3)\ii$ & 
		  $-0.00(2)-0.06(6)\ii$ & 
		  $292.55(8)+506.7(1)\ii$ \\
	$5$ & 
		  $0.002(6)-0.00(1)\ii$ & 
		  $-20.384(4)+154.83(3)\ii$ & 
		  $-0.00(1)-0.01(7)\ii$ & 
		  $292.55(8)+506.7(1)\ii$ \\
	$17$ & 
		   $0.1(3)-0.7(3)\ii$ & 
		   $-156(1)-21(1)\ii$ & 
		   $-0(4)-2(4)\ii$ & 
		   $3.0(1)\times10^{2}+5.1(2)\times10^{2}\ii$ \\
	$28$ & 
		   $0.01(1)+0.005(6)\ii$ & 
		   $20.378(5)-154.83(3)\ii$ & 
		   $-0.06(6)+0.02(2)\ii$ & 
		   $292.53(8)+506.6(1)\ii$ \\
	$29$ & 
		   $0.00(1)+0.001(7)\ii$ & 
		   $20.388(3)-154.86(3)\ii$ & 
		   $-0.02(7)+0.00(1)\ii$ & 
		   $292.61(7)+506.8(1)\ii$ \\
	\hline
    correct & $0$ &
    		$-20.382 + 154.815\ii$ &
    		$0$ & 
    		$292.500 + 506.625\ii$ \\
    \hline
    \end{tabular}}
   	\caption{Same as \cref{tab:quartic_1D_low} but for larger $n$.}
    \label{tab:quartic_1D_high}
\end{table*} 

Hence, as anticipated above, the kernels considered here indeed cover three different scenarios: correct convergence, incorrect convergence due to contributions from unwanted integration cycles, and, as shall be discussed in more detail below, incorrect convergence due to slowly decaying distributions. Knowing the exact solutions, such a conclusion is not hard to draw. What is less clear, however, is whether or not the correctness criteria defined in \cref{sec:criteria} are capable of predicting this outcome as well. This question will be discussed in the remainder of this subsection.

\paragraph{Dyson--Schwinger equations}
The Dyson--Schwinger equations \eqref{eq:dse} for \eqref{eq:quartic_1D} and $\obs(z)=z^n$ read
\begin{equation}\label{eq:quartic_1D_dse}
	\langle Az^n\rangle = 0\;, \quad
	\textnormal{with} \quad
	Az^n = nz^{n-1} - \lambda z^{n+3}\;.
\end{equation}
The complex-Langevin results $\langle Az^n\rangle_\CL$ for various (rather large) $n$ and different kernel parameters $m$ are shown in \cref{tab:quartic_1D_dse}. One finds that \eqref{eq:quartic_1D_dse} is satisfied for all kernel parameters except $m=0$ and $m=1$; for the latter two, again, the complex Langevin evolution does not result in a usable signal. A similar analysis (for smaller $n$ but with analogous conclusions) was performed in \cite{MSS25}. 

\begin{table*}[t]
    \centering
    \renewcommand{\arraystretch}{1.2}
	\resizebox{\textwidth}{!}{
    \begin{tabular}{|c|cccc|}
    \hline
    $m$ & $\langle Az^{9}\rangle_\CL$ & $\langle Az^{10}\rangle_\CL$ & $\langle Az^{11}\rangle_\CL$ & $\langle Az^{12}\rangle_\CL$ \\
    \hline
	$0$ & $-5(5)\times10^{15}+3(3)\times10^{16}\ii$ &
		  $-4(4)\times10^{18}-2(2)\times10^{18}\ii$ & 
		  $5(5)\times10^{20}-5(5)\times10^{20}\ii$ & 
		  $5(5)\times10^{22}+9(9)\times10^{22}\ii$ \\
	$1$ & $-4(3)\times10^{4}-0(2)\times10^{4}\ii$ &
		  $1(4)\times10^{5}-1(5)\times10^{5}\ii$ & 
		  $-6(4)\times10^{6}+8(8)\times10^{6}\ii$ & 
		  $-2(1)\times10^{8}-8(8)\times10^{7}\ii$ \\
	$2$ & $-0.000(3)-0.000(2)\ii$ &
		  $-0.006(9)+0.008(4)\ii$ & 
		  $-0.00(2)-0.003(9)\ii$ & 
		  $0.01(4)+0.04(3)\ii$ \\
	$4$ & $0.003(3)+0.002(2)\ii$ &
		  $-0.000(7)-0.002(1)\ii$ & 
		  $0.01(2)+0.000(3)\ii$ & 
		  $-0.01(3)-0.00(2)\ii$ \\
	$5$ & $0.003(3)+0.002(2)\ii$ &
		  $-0.000(7)-0.000(1)\ii$ & 
		  $0.01(2)-0.002(2)\ii$ & 
		  $0.01(3)-0.00(2)\ii$ \\
	$17$ & $0.14(8)+0.0(1)\ii$ &
		   $-0.6(3)+0.2(3)\ii$ & 
		   $-0(1)+1(1)\ii$ & 
		   $-2(4)+0(4)\ii$ \\
	$28$ & $-0.000(3)-0.001(2)\ii$ &
		   $0.000(2)-0.004(7)\ii$ & 
		   $0.01(2)+0.001(3)\ii$ & 
		   $0.02(2)+0.02(3)\ii$ \\
	$29$ & $0.005(3)+0.003(2)\ii$ & 
		   $0.001(2)-0.006(8)\ii$ & 
		   $-0.02(2)+0.003(2)\ii$ & 
		   $0.01(2)+0.03(3)\ii$ \\
    \hline
    \end{tabular}}
    \caption{Validity of the Dyson--Schwinger equations \eqref{eq:quartic_1D_dse} in complex-Langevin simulations of the model \eqref{eq:quartic_1D} for various values of $n$ and  different kernels (given by \eqref{eq:quartic_1D_kernel} with $\phi=\frac{m}{48}\pi$). The parentheses indicate the statistical uncertainties rounded to their respective first significant digits.}
   \label{tab:quartic_1D_dse}
\end{table*}

Notice that the signal-to-noise issue is rather intricate. While $m=2$, $4$, $5$, $28$, and $29$ all show statistical uncertainties of the same order of magnitude for given $n$, the errors become successively larger for $m=17$, $1$, and $0$. This behavior is also seen in \cref{tab:quartic_1D_high} and, strictly speaking, the expectation values $\langle Az^n\rangle_{\CL}$ are consistent with zero for all $n$ and $m$ shown in \cref{tab:quartic_1D_dse} within their respective uncertainties. This, however, implies that one would have to define a certain threshold on the estimators of $\langle Az^n\rangle_{\CL}$, beyond which one considers the results unacceptable and the Dyson--Schwinger equations no longer well defined (and, thus, violated). While it is reasonable to choose this threshold such that it excludes $m=0$ and $m=1$, for $m=17$ the situation is less clear. Depending on this definition, one might be able to predict incorrect convergence for $m=17$ from \cref{tab:quartic_1D_dse} as well.

Either way, the conclusion that can be drawn from \cref{tab:quartic_1D_dse} is that the Dyson--Schwinger equations are satisfied for various kernels even though only a subset of them produce correct results. This demonstrates that the conditions \eqref{eq:dse} on their own do not constitute a reliable correctness criterion, which confirms the theorem \eqref{eq:sase_theorem}, stating that the validity of the Dyson--Schwinger equations merely implies that a linear combination of integration cycles contributes to the complex Langevin dynamics. No such statement can be made for $m=0$ and $m=1$ (and, possibly, $m=17$), where the Dyson--Schwinger equations are affected by strong fluctuations.

\begin{figure}[t]
	\centering
	\includegraphics[scale=0.6]{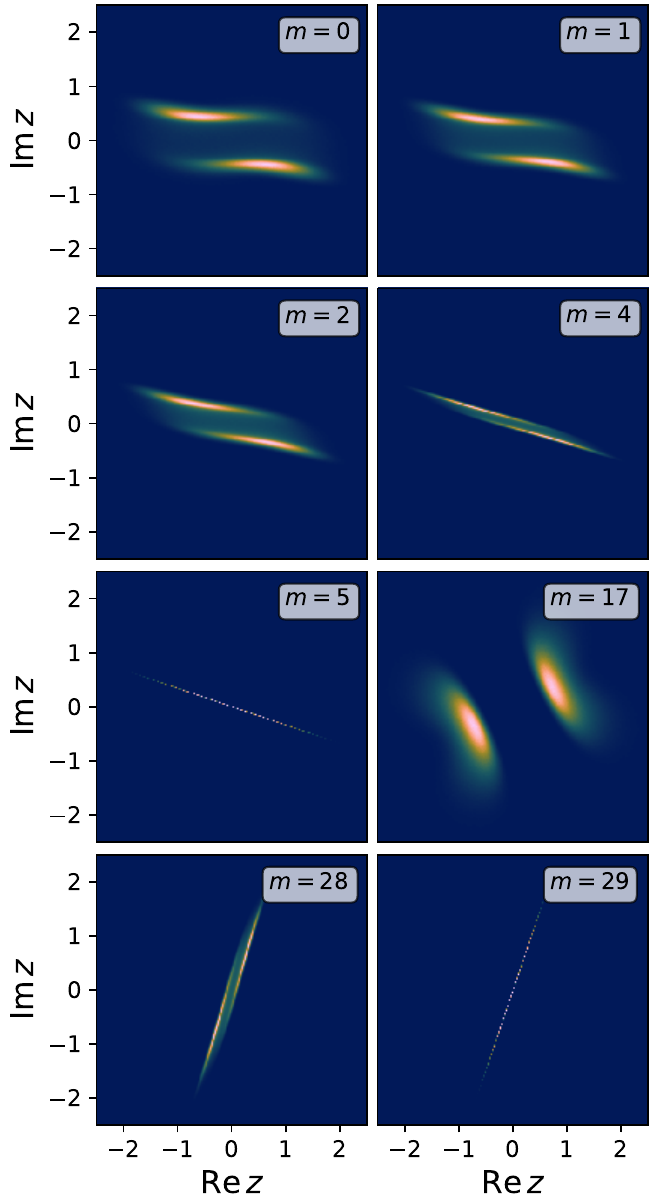}
	\caption{Histograms in the complex plane resulting from complex Langevin simulations of the model \eqref{eq:quartic_1D} for different kernels (given by \eqref{eq:quartic_1D_kernel} with $\phi=\frac{m}{48}\pi$). Brighter regions indicate a higher probability. The plots show sums over multiple independent runs.}
	\label{fig:quartic_1D_histogram_2D}
\end{figure}

\paragraph{Histograms}
Another straightforward analysis concerns histograms of the dynamical degrees of freedom in the complex plane, as they approximate the distribution $P$ for sufficiently large samples. Histograms of $z$ for the different choices of kernel are shown in \cref{fig:quartic_1D_histogram_2D}. A few observations, already made in \cite{HMS25}, should be mentioned. First of all, the distributions for $m=4$ and $m=28$, and those for $m=5$ and $m=29$, are perfect $90$-degree rotations of one another. Their respective decay behavior, thus, must be equivalent (up to rotations in the complex plane). Second, for $m=17$ the distribution consists of two disconnected parts. This is due to a violation of ergodicity, giving rise to different equilibrium distributions for different initial conditions. The plots in \cref{fig:quartic_1D_histogram_2D} are obtained by summing over multiple runs, which is why both equilibrium distributions are visible here. Finally, the distributions for $m=0$, $1$, and $2$ only deviate slightly from one another. Nonetheless, only the latter gives rise to correct results.

\begin{figure}[t]
	\centering
	\includegraphics[scale=0.5]{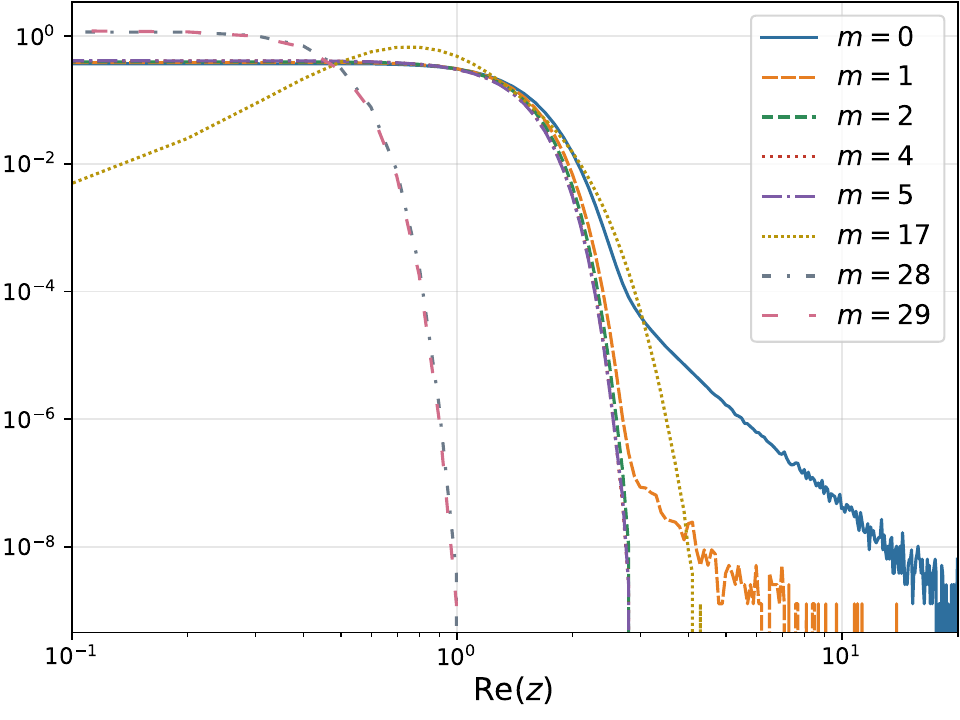}
	\caption{Projection of the data in \cref{fig:quartic_1D_histogram_2D} onto the real axis via integrating over the imaginary axis. Both axes are logarithmic. An analogous projection onto the imaginary axis looks similar.}
	\label{fig:quartic_1D_histogram_1D}
\end{figure}

One cannot expect two-dimensional visualizations like \cref{fig:quartic_1D_histogram_2D} to allow one to make definite conclusions concerning the decay behavior of the distributions. For this reason, a projection of the two-dimensional histograms onto the real axis (i.e., the integral of the distribution over the imaginary axis) is shown in \cref{fig:quartic_1D_histogram_1D}. From this figure, it is evident that the decay behavior for $m=0$ and $m=1$ differs substantially from that for the other kernels. Indeed, for the other values of $m$ the decay is clearly faster than algebraic, while for $m=0$ and $m=1$ the same cannot be said. By the conventional argument for correctness \cite{ASS10,AJS11}, one is thus led to conclude that $m=0$ and $m=1$ produce incorrect results, which is true according to \cref{tab:quartic_1D_low,tab:quartic_1D_high}. For $m=17$, $28$, and $29$, however, the decay is faster than polynomial as well, wrongly indicating correct convergence. This confirms that the histogram analysis is insensitive to contributions from unwanted integration cycles in the model \eqref{eq:quartic_1D}.

\begin{figure}[t]
	\centering
	\includegraphics[scale=0.6]{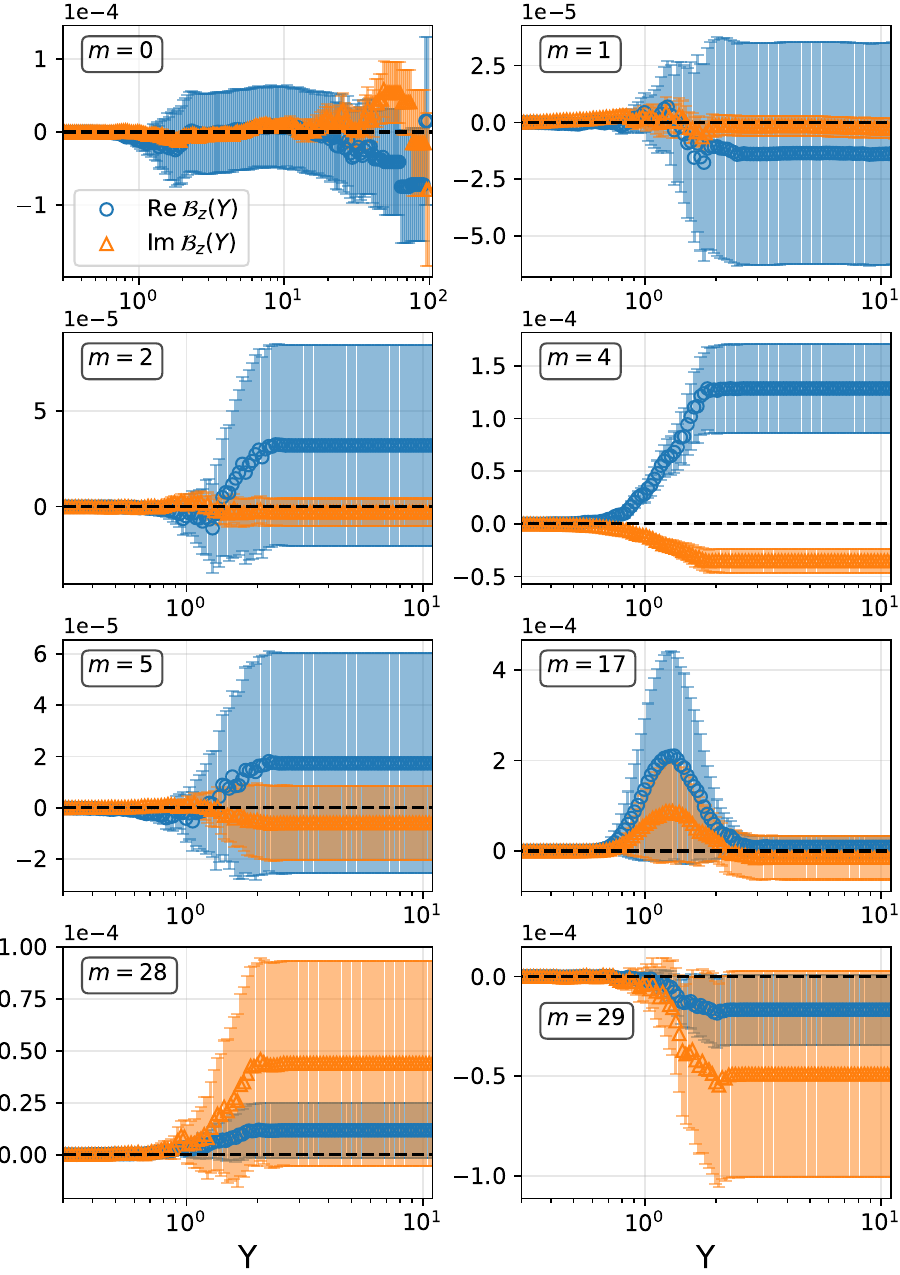}
	\caption{Boundary-term observable \eqref{eq:quartic_1D_boundary_term} for $\obs(z)=z$ in complex Langevin simulations of the model \eqref{eq:quartic_1D} as a function of the cutoff $Y$ for different kernels (given by \eqref{eq:quartic_1D_kernel} with $\phi=\frac{m}{48}\pi$). The horizontal axes are logarithmic, the data points for $\mathrm{Im}\,B_z(Y)$ are slightly offset horizontally for better visibility, and the dashed horizontal lines mark zero. Note the different scales on the axes.}
	\label{fig:quartic_1D_boundary_terms_1}
\end{figure}

\begin{figure}[t]
	\centering
	\includegraphics[scale=0.6]{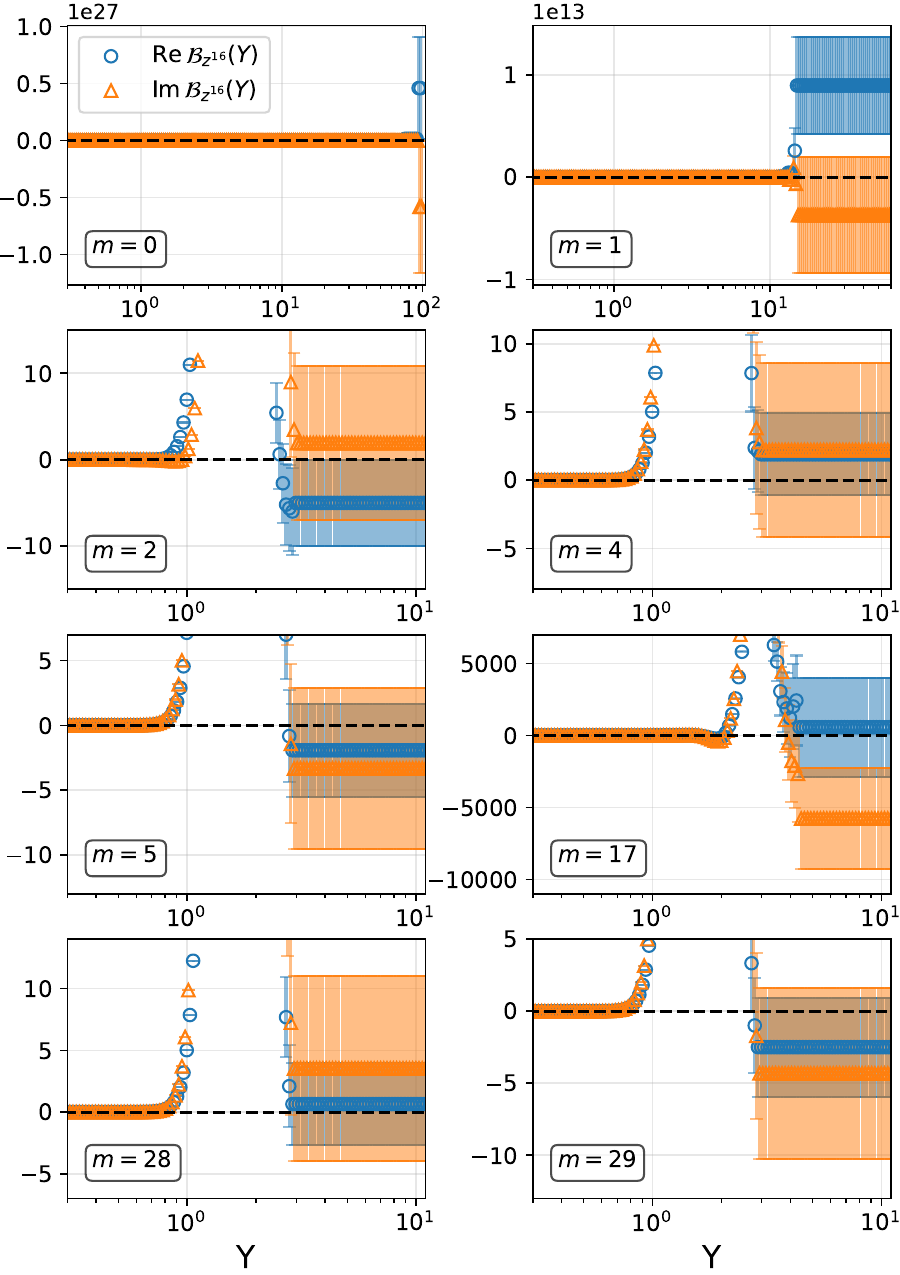}
	\caption{Similar as \cref{fig:quartic_1D_boundary_terms_16} but for $\obs(z)=z^{16}$.}
	\label{fig:quartic_1D_boundary_terms_16}
\end{figure}

\paragraph{Boundary terms}
The next criterion is based on the boundary-term observables in \eqref{eq:boundary_terms}. For $\obs(z)=z^n$, one finds
\begin{equation}\label{eq:quartic_1D_boundary_term}
	\mathcal{B}_{z^n}(Y) = nH^2\left\langle\Theta(Y-\vert z\vert)\left((n-1)z^{n-2}-\lambda z^{n+2}\right)\right\rangle\;.
\end{equation}
In principle, one would have to investigate the dependence of this quantity on the cutoff $Y$ for each observable separately. Here, only one the lowest and highest powers measured, $n=1$ and $n=16$, are considered and their boundary-term observables are shown in \cref{fig:quartic_1D_boundary_terms_1,fig:quartic_1D_boundary_terms_16}, respectively. 

\Cref{fig:quartic_1D_boundary_terms_1} reveals three different behaviors: For $m=1$, $2$, $5$, $17$, $28$, and $29$, the boundary term observable exhibits a plateau consistent with zero, indicating that boundary terms at infinity are absent and the formal proof of correctness \cite{ASS10,AJS11} applies. In fact, this plateau persists in the limit $Y\to\infty$, which means that all contributions to $\mathcal{B}_z$ are included already for moderate values of the cutoff.  For $m=0$, on the other hand, the situation is different. While there are signs of a plateau consistent with zero for intermediate values of $Y$, it is far less pronounced than for the other kernels and disappears again upon increasing $Y$. For very large cutoffs, $Y\geq100$, one finds another plateau, which is now stable as $Y\to\infty$. Thus, no definite conclusions can be made for this kernel. Finally, and perhaps most curiously, for $m=4$ the small deviation from correct results observed in \cref{tab:quartic_1D_low} is reflected also on the level of the boundary-term observables, which show a plateau, but at a value different from zero. The boundary-term analysis thus appears to be capable of resolving this minor discrepancy, which, for example, could not be seen in the histogram criterion above; recall, however, the results of \cref{app:resolution}.

\Cref{fig:quartic_1D_boundary_terms_16} shows a rather different picture. There, it becomes quite clear that $m=0$ and $m=1$ cannot give correct results due to the large fluctuations observed. For most other kernels, however, the boundary-term observable assumes a plateau at a finite value consistent with zero, the exception being $m=17$; the fluctuations for this kernel lie somewhat between the convergent and divergent ones, complicating the analysis. Crucially, though, the boundary-term criterion cannot detect incorrectness for $m=28$ and $m=29$, similar to the histogram criterion.

\paragraph{Convergence conditions}
The convergence conditions in \eqref{eq:convergence_conditions} are obtained from the boundary term observables \eqref{eq:quartic_1D_boundary_term} in the limit $Y\to\infty$. From the behavior of \cref{fig:quartic_1D_boundary_terms_1,fig:quartic_1D_boundary_terms_16} at large $Y$, one concludes that the convergence conditions are fulfilled in all cases, except for $\obs(z)=z$ and $m=4$. Note, however, that the strong fluctuations observed for $\obs(z)=z^{16}$ and $m=0$, $1$, and $17$ again require some more care. While boundary terms are known to be susceptible to signal-to-noise issues for large values of the cutoff \cite{SSS20}, a diverging value for $\langle L_c\obs\rangle$ should still be interpreted as incorrect convergence. Hence, for the cases studied here, the convergence conditions are equivalent to the boundary-term analysis. Note that this is not always the case; see, e.g., \cite{SSS20} and \cref{sec:quartic_4D} below.

\begin{figure}[t]
	\centering
	\includegraphics[scale=0.5]{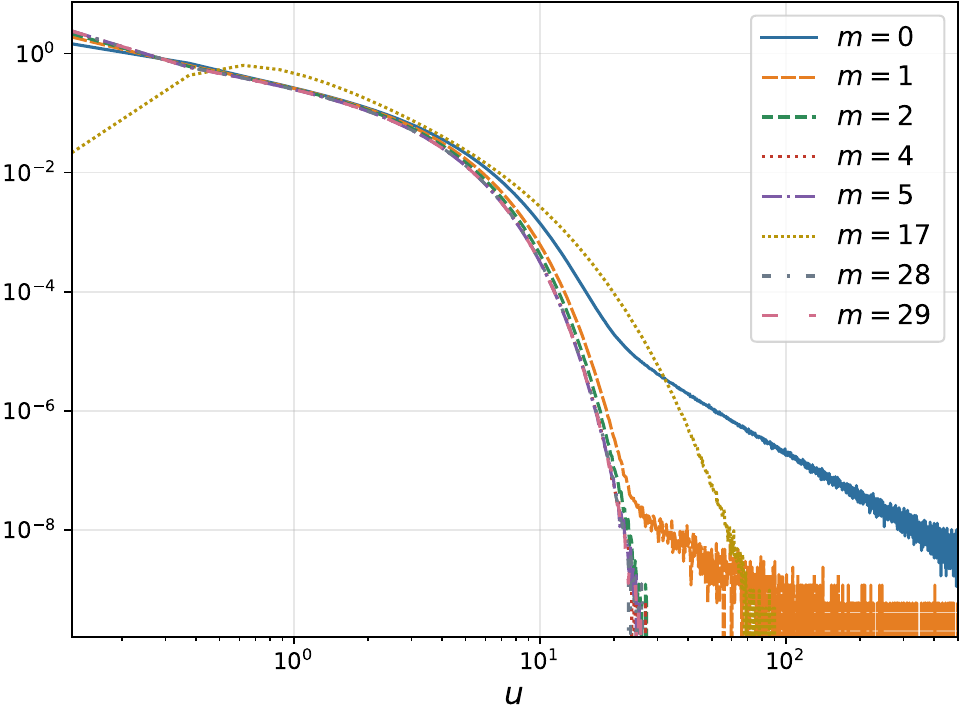}
	\caption{Distribution of the magnitude $u=\vert D\vert$ of the drift term $D$ in \eqref{eq:drift} of the model \eqref{eq:quartic_1D} in complex Langevin simulations with different kernels (given by \eqref{eq:quartic_1D_kernel} with $\phi=\frac{m}{48}\pi$). Both axes are logarithmic.}
	\label{fig:quartic_1D_drift}
\end{figure}

\paragraph{Drift criterion}
Histograms of the drift term magnitude $u(z)=\vert D(z)\vert=\vert z^3\vert$ (note that $\vert H\vert=1$), for the different kernels are shown in \cref{fig:quartic_1D_drift}. Similarly to \cref{fig:quartic_1D_histogram_1D}, one observes a clear difference between $m=0$, $1$, and all other kernels. For the former two, the decay behavior is certainly not exponential, from which one would conclude that results must be incorrect. For $m=17$, the behavior is largely consistent with an exponential (or faster) decay, even though there are traces of an algebraic tail in the distribution. For this kernel, the situation is thus not entirely clear from studying the drift distribution, akin to the previous correctness criteria. It should be mentioned, however, that the distribution for $m=17$ exhibits better qualitative agreement with $m=2$, $4$, $5$, $28$, and $29$, from which one could (erroneously) conclude that it leads to correct results. For $m=28$ and $29$ the exponential decay is more evident. Thus, while the drift criterion can detect incorrect convergence due to slow decay, it cannot detect contributions from unwanted integration cycles in the model \eqref{eq:quartic_1D}. In  \cref{app:kernel_1D}, it is argued why this must be the case, at least for $m=28$ and $29$.

\paragraph{Observable bounds}
In \cite{MSS25,MSS25p}, it was demonstrated that the criterion based on the bounds \eqref{eq:observable_bounds} is capable of detecting incorrect convergence due to unwanted integration cycles -- assuming an appropriately chosen control observable $\obs$. Here, this analysis shall be extended. In particular, the predictive power of the criterion for different control observables, norms (i.e., $p$ in \eqref{eq:observable_bounds}), choices of $w$ and $\rho_r$ in \eqref{eq:observable_bounds_aux}, and deformations $\mathcal{M}_r$ of the real integration cycle in \eqref{eq:p_norms} is still poorly understood. One should stress once more that this criterion can only make conclusions about overall (non-)convergence, i.e., it is not sensitive to individual observables, in contrast to, e.g., the boundary-term criterion.

Since the first part of the observable-bound criterion requires the validity of the Dyson--Schwinger equations, the findings of \cref{tab:quartic_1D_dse} that the kernel parameters $m=0$ and $m=1$ give incorrect results carry over directly, such that one may focus on the remaining ones. Another thing to note is that for the model \eqref{eq:quartic_1D} the two spaces of observables $\mathfrak{A}$ and $\mathcal{H}$ are identical, such that one must restrict to polynomial observables. In \cite{MSS25}, it was discovered that the polynomial
\begin{equation}\label{eq:control_observable}
	\obs(z) = \frac{\lambda}{4}z^4 - \frac{\sqrt{\lambda}}{2}z^2
\end{equation}
is a suitable control observable in the model \eqref{eq:quartic_1D}, capable of detecting incorrect convergence via the violation of the bounds \eqref{eq:observable_bounds}. However, \cite{MSS25} made use of only part of the freedom one has when applying the observable-bound criterion, considering only \eqref{eq:control_observable} as a control observable, one value of $p$ (namely $p=1$), one integration contour $\mathcal{M}_r$, and a single decomposition of $\rho$ into $\rho_r$ and $w$. While different values of $p$ were investigated in \cite{MSS25p}, here, these studies shall be extended to explore this freedom in some more detail. It should be stressed that one could at best hope to gain some heuristic insight by such an investigation; the optimal choice of bound will -- in general -- strongly depend on the system under study. Moreover, it is worth noting once more that at the time of writing no workable guiding principle exists for choosing the above parameters in an optimal way. 

\begin{figure}[t]
	\centering	
	\includegraphics[scale=0.5]{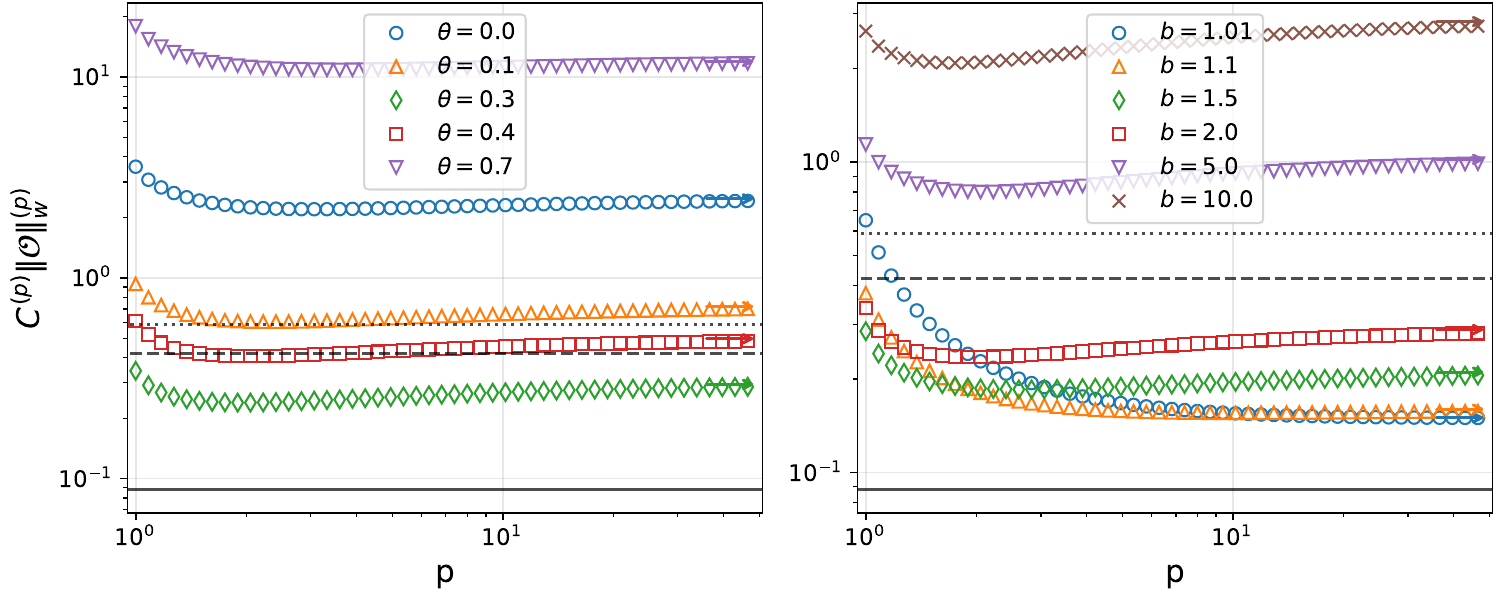}
	\caption{Observable bounds \eqref{eq:observable_bounds} in the model \eqref{eq:quartic_1D} as a function of the parameter $p$. The chosen control observable is \eqref{eq:control_observable} and the arrows indicate the respective $p\to\infty$ limits. The solid, dashed and dotted lines indicate the left-hand side of \eqref{eq:observable_bounds} for $m=2$, $4$, $5$, $m=17$, and $m=28$, $29$, respectively. The corresponding uncertainties are negligible. All axes are logarithmic. \emph{(left)}: Different integration contours $\mathcal{M}_r$ resulting from a rotation of the real axis via \eqref{eq:tilted_contour} with $b=2.0$ in \eqref{eq:rho_decomposition_2}. \emph{(right)}: Different decompositions \eqref{eq:rho_decomposition_2} with $\theta=\frac{\ln\lambda}{4\ii}$ in \eqref{eq:tilted_contour}.}
	\label{fig:quartic_1D_bounds_vs_p}
\end{figure}

The observable \eqref{eq:control_observable} shall serve as the starting point for the subsequent analysis. Moreover, a straightforward decomposition of $\rho$ is given by
\begin{equation}\label{eq:rho_decomposition_1}
	w(z) = e^{-\frac{S}{2}}\;, \quad 
	\rho_r(z) = \frac{e^{-\frac{S}{2}}}{Z}\;.
\end{equation}
The first goal shall be to use \eqref{eq:control_observable} and \eqref{eq:rho_decomposition_1} to study different contours $\mathcal{M}_r$ and values of $p$. For instance, one may parametrize the integration contour as
\begin{equation}\label{eq:tilted_contour}
	\mathcal{M}_r(\theta) = e^{-\ii\theta}\mathbb{R}\;, \quad \theta\in[0,2\pi)\;,
\end{equation}
with the requirement that $\mathrm{Re}(\lambda e^{-4\ii\theta})>0$ to ensure convergence of the integral in \eqref{eq:p_norms}. The $p$-dependence of the right-hand side of \eqref{eq:observable_bounds} for different values of $\theta$ is shown in \cref{fig:quartic_1D_bounds_vs_p} (left). A first observation that can be made is that all curves assume a minimum at some value of $p$, which, however, has a mild dependence on $\theta$. For larger $p$,  $C^{(p)}\Vert\obs\Vert_w^{(p)}$ converges to its $p\to\infty$ limit monotonically. The lowest bound in \cref{fig:quartic_1D_bounds_vs_p} (left) is achieved by the choice $\theta=0.3$, but it turns out that an even lower bound can be found using $\theta=\frac{\ln\lambda}{4\ii}\approx0.327$, for which $\lambda e^{-4\ii\theta}=1$. On such a contour, the action $S$ is entirely real, which might hint at what could be optimal contours $\mathcal{M}_r$ in \eqref{eq:p_norms} in more general scenarios. This is also the choice of contour employed in the previous studies \cite{MSS25,MSS25p}. In the remainder of this subsection, $\theta=\frac{\ln\lambda}{4\ii}$ shall be employed exclusively, as it was found to produce the lowest bounds in every case investigated.

Another modification one may test is the use of different decompositions of $\rho$. Instead of \eqref{eq:rho_decomposition_1}, consider, for instance, the parametrization
\begin{equation}\label{eq:rho_decomposition_2}
	w(z) = e^{-\frac{S}{b}}\;, \quad 
	\rho_r(z) = \frac{e^{-\frac{S(b-1)}{b}}}{Z}\;,
\end{equation}
depending on a real variable $b>1$. The $p$-dependence of the bounds \eqref{eq:observable_bounds} for different choices of $b$ is shown in \cref{fig:quartic_1D_bounds_vs_p} (right). It is seen that, quite generally, the bounds decrease with decreasing $b$. While the curves still attain a minimum at a finite value of $p$, for low $b$ they become increasingly flat. In particular, for the smallest $b$ considered ($b=1.01$), the minimum is well approximated by its $p\to\infty$ limit. Indeed, this choice gives a lower bound ($C^{(\infty)}\Vert\obs\Vert_w^{(\infty)}\approx0.15$) than $b=2$ ($C^{(p)}\Vert\obs\Vert_w^{(p)}\approx0.23$ at the minimum) considered above and in \cite{MSS25,MSS25p}. For comparison, the left-hand side of \eqref{eq:observable_bounds} reads
\begin{equation}
	\vert\langle\obs\rangle_\CL\vert\approx
	\begin{cases}
		0.09 \quad \textnormal{for} \quad m=2, 4\ \textnormal{and}\ 5\;,\\
		0.42 \quad \textnormal{for} \quad m=17\;, \\
		0.59
		\quad \textnormal{for} \quad m=28 \ \textnormal{and} \ 29\;.
	\end{cases}
\end{equation}
These values are indicated by the horizontal lines in \cref{fig:quartic_1D_bounds_vs_p}. Both of the aforementioned bounds thus rule out $m=17$, $28$, and $29$. Note, however, that not all the bounds in \cref{fig:quartic_1D_bounds_vs_p} are capable of excluding these kernels. This makes clear that the appropriate choice of $\mathcal{M}_r$, $w$, and $\rho_r$ is crucial. In practice, it is of course desirable to find as low a bound as possible for a given observable.

Finally, one may study the bounds resulting from different control observables. It should be mentioned that the qualitative behavior seen in \cref{fig:quartic_1D_bounds_vs_p} is the same for all observables considered, even though the values of the left- and right-hand sides of \eqref{eq:observable_bounds} are different for different $\obs$. This leads one to conclude that the best bound within the parametrizations considered should be given by $\theta=\frac{\ln\lambda}{4\ii}$ in \eqref{eq:tilted_contour} and $b=1.01$ in \eqref{eq:rho_decomposition_2}. For this choice, a comparison between the left- and right-hand sides of \eqref{eq:observable_bounds} for different observables and kernels is shown in \cref{tab:bound_observables}. As can be seen, out of the nine different observables considered, four are capable of excluding $m=17$, $28$, and $29$ as valid kernels, while the others are not. Also note that whenever $m=17$ is excluded, then so are $m=28$ and $m=29$. The chosen control observable in \eqref{eq:observable_bounds} thus plays an essential role in the analysis.

\begin{table*}[t]
    \centering
    \begin{tabular}{|c|c|ccc|}
    \hline
    \multirow{2}{*}{$\obs(z)$} & \multirow{2}{*}{$C^{(\infty)}\Vert\obs\Vert^{\infty}$} & & $\vert\langle\obs(z)\rangle_\CL\vert$ & \\
    & & $m=2$, $4$, $5$ & $m=17$ & $m=28$, $29$ \\
    \hline
    $\frac{\lambda}{4}z^4-\frac{\sqrt{\lambda}}{2}z^2$ & 
    	$0.150$ & $0.088$ & $0.420$ & $0.588$ \\
    $\frac{\lambda}{4}z^4+\frac{\sqrt{\lambda}}{2}z^2$ & 
    	$0.594$ & $0.588$ & $0.420$ & $0.088$ \\
    $\frac{1}{4}z^4-\frac{1}{2}z^2$ & 
    	$0.253$ & $0.207$ & $0.529$ & $0.558$ \\
    $z^8-z^6$ & 
    	$3.798$ & $3.609$ & $6.440$ & $6.724$ \\
    $z^8+z^6$ & 
    	$6.877$ & $6.723$ & $4.096$ & $3.609$ \\
    $\frac{\lambda}{4}z^4-\frac{\sqrt{\lambda}}{2}z^2+z$ & 
    	$0.699$ & $0.088$ & $0.420$ & $0.588$ \\
    $\frac{\lambda}{4}z^4-\frac{\sqrt{\lambda}}{2}z^2-z$ & 
    	$0.699$ & $0.088$ & $0.421$ & $0.588$ \\
    $z^6+\frac{\lambda}{4}z^4-\frac{\sqrt{\lambda}}{2}z^2$ & 
    	$2.127$ & $2.063$ & $2.398$ & $2.318$ \\
    $-z^6+\frac{\lambda}{4}z^4-\frac{\sqrt{\lambda}}{2}z^2$ & 
    	$2.089$ & $1.996$ & $1.682$ & $1.883$ \\
    \hline
    \end{tabular}
   	\caption{Comparison between the left- and right-hand sides of \eqref{eq:observable_bounds} for different observables $\obs$. The first column defines the observables and the second one gives the bounds, i.e., the right-hand sides of \eqref{eq:observable_bounds} employing the `optimal' choices for $p$, $w$, $\rho_r$, and $\mathcal{M}_r$ mentioned in the main text. The remaining three columns show the left-hand sides of \eqref{eq:observable_bounds} for different choices of kernel, which are split into three groups between which results differ. No statistical uncertainties are given here, as they would affect at most the last respective shown digit.}
    \label{tab:bound_observables}
\end{table*}

\begin{figure}[t]
	\centering
	\includegraphics[scale=0.5]{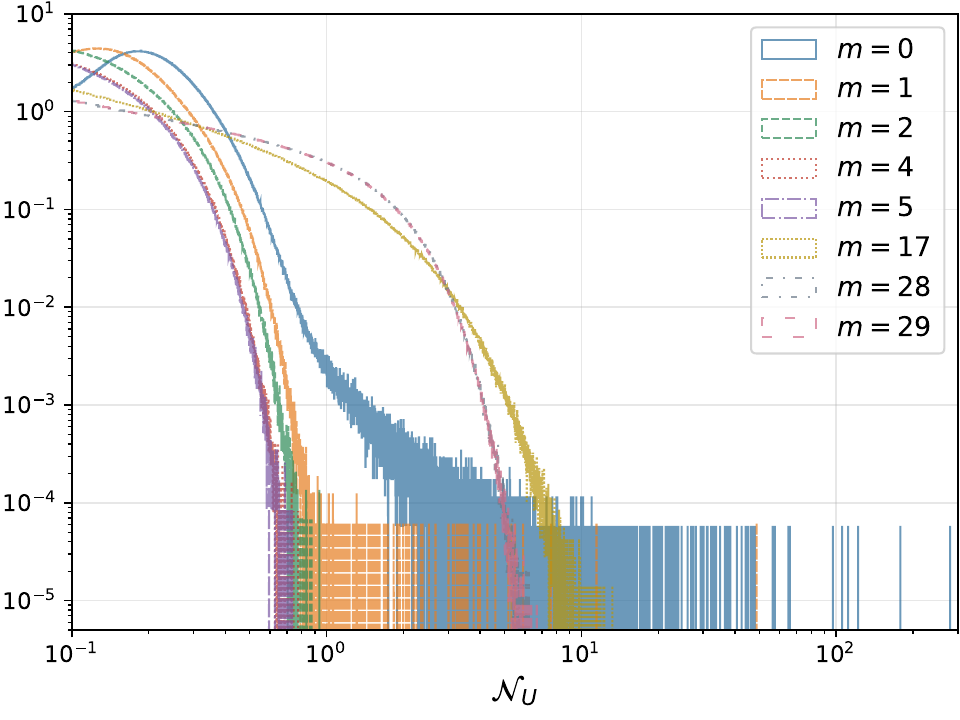}
	\caption{Histograms of $\mathcal{N}_U=(\imag z)^2$ in complex Langevin simulations of the model \eqref{eq:quartic_1D} with different kernels  (given by \eqref{eq:quartic_1D_kernel} with $\phi=\frac{m}{48}\pi$). Only the restricted ensemble is used. Both axes are logarithmic.}
	\label{fig:quartic_1D_unitarity_norm}
\end{figure}

\paragraph{Unitarity norm}
The next correctness criterion is a more heuristic one. It is based on the `unitarity norm' \eqref{eq:unitarity_norm}, given by the square of $\imag z$. The naive expectation is that if this quantity grows too large over the course of a simulation, the latter samples irrelevant regions of the complexified field space and thus produces incorrect results -- at least this is the case in the context of gauge theories \cite{Sex19}. In the present subsection, the appearance of large unitarity norms in the model \eqref{eq:quartic_1D} shall be investigated. To quantify this appearance, histograms of $(\imag z)^2$ are computed, albeit only on the restricted ensemble. The results are shown in \cref{fig:quartic_1D_unitarity_norm}.

One observes the largest unitarity norms for $m=0$ and (taking outliers into account) $m=1$. Note that in those two cases the decay of the histograms towards infinity is also much slower than exponential, in contrast to the other kernels, even though an exponential decay of $\mathcal{N}_U$ itself is not required by any known theorem. Moreover, one observes the presence of large unitarity norms for $m=17$, $28$, and $29$ as well. Thus, perhaps surprisingly, this criterion appears to be perfectly capable of ruling out the kernels leading to incorrect results with very little computational effort. Still, it is not clear a priori which values of $\mathcal{N}_U$ should be considered `too large'. From \cref{fig:quartic_1D_unitarity_norm}, it would appear that the presence of unitarity norms larger than $1$ could already indicate incorrect convergence.

\paragraph{Configurational temperature}
The final correctness criterion to be investigated in the context of the model \eqref{eq:quartic_1D} is based on the (inverse) configurational temperature defined in \eqref{eq:configurational_temperature}. For \eqref{eq:quartic_1D}, it reads
\begin{equation}\label{eq:quartic_1D_configurational_temperature}
	\tilde{\beta} = 
	\left\langle-\frac{3}{\lambda z^4}\right\rangle = 
	-\frac{3}{4}\left\langle\frac{1}{S(z)}\right\rangle\;,
\end{equation}
with $S(z)$ denoting the action in \eqref{eq:quartic_1D}. This observable has a singularity at $z=0$, implying that the evaluation of \eqref{eq:quartic_1D_configurational_temperature} according to \eqref{eq:cycle_integral} with $\gamma_i$ being the real (or imaginary) axis, diverges. Thus, it is questionable whether the configurational-temperature criterion is even applicable in the case at hand, since the complex Langevin simulations for all considered kernels apart from $m=17$ sample the point $z=0$ (c.f. \cref{fig:quartic_1D_histogram_2D}).

Indeed, the observable $\tilde{\beta}$ is found to diverge in complex Langevin simulations with all employed kernels apart from $m=17$. For the latter kernel, $\tilde{\beta}_\CL$ is consistent with $1$. Curiously, this would imply correct convergence according to the criterion, which, however, is known not to be true from \cref{tab:quartic_1D_low,tab:quartic_1D_high}. The fact that one even measures a finite value with this kernel is solely due to the ergodicity problem arising because the trajectories are being pushed away from the origin. Moreover, the observations that the configurational temperature gives a false positive result for $m=17$ and that $\tilde{\beta}$ is not a well-defined finite observable for the other choices of kernel lead one to conclude that the criterion cannot be applied reliably within the model \eqref{eq:quartic_1D}. However, this is not surprising, as a one-variable model can certainly not be expected to satisfy the assumption of a thermodynamic limit underlying the configurational-temperature criterion.

\paragraph{Summary}
This concludes the analysis of the different correctness criteria of \cref{sec:criteria} in the context of the one-dimensional quartic model \eqref{eq:quartic_1D}. To summarize, it can be said that all criteria apart from the configurational temperature are capable of distinguishing correct convergence from incorrect convergence due to slowly-decaying distributions, i.e., boundary terms. However, no criterion except the observable bounds \eqref{eq:observable_bounds} and (under certain assumptions) the unitarity norm \eqref{eq:unitarity_norm} can detect incorrectness due to unwanted integration cycles. Whether or not these conclusions are also true when considering different models shall be the subject of discussion in the subsections that follow.

\subsection{One-pole model}\label{sec:one_pole}
The next theory to be investigated is the so-called one-pole model, defined by the density
\begin{equation}\label{eq:one_pole}
	\rho(z) = \frac{1}{Z}(z-z_0)^{n_p}e^{-\beta z^2}\;, \quad Z = \int dz (z-z_0)^{n_p}e^{-\beta z^2}\;,
\end{equation}
where $z_0$ is a complex number, $n_p>0$ is an integer and $\beta$ is a real parameter. The zero of $\rho(z)$ at $z=z_0$ corresponds to a pole in the drift term (hence the name). The model \eqref{eq:one_pole} and variants thereof have been studied extensively in the complex-Langevin literature \cite{Sal93,FOS94,NS15,ASS17,SS19,MSS25} and it is particularly interesting in the context of QCD, as it captures in a simple way some of the properties of the meromorphic QCD drift term \cite{ASS17}. 

In this work, the parameters $z_0$ and $n_p$ are fixed to $z_0=\ii$ and $n_p=2$, respectively, leaving $\beta$ as the only free parameter. For this setting, it was observed in \cite{ASS17} that complex Langevin dynamics (without a kernel) produce expectation values that agree with exact solutions as long as $\beta$ is sufficiently large ($\beta\gtrsim3.2$). In contrast, severe discrepancies were found for small $\beta$ $(\beta\leq1.6$). These findings were explained by the distributions of $z$ forming strips in the complex plane, bounding $\imag\,z$: For sufficiently large $\beta$, the drift-term pole lies outside of these strips, whereas the same cannot be said for, e.g., $\beta=1.6$ \cite{ASS17}. In this subsection, the model shall be revisited, previous investigations partially extended and additional analyses performed. As in \cref{sec:quartic_1D}, the main goal is to test the various correctness criteria of \cref{sec:criteria} with respect to their ability to predict the onset of incorrect convergence for small values of $\beta$. 

The simulations of \eqref{eq:one_pole} employ the naive update step \eqref{eq:euler_maruyama} and a maximum Langevin step size of $\eps_\mathrm{max}=10^{-5}$, while the remaining simulation setup is the same as in \cref{sec:quartic_1D}. Simulation results for the expectation values $\langle z^n\rangle$ for different values of $n$ and $\beta$ are compared with the exact results
\begin{align}\label{eq:one_pole_exact}
	\begin{aligned}
	\langle z^n\rangle_{\mathrm{exact}} = \frac{\beta^{-\frac{n}{2}}}{\sqrt{\pi}(1-2\beta)}\times\begin{cases}(1+n-2\beta)\Gamma\left(\frac{1+n}{2}\right)\quad &\textnormal{for even $n$}\;,\\
	-4\ii\sqrt{\beta}\,\Gamma\left(1+\frac{n}{2}\right) \quad &\textnormal{for odd $n$}\;,
	\end{cases}
	\end{aligned}
\end{align}
where $\Gamma$ denotes the usual gamma function, in \cref{tab:one_pole}. For $n\in\{3,4\}$ and $\beta\in\{1.6,3.2,4.8\}$ the results can be compared with \cite{ASS17} and one indeed finds agreement. Moreover, as was noted in \cite{ASS17,SS19}, the complex Langevin results reproduce the exact ones for sufficiently large $\beta$. However, deviations start to become statistically significant for $\beta\lesssim3.4$. It was argued in \cite{SS19,MSS25} that this disagreement is due to contributions from unwanted integration cycles. Note that with the amount of statistics going into the simulations, one finds a deviation for $\beta=3.2$ and $n=3$ that was not (as clearly) observed in earlier works. In fact, for large enough $n$ a disagreement can be seen even for $\beta=3.4$. It should be kept in mind, however, that with the unimproved update scheme \eqref{eq:euler_maruyama} deviations of the order of the maximum Langevin step size $\eps_{\max}=10^{-5}$ or smaller can be explained by finite-step-size effects.

\begin{table*}[t]
    \centering
    \renewcommand{\arraystretch}{1.2}
	\resizebox{\textwidth}{!}{
    \begin{tabular}{|c|cc|cc|cc|cc|}
    \hline
    $\beta$ & $-\ii\langle z^3\rangle_\CL$ & $-\ii\langle z^3\rangle_{\mathrm{exact}}$ & $\langle z^4\rangle_\CL$ & $\langle z^4\rangle_{\mathrm{exact}}$ & $-\ii\langle z^{15}\rangle_\CL$ & $-\ii\langle z^{15}\rangle_{\mathrm{exact}}$ & $\langle z^{16}\rangle_\CL$ & $\langle z^{16}\rangle_{\mathrm{exact}}$ \\
    \hline
    $1.6$ & $0.75616(1)$ & $0.852273$ & $0.21704(4)$ & $-0.239702$ & $638.4(5)$ & $536.311$ & $-416(1)$ & $-1156.42$ \\
    $2.4$ & $0.338364(6)$ & $0.328947$ & $0.01905(1)$ & $-0.006853$ & $19.56(2)$ & $18.1726$ & $-19.82(4)$ & $-23.0943$ \\
    $3.2$ & $0.173716(4)$ & $0.173611$ & $0.019028(5)$ & $0.018989 $ & $1.711(2)$ & $1.70701$ & $-1.411(3)$ & $-1.41362$ \\
    $3.4$ & $0.152134(4)$ & $0.152130$ & $0.020136(4)$ & $0.020135$ & $1.040(1)$ & $1.03968$ & $-0.783(1)$ & $-0.779758$ \\
    $3.5$ & $0.142858(3)$ & $0.142857$ & $0.020404(4)$ & $0.020408$ & $0.822(1)$ & $0.820449$ & $-0.586(2)$ & $-0.586035$ \\
    $3.6$ & $0.134409(3)$ & $0.134409$ & $0.020537(3)$ & $0.020535$ & $0.6515(7)$ & $0.651883$ & $-0.4452(8)$ & $-0.443643$ \\
    $4.0$ & $0.107146(2)$ & $0.107143$ & $0.020091(3)$ & $0.020089$ & $0.2773(4)$ & $0.276160$ & $-0.1544(5)$ & $-0.155340$ \\
    $4.8$ & $0.072676(1)$ & $0.072674$ & $0.017413(2)$ & $0.017412$ & $0.06285(9)$ & $0.062732$ & $-0.02405(9)$ & $-0.024178$ \\
    \hline
    \end{tabular}}
    \caption{Comparison of expectation values $\langle z^n\rangle$ between complex-Langevin simulations of the model \eqref{eq:one_pole} with $z_0=\ii$, $n_p=2$, and various values of $\beta$, and exact results \eqref{eq:one_pole_exact} for different $n$. The parentheses indicate the statistical uncertainties rounded to their respective first significant digits. For even (odd) $n$, the imaginary (real) parts of $\langle z^n\rangle_\CL$ are found to be consistent with zero and are thus not shown.}
    \label{tab:one_pole}
\end{table*}

As in the previous subsection, it shall now be the goal to test which of the correctness criteria can produce the same conclusions as \cref{tab:one_pole} without comparing to exact results. Note that it is not the purpose of this work to determine the coefficients $a_i$ in \eqref{eq:sase_theorem} for the different values of $\beta$ considered, but only to detect incorrect convergence (from any origin).

\paragraph{Dyson--Schwinger equations}
The Dyson--Schwinger equations \eqref{eq:dse} of the model \eqref{eq:one_pole} (for observables $\obs(z)=z^n$ and with $z_0=\ii$, $n_p=2$) read 
\begin{equation}\label{eq:one_pole_dse}
	\langle A z^n\rangle = 0\;, \quad \textnormal{with} \quad
	Az^n = nz^{n-1} + \frac{2z^n}{z-\ii}-2\beta z^{n+1}\;.
\end{equation}
For $\beta=1.6$ and $n\leq8$, the validity of \eqref{eq:one_pole_dse} in complex Langevin simulations was studied previously in \cite{MSS25} and found to hold. Here, a broader range of both $\beta$ and $n$ is considered and the results are presented in \cref{tab:one_pole_dse}.

\begin{table*}[t]
    \centering
    \renewcommand{\arraystretch}{1.2}
	\resizebox{\textwidth}{!}{
    \begin{tabular}{|c|ccccc|}
    \hline
    $\beta$ & $10^5\times\langle Az\rangle_\CL$ &  $10^5\times\langle Az^3\rangle_\CL$ & $10^5\times\langle Az^4\rangle_\CL$ & $\langle Az^{14}\rangle_\CL$ & $\langle Az^{15}\rangle_\CL$ \\
    \hline
	$1.6$ & 
		    $-17(5)-1(3)\ii$ & 
		    $-5(6)-3(6)\ii$ & 
		    $0(1)\times10^{1}-2(1)\times10^{1}\ii$ & 
		    $-0.3(7)-1.7(8)\ii$ & 
		    $-3(2)+3(3)\ii$ \\
	$2.4$ & 
		    $-3(3)+1(3)\ii$ & 
		    $4(3)-5(4)\ii$ & 
		    $5(5)+3(5)\ii$ & 
		    $-0.05(3)-0.06(5)\ii$ & 
		    $-0.05(9)-0.1(1)\ii$ \\
	$3.2$ & 
	 	    $-6(3)+3(2)\ii$ & 
	 	    $-5(2)+3(2)\ii$ & 
	 	    $-1(2)-7(2)\ii$ & 
	 	    $-0.001(4)-0.007(6)\ii$ & 
	 	    $-0.001(8)+0.00(1)\ii$ \\
	$3.4$ & 
			$-1(3)+0(2)\ii$ & 
			$-0(2)+1(2)\ii$ & 
			$2(2)-2(2)\ii$ & 
			$0.001(2)+0.001(4)\ii$ & 
			$0.012(4)+0.014(8)\ii$ \\
	$3.5$ & 
			$2(2)-2(2)\ii$ & 
			$2(2)-2(2)\ii$ & 
			$0(2)+1(2)\ii$ & 
			$-0.002(3)-0.003(4)\ii$ & 
			$0.000(7)-0.006(9)\ii$ \\
	$3.6$ & 
			$-1(2)+1(2)\ii$ & 
			$-1(1)+0(1)\ii$ & 
			$-0(1)-2(1)\ii$ & 
			$-0.001(1)+0.002(3)\ii$ & 
			$0.006(3)+0.005(5)\ii$ \\
	$4.0$ & 
			$-2(2)+0(1)\ii$ & 
			$-1(1)-1(1)\ii$ & 
			$1(1)-2(1)\ii$ & 
			$-0.002(1)-0.005(2)\ii$ & 
			$-0.003(2)-0.003(4)\ii$ \\
	$4.8$ & 
			$-2(2)-1(1)\ii$ & 
			$-1(1)-0.9(8)\ii$ & 
			$-0.4(9)-0.8(9)\ii$ & 
			$-0.0002(2)-0.0005(4)\ii$ & 
			$-0.0004(5)-0.0004(7)\ii$ \\
    \hline
    \end{tabular}}
   	\caption{Validity of the Dyson--Schwinger equations \eqref{eq:one_pole_dse} in complex-Langevin simulations of the model \eqref{eq:one_pole} for various values of $n$ and $\beta$, assuming $\langle1\rangle_\CL=1$. The parentheses indicate the statistical uncertainties rounded to their respective first significant digits. Note that for small $n$, $\langle A z^n\rangle_\CL$ is rescaled by $10^5$.}
    \label{tab:one_pole_dse}
\end{table*}

For $n=1$, $3$, and $4$, the deviations of $\langle Az^n\rangle_\CL$ from zero are of the order of $\eps_{\max}$ for all values of $\beta$, indicating that the Dyson--Schwinger equations hold in these cases. For the higher powers, the differences become larger in magnitude, but can still be considered consistent with zero within the statistical uncertainties. Hence, one concludes that the Dyson--Schwinger equations are satisfied in the model \eqref{eq:one_pole}, in accordance with the findings of \cite{MSS25}. It should be mentioned, however, that the expectation values $\langle Az^n\rangle_\CL$ tend to become smaller in magnitude as $\beta$ is increased. Nonetheless, the results of \cref{tab:one_pole_dse} confirm once again the known fact that the validity of \eqref{eq:dse} is insufficient on its own to guarantee correct convergence.

\begin{figure}[t]
	\centering
	\includegraphics[scale=0.6]{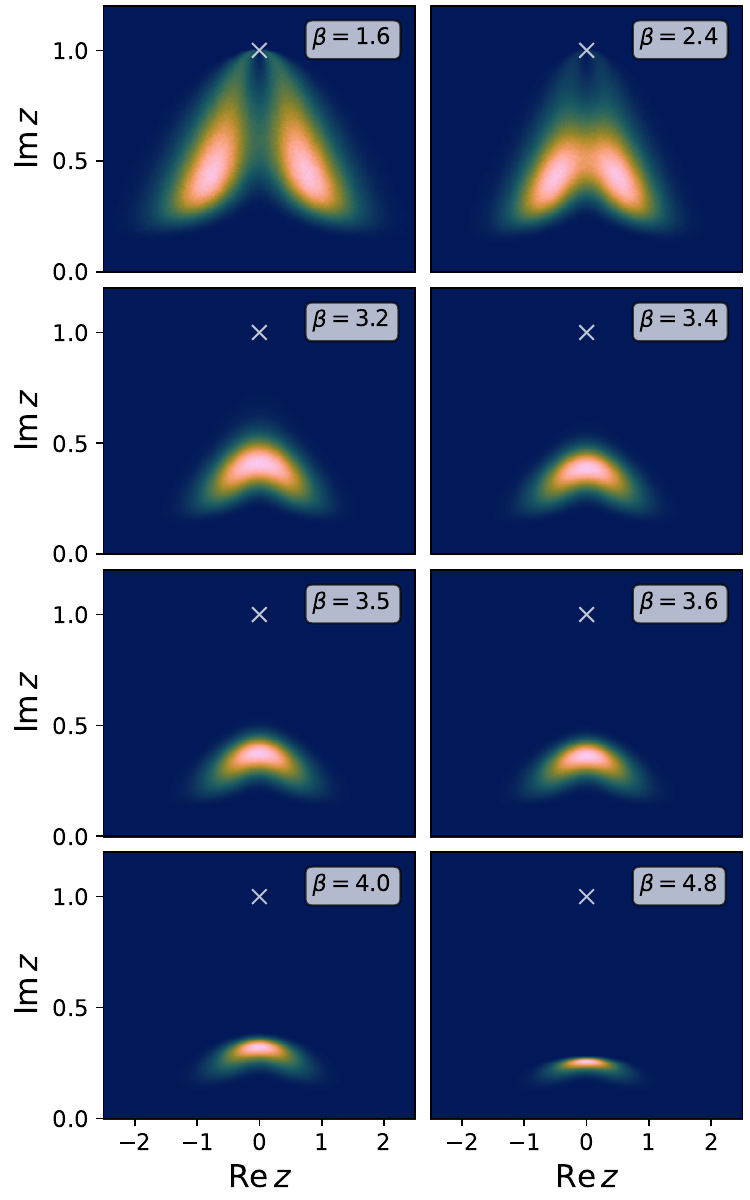}
	\caption{Histograms in the complex plane resulting from complex Langevin simulations of the model \eqref{eq:one_pole} for different values of $\beta$. Brighter regions indicate a higher probability and the crosses mark the pole position $z=\ii$. The plots show sums over multiple independent runs.}
	\label{fig:one_pole_histogram_2D}
\end{figure}

\paragraph{Histograms}
Histograms of the variable $z$ in the complex plane resulting from complex Langevin simulations of the model \eqref{eq:one_pole} with different values of $\beta$ are shown in \cref{fig:one_pole_histogram_2D}. As was noted in \cite{ASS17}, for small values of $\beta$, the equilibrium distribution approaches the drift term pole at $z=\ii$. In fact, it was first realized in \cite{NS15},\footnote{Another important finding of \cite{NS15} was that the presence of branch cuts of the logarithm in the action ($S=-\ln\,\rho$) of theories like \eqref{eq:one_pole} are not problematic in general.} by studying radial distributions around the poles of the drift term, that an insufficient suppression of the equilibrium distribution near such poles is problematic and a central issue causing wrong convergence.

\begin{figure}[t]
	\centering
	\includegraphics[scale=0.6]{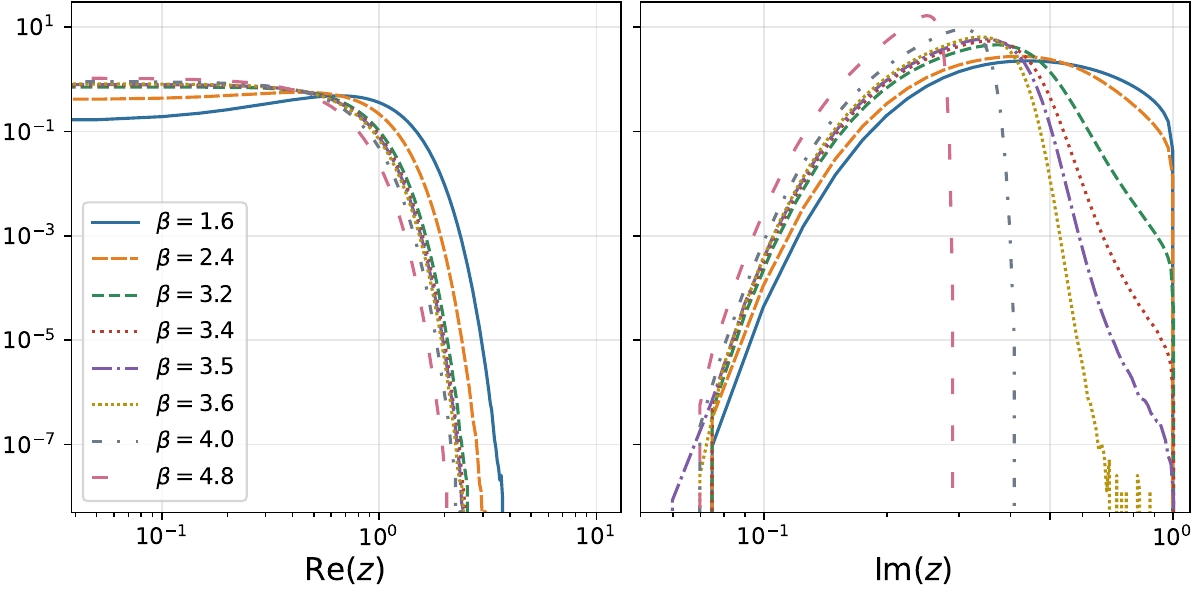}
	\caption{Projection of the data in \cref{fig:one_pole_histogram_2D} onto the real (left) and imaginary (right) axis. All axes are logarithmic.}
	\label{fig:one_pole_histogram_1D}
\end{figure}

As $\beta$ increases, on the other hand, the distribution moves away from the pole. A more detailed picture emerges after considering projections of the distributions onto the real and imaginary axis, respectively, which are shown in \cref{fig:one_pole_histogram_1D}. Clearly, the decay towards infinity is under control in all directions. What is equally important, however, is the decay behavior near the pole at $z=\ii$, which is best investigated in the right plot of \cref{fig:one_pole_histogram_1D}. There, it is seen that the distribution lies very far away from the pole for $\beta=4.8$ and $\beta=4.0$, still reasonably far away for $\beta=3.6$, but is nonzero close to $z=\ii$ for all other values of $\beta$. The qualitative decay behavior towards $z=\ii$ is comparable in all of the latter cases. From this, one may conclude that complex Langevin simulations can become problematic for $\beta\leq3.5$. However, clear indications for incorrect convergence in \cref{tab:one_pole} are only observed for $\beta\leq3.4$.

These findings can be interpreted in two ways: The first possibility is that the histogram-based criterion is very sensitive, meaning that one should indeed find incorrect convergence already for values of $\beta$ as high as $\beta=3.5$, contrary to what was observed in \cite{ASS17}. In this case, the available statistics going into \cref{tab:one_pole} or the employed maximum Langevin step size $\eps_{\max}$ might be insufficient to make this discrepancy visible. It could also be that a significant disagreement only emerges for even higher powers of $z$. The other interpretation is that the histogram criterion could be more subtle than what is depicted here. While, according to \cref{fig:one_pole_histogram_1D}, the distributions are clearly non-zero in the close vicinity of $z=\ii$ for $\beta\leq3.5$, the probability to find configurations near the pole decreases for increasing $\beta$. A sufficiently small probability near the pole might again be a sign for correct convergence. This second possibility, however, appears to be less likely, especially in the light of the drift criterion discussed below. It should also be noted that the histograms of $\imag\,z$ are obtained by integrating over $\real\,z$, i.e., they do not probe the pole directly.

The above discussion implies that in the model \eqref{eq:one_pole} one is confronted with a situation in which the histogram criterion predicts incorrect convergence for parameters for which, nonetheless, complex Langevin results agree with exact solutions within the available statistics. While the latter is much better than what can usually be obtained in realistic simulations, upon further increasing the statistics a significant deviation is expected to become visible nonetheless. Moreover, while correctness within the statistical uncertainties is -- in principle -- sufficient for practical applications, such a situation is still problematic. This is because one has no way of knowing beforehand how much statistics is necessary to make the discrepancy visible for a given set of parameters. Therefore, in practice it is advisable to refrain from considering parameters that lead to nonzero probability densities close to poles of the drift term in equilibrium.

\paragraph{Boundary terms}
In the model \eqref{eq:one_pole} with $z_0=\ii$ and $n_p=2$, the boundary-term observable \eqref{eq:boundary_terms} with a trivial kernel and for $\obs(z)=z^n$ with integer $n>0$ reads
\begin{equation}\label{eq:one_pole_boundary_terms}
	\mathcal{B}_{z^n}(Y) = \left\langle\Theta(Y-\vert z\vert)\left(n(n-1)z^{n-2} + \frac{2nz^{n-1}}{z-\ii} - 2n\beta z^n\right)\right\rangle\;.
\end{equation}
Note that this observable is sensitive to boundary terms at infinity only, while, as discussed before, one might be more concerned about boundary terms at the drift term pole $z=\ii$. Nevertheless, the latter are not discussed here because of the observation that boundary terms at poles should disappear in the equilibrium limit \cite{Sei20}. As the present discussion focuses on diagnostic tools that can be applied retroactively, i.e., after equilibration, boundary terms at poles are not considered here. Instead, it shall be of interest to investigate whether any statements at all can be made from studying boundary terms at infinity in the context of \eqref{eq:one_pole}.

\begin{figure}[t]
	\centering
	\includegraphics[scale=0.6]{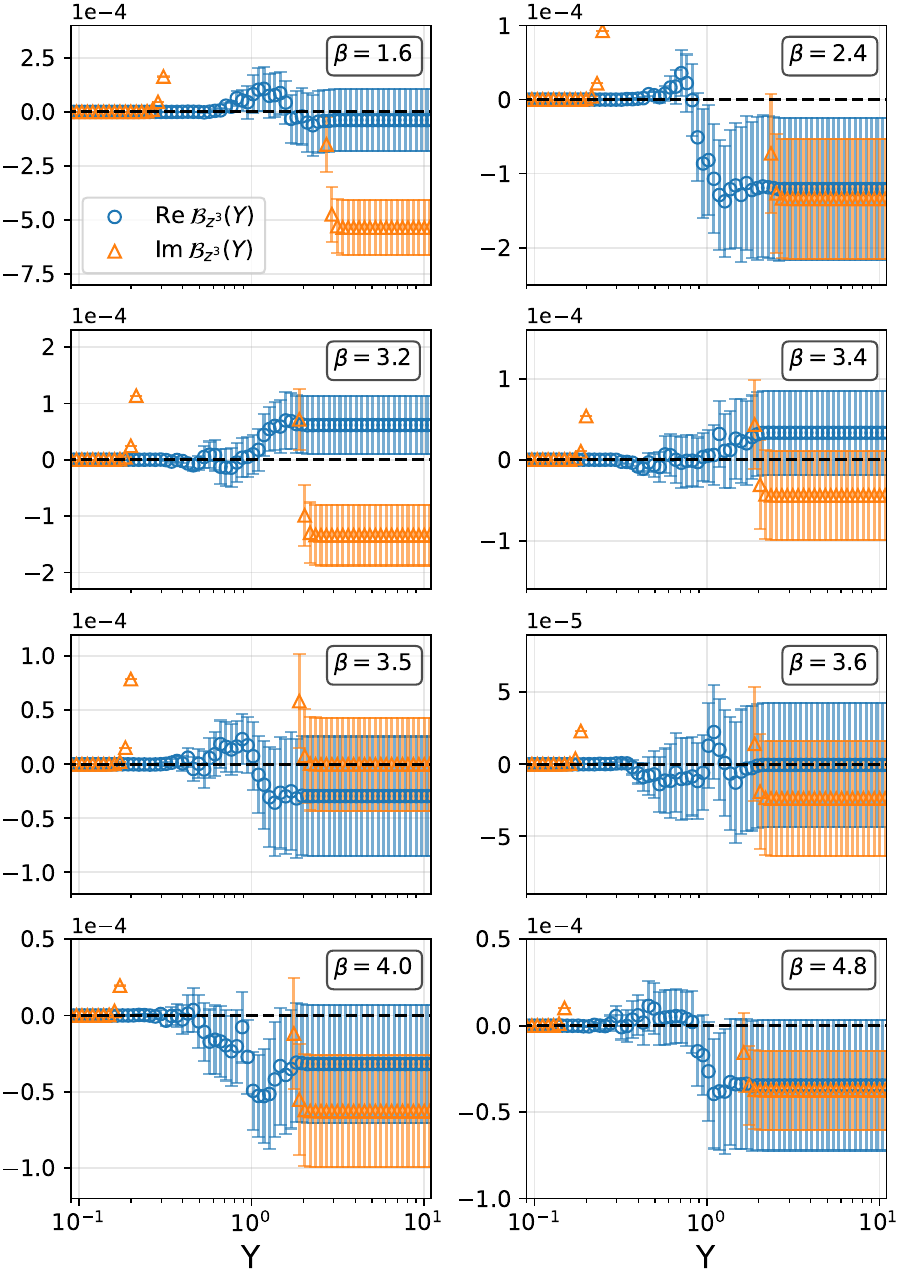}
	\caption{Boundary-term observable \eqref{eq:one_pole_boundary_terms} for $\obs(z)=z^3$ in complex Langevin simulations of the model \eqref{eq:one_pole} as a function of the cutoff $Y$ for different values of $\beta$. The horizontal axes are logarithmic and identical for each plot, the data points for $\mathrm{Im}\,\mathcal{B}_{z^3}(Y)$ are slightly offset horizontally for better visibility, and the dashed horizontal lines mark zero.}
	\label{fig:one_pole_boundary_terms_3}
\end{figure}

\begin{figure}[t]
	\centering
	\includegraphics[scale=0.6]{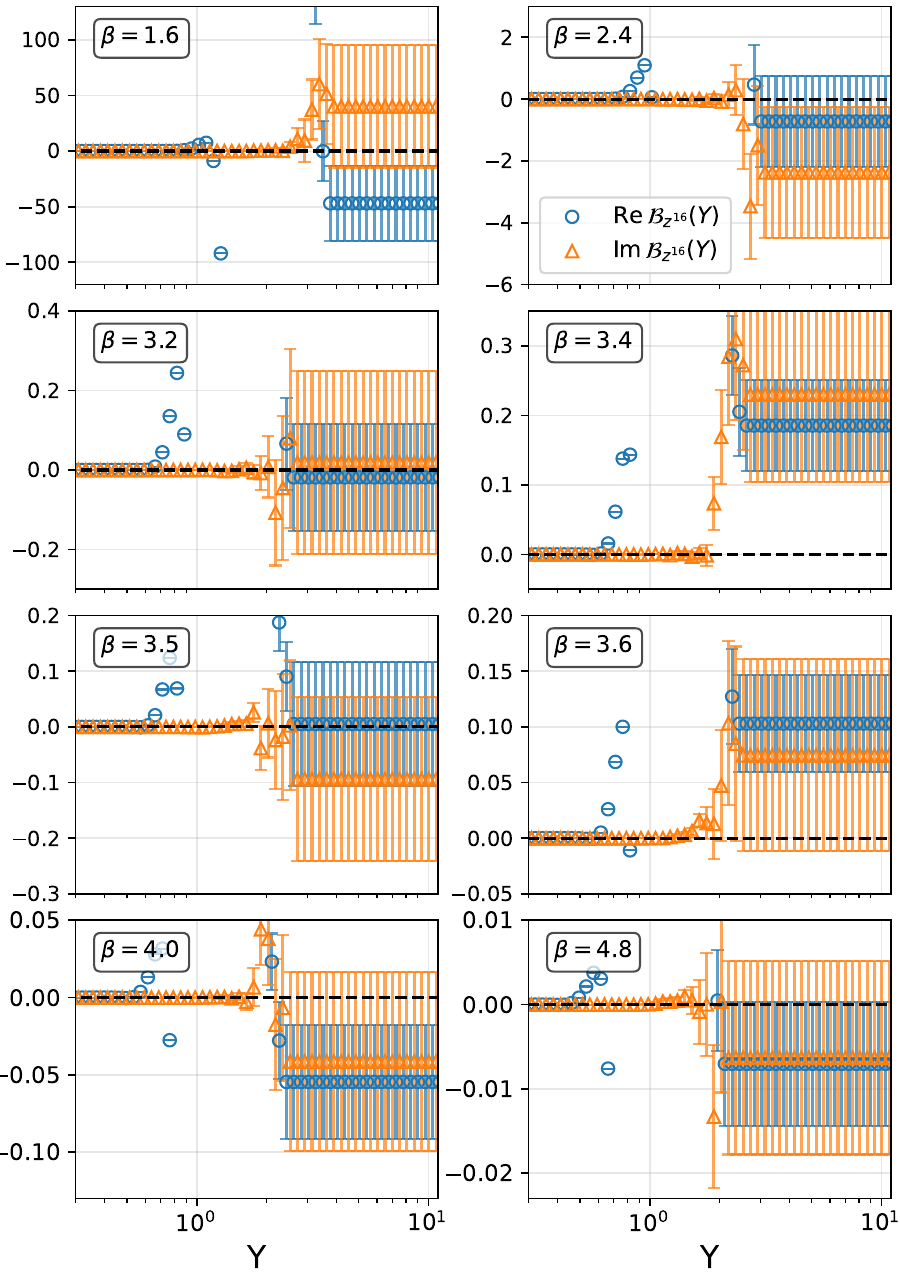}
	\caption{Similar as \cref{fig:one_pole_boundary_terms_3}, but for $\obs(z)=z^{16}$.}
	\label{fig:one_pole_boundary_terms_16}
\end{figure}

\begin{figure}[t]
	\centering
	\includegraphics[scale=0.5]{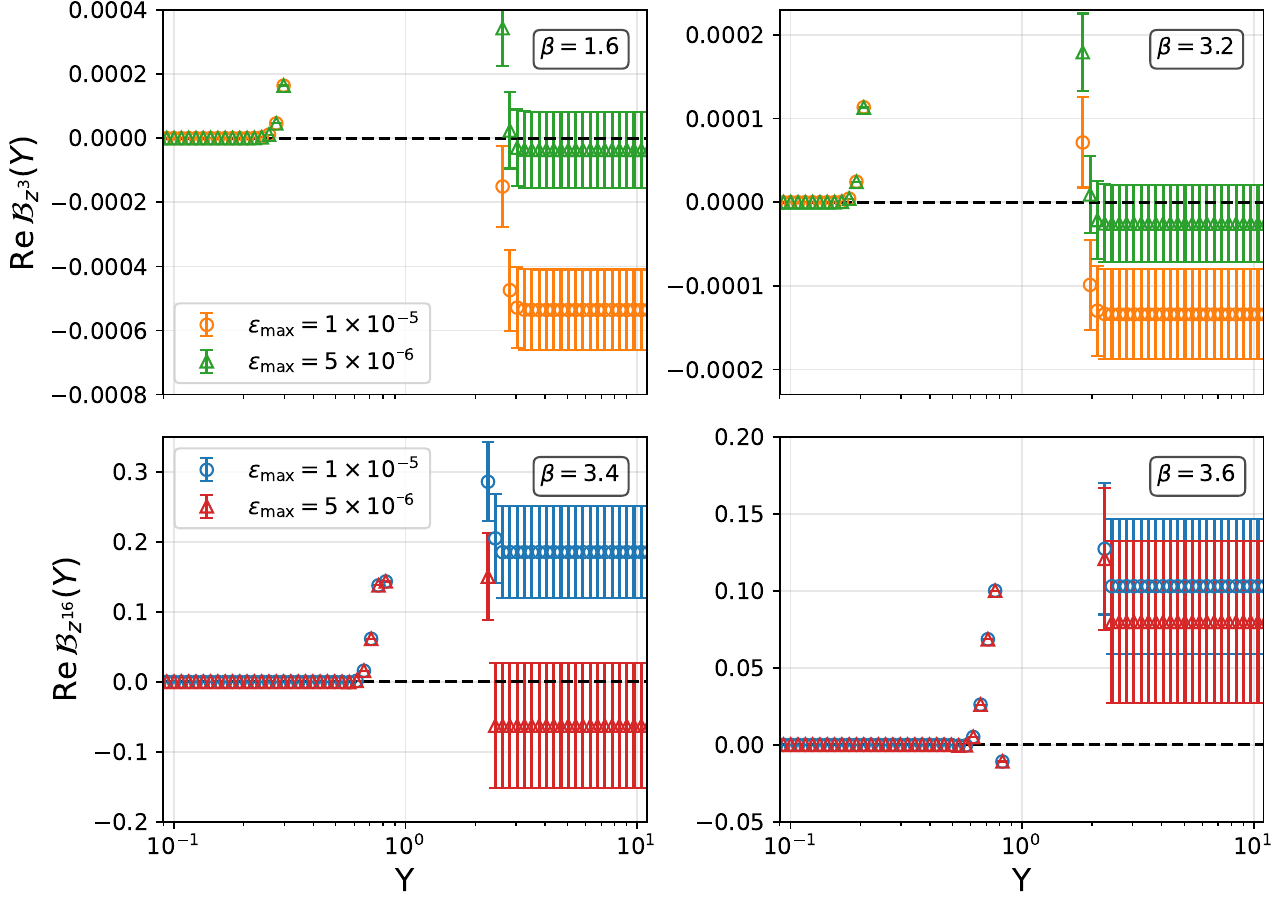}
	\caption{Boundary-term observable \eqref{eq:one_pole_boundary_terms} in complex Langevin simulations of the model \eqref{eq:one_pole} as a function of the cutoff $Y$ for different maximum Langevin-time step-sizes $\eps_{\max}$. The upper plots show $\imag\,\mathcal{B}_{z^3}(Y)$ for $\beta=1.6$ (left) and $3.2$ (right) and the lower plots show $\real\,\mathcal{B}_{z^{16}}(Y)$ for $\beta=3.4$ (left) and $3.6$ (right); these are the parameters for which descrepancies were found in \cref{fig:one_pole_boundary_terms_3,fig:one_pole_boundary_terms_16}. The horizontal axes are logarithmic and identical for each plot and the dashed horizontal lines mark zero.}
	\label{fig:one_pole_boundary_terms_continuum}
\end{figure}

The $Y$-dependence of \eqref{eq:one_pole_boundary_terms} for $\obs(z)=z^3$ and $z^{16}$ is shown in \cref{fig:one_pole_boundary_terms_3,fig:one_pole_boundary_terms_16}, respectively. In the former, one observes that the boundary-term observable assumes a plateau at rather small values for all $\beta$ considered. Indeed, the values of $\mathcal{B}_{z^3}$ on the plateaus are consistent with zero for most $\beta$. For $\beta=1.6$ and $3.2$, on the other hand, the deviations are more pronounced. It turns out that this discrepancy is caused by effects stemming from the finite maximum Langevin-time step-size $\eps_{\max}$. Indeed, as is demonstrated in \cref{fig:one_pole_boundary_terms_continuum}, upon decreasing $\eps_{\max}$ $\mathcal{B}_{z^3}(Y)$ is found to be consistent with zero on the plateaus for $\beta=1.6$ and $3.2$ as well. Thus, while not directly visible in \cref{fig:one_pole_boundary_terms_3}, boundary terms for $\obs(z)=z^3$ in the model \eqref{eq:one_pole} actually vanish for all values of $\beta$ considered. These findings are at odds with \cref{tab:one_pole}, in which one finds statistically significant deviations for this observable for all $\beta\leq3.2$. For the higher power of $z$ studied in \cref{fig:one_pole_boundary_terms_16}, the situation is very similar, the boundary-term observable agreeing with zero for most values of $\beta$, the exceptions being $\beta=3.4$ and $3.6$. Once again, these deviations vanish as $\eps_{\max}$ is decreased, as seen in \cref{fig:one_pole_boundary_terms_continuum}.

These observations hint at the conclusion already drawn in \cite{ASS17,SS19} that the incorrect convergence of complex Langevin simulations of the model \eqref{eq:one_pole} at low $\beta$ is not due to boundary terms but instead due to unwanted integration cycles. The boundary-term observable \eqref{eq:one_pole_boundary_terms} is thus incapable of detecting this wrong convergence. However, an important insight one has to take into account is that boundary terms can be quite sensitive to finite-step-size effects.

\paragraph{Convergence conditions}
The boundary-term observables shown in \cref{fig:one_pole_boundary_terms_3,fig:one_pole_boundary_terms_16,fig:one_pole_boundary_terms_continuum} are all well behaved as the cutoff increases. In particular, they exhibit a pronounced single plateau that can be extrapolated to the limit $Y\to\infty$. Thus, the convergence conditions \eqref{eq:convergence_conditions} in the model \eqref{eq:one_pole} are fulfilled for all $\beta$, provided $\eps_{\max}$ is sufficiently small. In other words, the conclusions drawn from \eqref{eq:convergence_conditions} are once again identical to those of the boundary-term analysis.

\begin{figure}[t]
	\centering
	\includegraphics[scale=0.5]{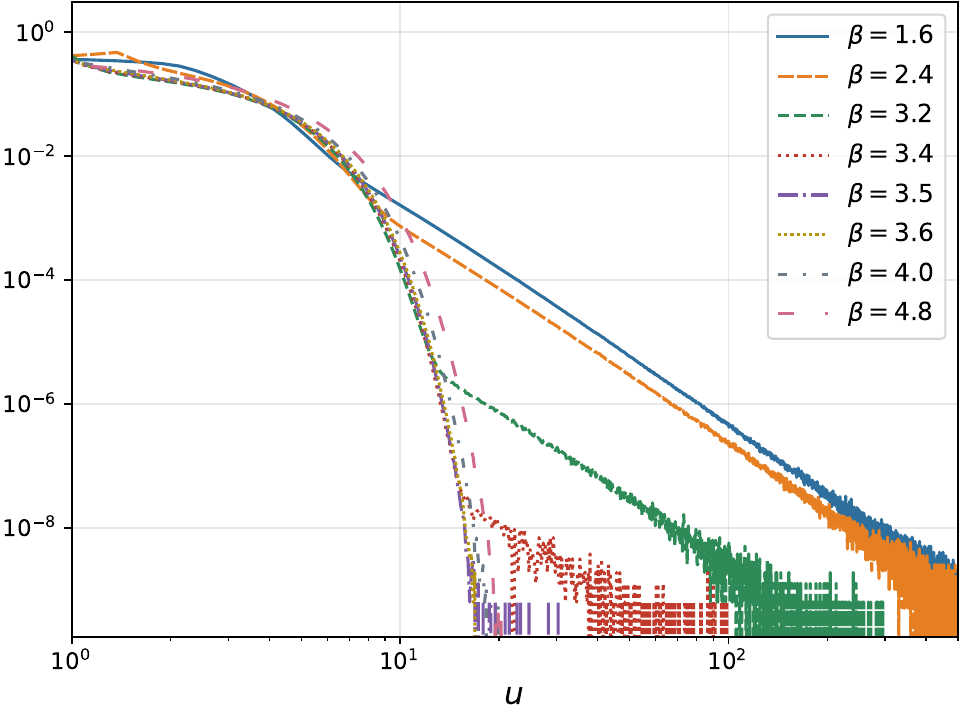}
	\caption{Distribution of the magnitude $u=\vert D\vert$ of the drift term $D$ in \eqref{eq:one_pole_drift} of the model \eqref{eq:quartic_1D} in complex Langevin simulations with different values of $\beta$. Both axes are logarithmic.}
	\label{fig:one_pole_drift}
\end{figure}

\paragraph{Drift criterion}
Histograms of the magnitude $u(z)=\vert D(z)\vert$ of the drift term of \eqref{eq:one_pole},
\begin{equation}\label{eq:one_pole_drift}
	 D(z) = \frac{2}{z-\ii}-2\beta z\;,
\end{equation}
for different values of $\beta$ are shown in \cref{fig:one_pole_drift}. The decay behavior of the distributions shows a clear tendency: For sufficiently small $\beta$, the decay is algebraic, while it becomes exponential or faster for $\beta\gtrsim3.6$. Again, it is difficult to draw definite conclusions, since the correct determination of the decay properties relies on the adequate sampling of a low-probability region of field space. However, with the available statistics, \cref{fig:one_pole_drift} suggests correct convergence as long as $\beta\geq3.6$. This is unexpected for various reasons.

First of all, while clear deviations between simulation and exact results in \cref{tab:one_pole} can be seen for $\beta\leq3.4$, the same cannot be said for $\beta=3.5$. Thus, if one were to believe the predictions of the drift criterion, one would expect to find discrepancies to appear also for that value of $\beta$, but only upon increasing the statistics and/or considering even larger powers $z^n$. The same conclusion was drawn in the histogram analysis of \cref{fig:one_pole_histogram_1D}.

Second, according to \cite{ASS17,SS19,MSS25}, the incorrect convergence in \eqref{eq:one_pole} is believed to be due to contributions from unwanted integration cycles and not due to slowly decaying distributions, i.e., boundary terms at infinity. However, the discussion of the one-dimensional quartic model \eqref{eq:quartic_1D} in \cref{fig:quartic_1D_drift} and \cref{app:kernel_1D} demonstrates that the drift criterion cannot be sensitive to integration cycles in that model, providing a counterexample. The situation encountered in the present case appears to be different. Here, if the incorrect convergence for sufficiently small $\beta$ is really due to unwanted integration cycles, then the drift criterion can nevertheless detect it, suggesting that the argument of \cref{app:kernel_1D} based on variable transformations cannot be applied in the context of \eqref{eq:one_pole}. This might have the important consequence that the drift and histogram criteria are more powerful than what is anticipated from their application in \eqref{eq:quartic_1D}. After all, due to its sheer simplicity the latter might exhibit intricacies that are not observed in more realistic models. If this were true, it might greatly simplify arguments for correctness of complex Langevin simulations of, say, QCD. It should be noted, however, that the drift criterion can also give false negative answers in some situations, suggesting incorrect convergence in cases when complex Langevin results agree with exact solutions within the statistical uncertainties, as outlined in \cref{app:drift_criterion}. This appears to be especially prominent in models with compact degrees of freedom.

\paragraph{Observable bounds}
As found in \cite{MSS25}, the study of the observable bounds in \eqref{eq:one_pole} is less straightforward than in \eqref{eq:quartic_1D}. Indeed, with the choice of parameters of \cite{MSS25}, the simplest control observable violating the bounds \eqref{eq:observable_bounds} for $\beta=1.6$ was found to be a polynomial of $12$-th degree:
\begin{align}\label{eq:one_pole_control_observable}
	\begin{aligned}
		\obs(z) =\ 210 - &1213\ii z - 3505z^2 + 6331\ii z^3 + 8583 z^4 - 8175\ii z^5 - 6825 z^6 + \\ &3813\ii z^7 + 2204 z^8 - 698\ii z^9 - 300 z^{10} + 42.2\ii z^{11} + 14.2z^{12}\;.
	\end{aligned}
\end{align}
In the present work, it shall be the goal to extend the study of \cite{MSS25} in order to find a simpler setup in which the observable-bound criterion is capable of detecting incorrect convergence. Since, according to \cref{tab:one_pole_dse}, the Dyson--Schwinger equations are fulfilled for all $\beta$ considered, one may focus on the bounds in \eqref{eq:observable_bounds}. On the one hand, it shall be of interest to obtain lower bounds for $\beta=1.6$ than those found in \cite{MSS25}. On the other hand, since all $\beta\lesssim3.5$ give rise to incorrect results according to \cref{tab:one_pole}, it would be important to find observables violating the bounds for this entire range of $\beta$.

To this end, the density $\rho$ in \eqref{eq:one_pole} is decomposed into $w$ and $\rho_r$ as
\begin{equation}\label{eq:one_pole_rho_decomposition}
	w(z) = (z-z_0)^le^{-\frac{\beta z^2}{b}}\;,\quad \rho_r(z) = \frac{1}{Z}(z-z_0)^{n_p-l}e^{-\frac{\beta z^2(b-1)}{b}}\;,
\end{equation}
parametrized by $b>1$ and $0\leq l\leq n_p$. As before, $z_0=\ii$ and $n_p=2$. In \cite{MSS25}, \eqref{eq:one_pole_control_observable} was found to violate the bounds \eqref{eq:observable_bounds} for $\beta=1.6$ using $l=0$, $b=2$, $p=1$ in \eqref{eq:observable_bounds_aux}, and $\mathcal{M}_r=\mathbb{R}$ in \eqref{eq:p_norms}. Indeed, with this choice of parameters one finds $\left\vert\langle\obs(z)\rangle_\CL\right\vert\approx902$, while the right-hand side of \eqref{eq:observable_bounds} gives $C^{(1)}\Vert\obs\Vert_w^{(1)}\approx702$. Since in this setup the bounds are violated for $\beta=1.6$, \eqref{eq:one_pole_control_observable} represents a reasonable starting point in the search for better bounds.

\begin{figure}[t]
	\centering
	\includegraphics[scale=0.5]{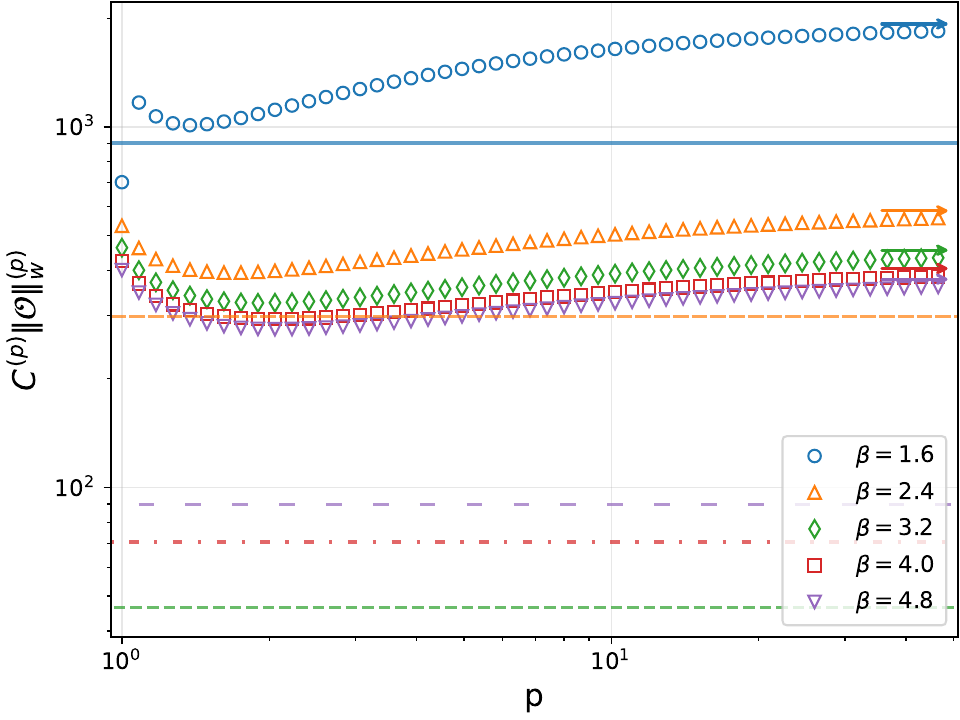}
	\caption{Observable bounds \eqref{eq:observable_bounds} in the model \eqref{eq:one_pole} as a function of the parameter $p$ in \eqref{eq:observable_bounds_aux} for the decomposition $l=0$, $b=2$ in  \eqref{eq:one_pole_rho_decomposition} and different values of $\beta$. The chosen control observable is \eqref{eq:one_pole_control_observable} and the arrows indicate the respective $p\to\infty$ limits. The horizontal lines correspond to the left-hand side of \eqref{eq:observable_bounds} for (from top to bottom) $\beta=1.6$, $2.4$, $4.8$, $4.0$, and $3.2$ respectively. Corresponding results for $\beta=3.4$, $3.5$, and $3.6$ are similar to $\beta=3.2$ and hence not shown. Both axes are logarithmic.}
	\label{fig:one_pole_bound_betas}
\end{figure}

An obvious question is whether the same setup also leads to violations of the bounds for other values of $\beta$. To this end, the left- and right-hand sides of \eqref{eq:observable_bounds} are compared with one another for different $\beta$ as a function of $p$ in \cref{fig:one_pole_bound_betas}. A number of conclusions can be drawn already from this figure. First of all, the behavior of the bounds $C^{(p)}\Vert\obs\Vert_w^{(p)}$ for $\beta=1.6$ is rather peculiar, showing a jump for $p\gtrsim1$, which is not observed for any other $\beta$. For $\beta>1.6$, instead, minimal bounds are attained at some finite value $p>1$, similar to the situation in the model \eqref{eq:quartic_1D} studied in \cref{fig:quartic_1D_bounds_vs_p}. Second, only for $p\approx1$ are the bounds \eqref{eq:observable_bounds} violated for $\beta=1.6$. Finally, and most importantly, no bound violation is observed for any value of $p$ when $\beta>1.6$. Thus, the setup of \cite{MSS25} does not extend in a straightforward way to general $\beta$. Given that \eqref{eq:one_pole_control_observable} was specifically tailored for $\beta=1.6$, this is not too surprising. However, it also makes clear the difficulty of making general statements about any universal behavior of the observable bounds. For instance, for different values of $\beta$ observables violating the bounds \eqref{eq:observable_bounds} might look entirely different. It should be mentioned that it is possible to find somewhat lower bounds with the control observable \eqref{eq:one_pole_control_observable}. For instance, with $l=2$, $b=1.01$, and $p=\infty$, one finds $C^{(\infty)}\Vert\obs\Vert_w^{(\infty)}\approx639$. Within the given parametrization, a much lower bound for \eqref{eq:one_pole_control_observable} seems unlikely to exist.

One should keep in mind that in the model \eqref{eq:one_pole} one is not restricted to polynomial observables $\obs\in\mathfrak{A}$ in \eqref{eq:observable_bounds}, but may consider more general observables in the extended space $\mathcal{H}$ as well. Indeed, it turns out that it is much easier to find suitable control observables in this extended space. For instance, with the choice
\begin{equation}
	\obs(z) = \frac{z^2}{z-\ii}
\end{equation}
one obtains 
\begin{equation}
	\vert\langle\obs\rangle_\CL\vert\approx0.685\;,\quad
	C^{(\infty)}\Vert\obs\Vert_w^{(\infty)}\approx0.633
	\quad \textnormal{for} \quad 
	\beta=1.6\;,
\end{equation}
and
\begin{equation}
	\vert\langle\obs\rangle_\CL\vert\approx0.342\;,\quad
	C^{(\infty)}\Vert\obs\Vert_w^{(\infty)}\approx0.337
	\quad \textnormal{for} \quad 
	\beta=2.4\;,
\end{equation}
both with negligible statistical uncertainties. To obtain these bounds, again $b=1.01$ and $l=2$ was used. For larger values of $\beta$, on the other hand, this control observable fails to violate the bounds as well. However, the validity of \eqref{eq:observable_bounds} does not imply that the complex Langevin results are correct. Rather, it simply again becomes very difficult to find control observables violating \eqref{eq:observable_bounds}. Given how numerically close the complex Langevin and exact results are for $\beta\geq3.2$, this is not unexpected. At the time of writing, no observable $\obs\in\mathcal{H}$ could be found that would violate \eqref{eq:observable_bounds} for $\beta\geq3.2$. Note, however, that in order to find such an observable, one could at best make educated guesses, but a concrete guiding principle is still not available.

\begin{figure}[t]
	\centering
	\includegraphics[scale=0.5]{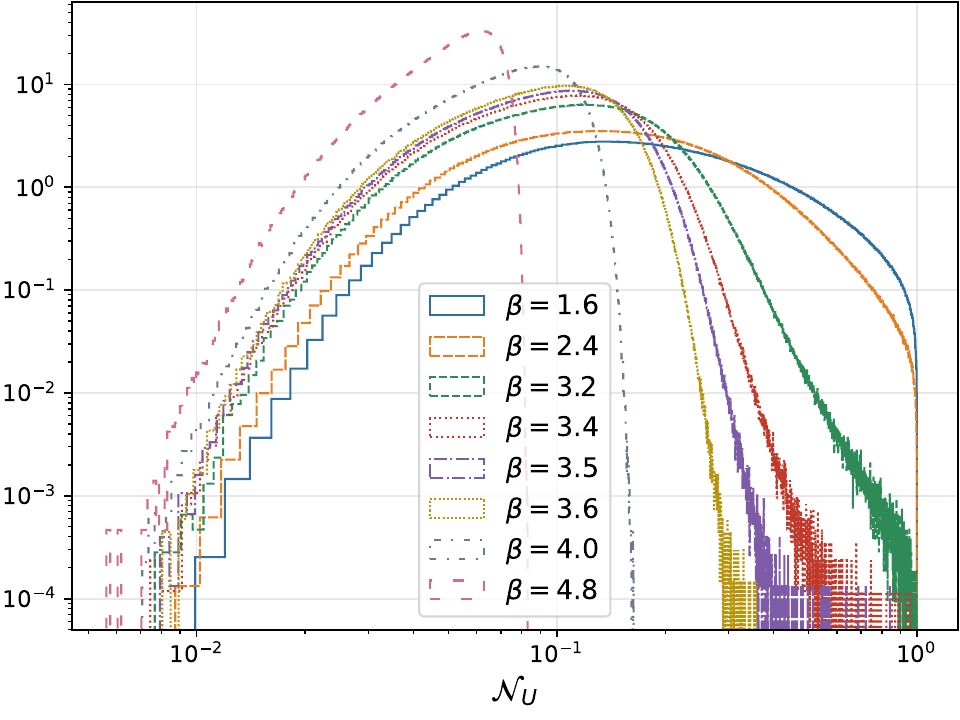}
	\caption{Histograms of $\mathcal{N}_U=(\imag z)^2$ in complex Langevin simulations of the model \eqref{eq:one_pole} with different values of $\beta$. Only the restricted ensemble is used. Both axes are logarithmic.}
	\label{fig:one_pole_unitarity_norm}
\end{figure}

\paragraph{Unitarity norm}
Similar to the previous subsection, the evolution of the unitarity norm \eqref{eq:unitarity_norm} in complex Langevin simulations of the model \eqref{eq:one_pole} was investigated and histograms of $\mathcal{N}_U$ for the different values of $\beta$ are shown in \cref{fig:one_pole_unitarity_norm}. Since the distribution of $z$ is confined to a strip in the complex plane, the distribution of $\mathcal{N}_U=(\imag z)^2$ is confined as well. It should come as no surprise that \cref{fig:one_pole_unitarity_norm} is very similar to \cref{fig:one_pole_histogram_1D} (right). The interpretation of the former, however, is different. For the unitarity-norm criterion, one studies the appearance of large values of $\mathcal{N}_U$ in a simulation and no statements can be made from the (non-)vanishing of its distribution close to the drift term pole at $z=\ii$, i.e., $\mathcal{N}_U=1$. What can be concluded from \cref{fig:one_pole_unitarity_norm} is that the unitarity norm never grows larger than $1$ for any $\beta$ and that larger values of $\beta$ tend to produce overall smaller unitarity norms. While the behavior of the unitarity norm is indeed correlated with the correctness or incorrectness of simulations with different $\beta$, this observation is again difficult to quantify.

\paragraph{Configurational temperature}
Finally, the (inverse) configurational temperature observable \eqref{eq:configurational_temperature} in the model \eqref{eq:one_pole} with $z_0=\ii$ and $n_p=2$ reads
\begin{equation}\label{eq:one_pole_configurational_temperature}
	\tilde{\beta} = -\frac{1}{2}\left\langle\frac{1+(z-\ii)^2\beta}{(1-z(z-\ii)\beta)^2}\right\rangle\;.
\end{equation}
Note that $\tilde{\beta}$ is a different quantity than the free parameter $\beta$. As compared to \eqref{eq:quartic_1D_configurational_temperature}, this observable does not have a pole on $\mathbb{R}$, implying that its computation along the real integration cycle is well defined. Indeed, one finds the exact result $\tilde{\beta}_{\mathrm{exact}}=1$ for all $\beta>0$. This, in turn, makes the applicability of the configurational-temperature criterion much more plausible in the context of \eqref{eq:one_pole}.

\begin{figure}[t]
	\centering
	\includegraphics[scale=0.5]{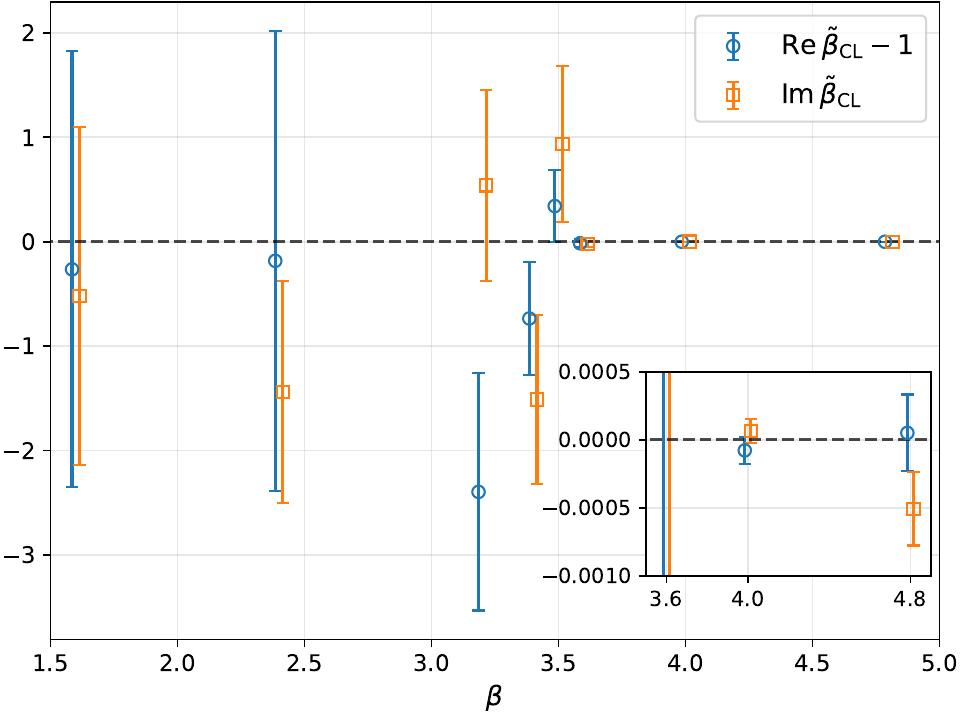}
	\caption{(Inverse) configurational temperature observable \eqref{eq:one_pole_configurational_temperature} in complex Langevin simulations of the model \eqref{eq:one_pole} as a function of $\beta$. The inset shows a close-up view of the largest values of $\beta$ considered.}	\label{fig:one_pole_configurational_temperature}
\end{figure}

The observable \eqref{eq:one_pole_configurational_temperature} is shown as a function of $\beta$ in \cref{fig:one_pole_configurational_temperature}. For small values of $\beta$, one observes large fluctuations, which, however, decrease with larger $\beta$, while the largest deviations from $\tilde{\beta}=1$ appear for intermediate $\beta$. It is not entirely clear how to interpret \cref{fig:one_pole_configurational_temperature}. On the one hand, the configurational-temperature observable is consistent with $\tilde{\beta}=1$ for almost all values of $\beta$, which would -- in principle -- be a signal for correct convergence. On the other hand, the large fluctuations give reason to doubt this assertion. What is certainly true is that for $\beta\geq3.6$, where fluctuations are small and $\tilde{\beta}$ is consistent with unity, the configurational-temperature criterion predicts correctness. A consistent picture with \cref{fig:one_pole_histogram_1D,fig:one_pole_drift} emerges if one interprets large fluctuations in $\tilde{\beta}$ as signs for incorrectness. This conclusion would, again, suggest that for the value $\beta=3.5$ a discrepancy would have to appear at some point. One should also mention that the fluctuations are caused by the distributions in \cref{fig:one_pole_histogram_2D} approaching the poles of \eqref{eq:one_pole_configurational_temperature}.

\paragraph{Summary}
The analysis of the one-pole model \eqref{eq:one_pole} from the point of view of the different correctness criteria of \cref{sec:criteria} can be summarized as follows: for $z_0=\ii$ and $n_p=2$, complex Langevin dynamics produce incorrect results for small values of $\beta$, namely $\beta\lesssim3.4$. This range of incorrect convergence can be detected rather well with the histogram and drift criteria. The unitarity-norm and configurational-temperature criteria allow for analogous conclusions, albeit only under some additional assumptions (a heuristic upper limit for $\mathcal{N}_U$ beyond which results are incorrect in the former and strong fluctuations in $\tilde{\beta}$ indicating incorrectness in the latter case). The rigorous correctness criterion based on observable bounds can unambigously detect incorrect convergence for the smallest values of $\beta$ but fails for higher ones due to the difficulty of finding appropriate control observables. Boundary terms at infinity vanish, erroneously signalling correct convergence. For this conclusion, however, a very fine discretization of the Langevin-time evolution was necessary. Similar statements can be made using the convergence conditions or the Dyson--Schwinger equations. That the latter three correctness criteria cannot predict incorrect convergence within \eqref{eq:one_pole} at all is, to the best of current knowledge, due to the fact that it is caused by contributions from unwanted integration cycles.

\subsection{One-site Hubbard model}\label{sec:hubbard}
Another simple but interesting model that suffers from a sign problem but can be solved analytically is the one-site Hubbard model, given by the density \cite{THH16}
\begin{equation}\label{eq:hubbard}
	\rho(z) = \frac{1}{Z}\left(1+e^{\beta\left(\ii z+\mu+\frac{U}{2}\right)}\right)^2e^{-\frac{\beta}{2U}z^2}\;, \quad Z = \int dz \left(1+e^{\beta\left(\ii z+\mu+\frac{U}{2}\right)}\right)^2e^{-\frac{\beta}{2U}z^2}\;,
\end{equation}
with a real parameter $\mu$ and $\beta$, $U>0$. It is of relevance in the context of condensed-matter physics, arising in the strong-coupling limit of the conventional Hubbard model. Interestingly, the drift term of \eqref{eq:hubbard} has infinitely many poles. The model \eqref{eq:hubbard} has been studied using complex Langevin dynamics before \cite{HHT16}, with the conclusion that the method -- when applied naively -- generally fails to reproduce the exact results. However, in the same reference a reweighting strategy was proposed with which correct results could be obtained. Other works have investigated \eqref{eq:hubbard} from the point of view of Lefschetz thimbles \cite{THH16,UV17} and contour deformations \cite{PP25}. In the present case, the complex Langevin equation shall be applied straightforwardly, i.e., without any reweighting and without a kernel. 

However, in this subsection a slightly more general (but unimproved) complex Langevin equation is employed, including an imaginary part for the noise variable. For a single degree of freedom and $H=1$, this generalization of \eqref{eq:euler_maruyama} reads 
\begin{equation}\label{eq:imaginary_noise}
	z^{(n+1)} = z^{(n)} + \eps\frac{1}{\rho(z)}\frac{\partial \rho(z)}{\partial z}\bigg\vert_{z=z^{(n)}} + \sqrt{\eps}\left(\sqrt{1+N_I}\eta_R^{(n)}+\ii \sqrt{N_I}\eta_I^{(n)}\right)\;,
\end{equation}
where $\eta_R$ and $\eta_I$ are independent real noise variables each following a Gaussian distribution (with mean zero and a variance of $2$) and $N_I$ is a tunable parameter quantifying the amount of imaginary noise. The reason for considering complex noise in the model \eqref{eq:hubbard} will become clear below. 

A particularly important observable within the theory \eqref{eq:hubbard} is the fermion number density $\langle n\rangle$, given by \cite{HHT16}
\begin{equation}\label{eq:hubbard_density}
	\langle n\rangle = \frac{1}{\beta}\left\langle\frac{1}{\rho}\frac{\partial \rho}{\partial\mu}\right\rangle = \imag\frac{\langle z\rangle}{U}\;.
\end{equation} 
In this work, following \cite{HHT16}, the parameters of \eqref{eq:hubbard} are set to $U=1$, $\beta=30$, and $\mu=0$, respectively, giving $\langle n\rangle=\imag\,\langle z\rangle$. In addition to this observable, higher powers of $z$ are measured as well. The simulations of \eqref{eq:hubbard} employ \eqref{eq:imaginary_noise} as the update prescription and a maximum Langevin step size of $\eps_{\mathrm{max}}=10^{-5}$. 

\begin{figure}[t]
	\centering
	\includegraphics[scale=0.5]{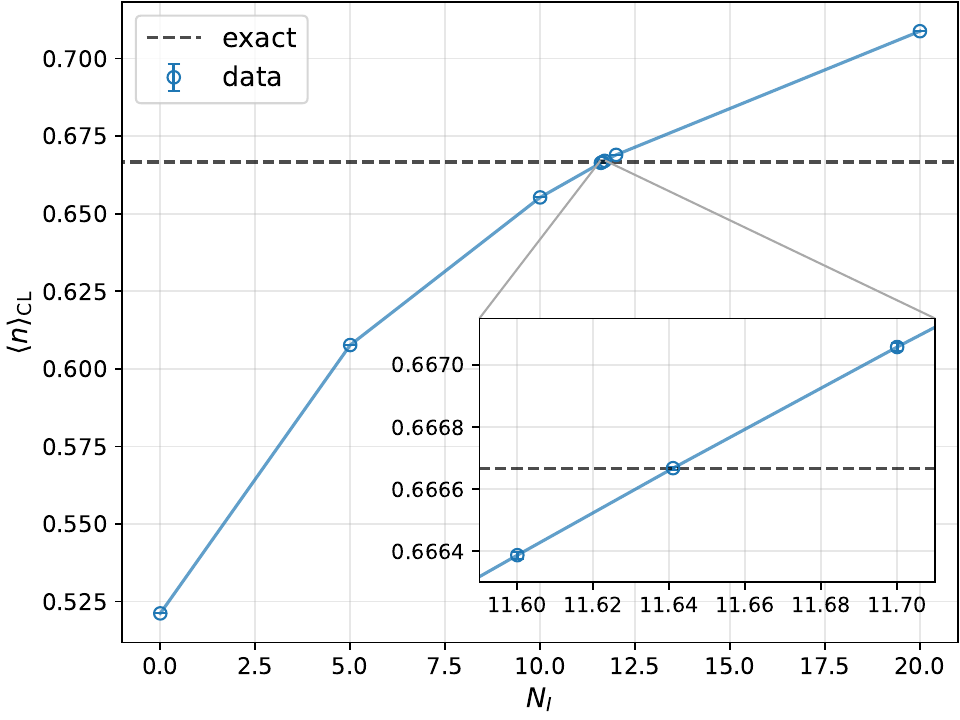}
	\caption{Fermion number density \eqref{eq:hubbard_density} in complex Langevin simulations of the model \eqref{eq:hubbard} with $U=1$, $\beta=30$, and $\mu=0$ as a function of the imaginary noise parameter $N_I$ in \eqref{eq:imaginary_noise}. The dashed line indicates the exact result $\langle n\rangle\approx\frac{2}{3}$ and the inset shows a close-up view around that value.}
	\label{fig:hubbard_imag_noise}
\end{figure}

To begin with, the fermion number density \eqref{eq:hubbard_density} is shown as a function of $N_I$ in \cref{fig:hubbard_imag_noise}. As can be seen, the imaginary-noise parameter can be tuned in such a way as to reproduce the exact result \cite{THH16}
\begin{equation}\label{eq:hubbard_exact}
	\langle n\rangle = \frac{2(e^{\beta\mu}+e^{\beta(2\mu-U)})}{1+2e^{\beta\mu}+e^{\beta(2\mu-U)}}\approx\frac{2}{3}
\end{equation}
within errors. More precisely, $\langle n\rangle_\CL=0.666668(4)$ for the tuned value $N_I=11.641$, with the real part of $\langle z\rangle_\CL$ being consistent with zero. A crucial observation, however, is the following: the fact that the complex Langevin result for $\langle n\rangle$ agrees with the exact solution for this value of $N_I$ is entirely misleading, since no other observable comes out correctly. This can be seen in \cref{tab:hubbard}, which compares complex Langevin and exact results for the tuned value $N_I=11.641$ and a few low powers of $z$.

\begin{table*}[t]
    \centering
    \begin{tabular}{|c|cc|}
    \hline
    $\obs(z)$ & $\langle\obs(z)\rangle_\CL$ & correct \\
    \hline
    $z$ & 
    	$6(7)\times10^{-6}+0.666668(4)\ii$ & 
    	$0.666667\ii$ \\
    $z^2$ & 
    	$-0.16231(1)+1(1)\times10^{-5}\ii$ & 
    	$-0.633333$ \\
    $z^3$ & 
    	$-2(2)\times10^{-5}+0.24414(2)\ii$ & 
    	$0.600000\ii$ \\
    $z^4$ & 
    	$-0.23308(3)-3(3)\times10^{-5}\ii$ & 
    	$0.536667$ \\
    \hline
    \end{tabular}
    \caption{Comparison of expectation values $\langle z^n\rangle$ between complex-Langevin simulations of the model \eqref{eq:hubbard} employing \eqref{eq:imaginary_noise} and $N_I=11.641$ and exact results. The parentheses indicate the statistical uncertainties rounded to their respective first significant digits.}
    \label{tab:hubbard}
\end{table*}

Such a scenario, while of limited use in practical applications, is an ideal testing ground for the various correctness criteria. In particular, it shall be interesting to investigate whether the criteria predict incorrect convergence at all, for all observables, or for powers of $z$ greater than one only. This question is discussed in the remainder of this subsection, focusing on $N_I=11.641$ and comparing with $N_I=0.0$, which produces incorrect results even for $\langle n\rangle$ according to \cref{fig:hubbard_imag_noise}.

\paragraph{Dyson--Schwinger equations}
The Dyson--Schwinger equations \eqref{eq:dse} of the model \eqref{eq:hubbard} read
\begin{equation}\label{eq:hubbard_dse}
	\langle Az^n\rangle = 0\;, \quad \textnormal{with} \quad
	Az^n = nz^{n-1} + \frac{2\ii\beta z^n}{1+e^{-\beta(\ii z+\mu+\frac{U}{2})}}-\frac{\beta z^{n+1}}{U}\;.
\end{equation}
Complex Langevin results $\langle A z^n\rangle_\CL$ for low powers $n$, computed in the restricted ensemble, are shown in \cref{tab:hubbard_dse}. One observes statistically significant violations of \eqref{eq:hubbard_dse} for $n=1$ and $n=2$ in the results based on purely real noise. For higher powers $n$ in that case, and for all $n$ in the case with $N_I\neq0$, however, the Dyson--Schwinger equations are fulfilled. If taken at face value, these findings would indicate that for $N_I=11.641$ the complex Langevin evolution is subject to contributions from unwanted integration cycles, causing the incorrect results. However, as argued below, statistics within the resticted ensemble is simply insufficient in this case to reveal the violations of \eqref{eq:hubbard_dse}. Indeed, this was confirmed explicitly by increasing the sample size. The corresponding results, however, are not shown here for reasons of consistency. 

\begin{table*}[t]
    \centering
    \begin{tabular}{|c|cc|}
    \hline
    	$n$ & $\langle Az^n\rangle_\CL$ ($N_I=0.0$) & $\langle Az^n\rangle_\CL$ ($N_I=11.641$) \\
    \hline
	$1$ & 
		  $0.034(2)+0.003(4)\ii$ & 
		  $-0.02(1)-0.01(1)\ii$ \\
	$2$ & 
		  $-0.002(2)+0.012(2)\ii$ & 
		  $-0.01(1)-0.00(2)\ii$ \\
	$3$ & 
		  $-0.002(1)-0.000(1)\ii$ & 
		  $-0.02(2)-0.03(2)\ii$ \\
	$4$ & 
		  $-0.0005(8)+0.0001(6)\ii$ & 
		  $0.03(4)-0.04(4)\ii$ \\
    \hline
    \end{tabular}
    \caption{Validity of the Dyson--Schwinger equations \eqref{eq:hubbard_dse} in complex-Langevin simulations of the model \eqref{eq:hubbard} for low powers $n$ and two values of $N_I$ in \eqref{eq:imaginary_noise}, computed in the restricted ensemble. The parentheses indicate the statistical uncertainties rounded to their respective first significant digits.}
    \label{tab:hubbard_dse}
\end{table*}

Another question concerns the dependence on the Dyson--Schwinger equations on the maximum Langevin-time step-size $\eps_{\max}$. Within the restricted ensemble, $\langle Az^n\rangle_{\CL}$ is found to still be consistent with zero upon decreasing $\eps_{\max}$, which, however, might again be due to the restricted ensemble not being large enough. Thus, within the available statistics it is difficult to make a definite statement.

\begin{figure}[t]
	\centering
	\includegraphics[scale=0.55]{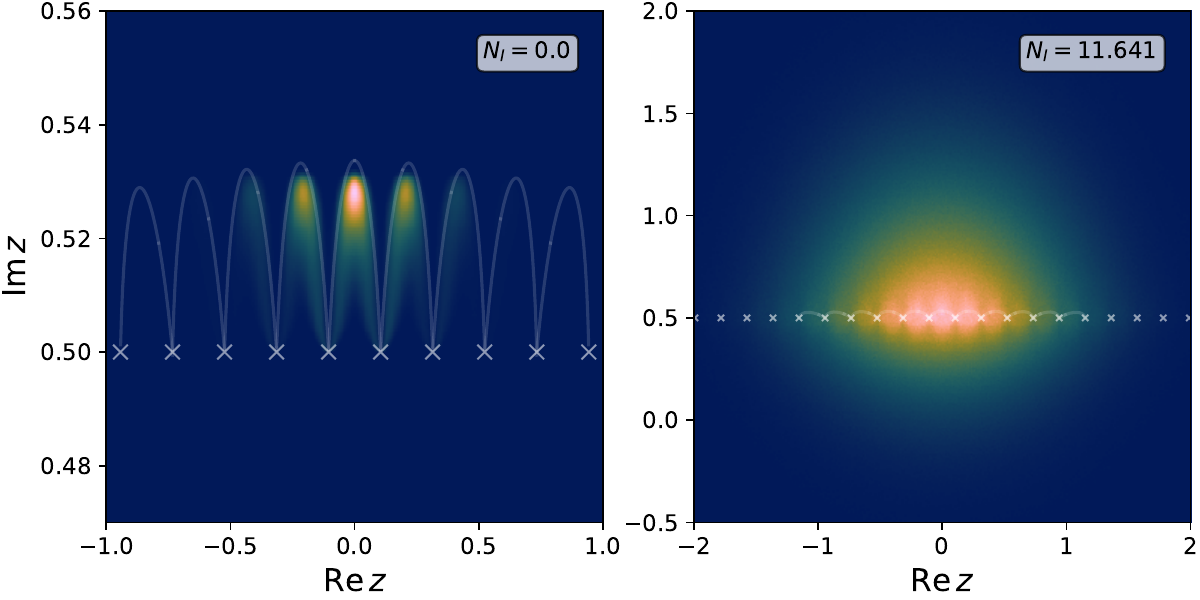}
	\caption{Histograms in the complex plane resulting from complex Langevin simulations of the model \eqref{eq:hubbard} with $U=1$, $\beta=30$, $\mu=0$ using the update prescription \eqref{eq:imaginary_noise} with two different $N_I$. The faint solid lines correspond to the Lefschetz thimbles closest to the origin. Brighter regions indicate a higher probability and the crosses mark the drift term poles. Note the different scales on the respective axes.}
	\label{fig:hubbard_histogram_2D}
\end{figure}

\begin{figure}[t]
	\centering
	\includegraphics[scale=0.5]{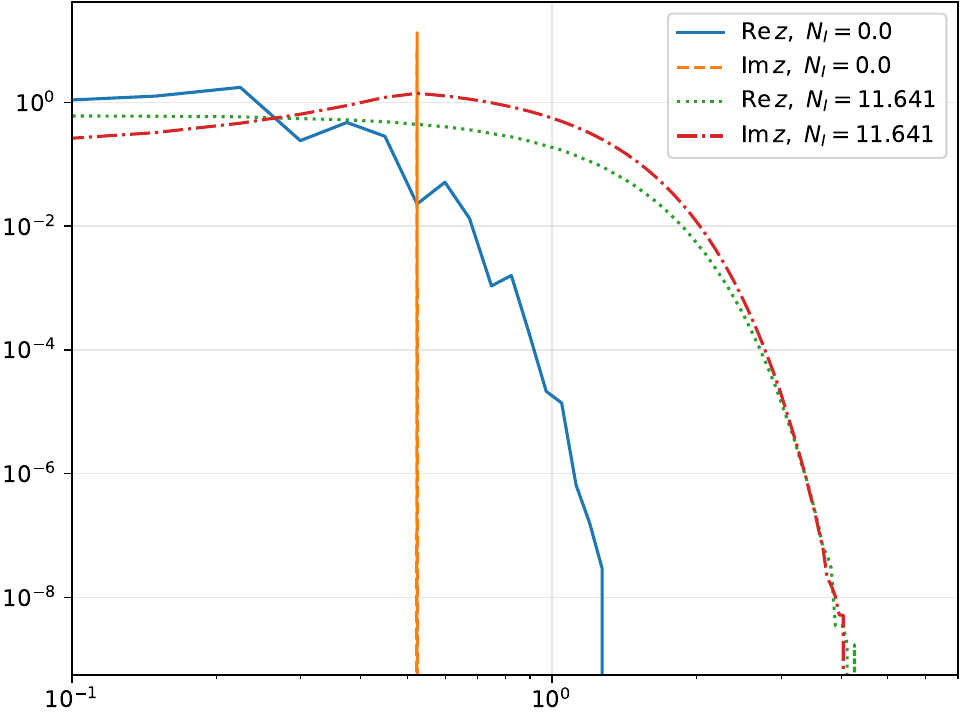}
	\caption{Projections of the left and right plots of \cref{fig:hubbard_histogram_2D} onto the respective real and imaginary axes. Both axes are logarithmic.}
	\label{fig:hubbard_histogram_1D}
\end{figure}

\paragraph{Histograms}
Histograms of $z$ in the complex plane for $N_I=0.0$ and $N_I=11.641$ are shown in the left and right plots of \cref{fig:hubbard_histogram_2D}, respectively. While in the former case, the distribution appears to roughly align with the Lefschetz thimbles closest to the origin \cite{THH16} (see also \cite{Aar13,ABS14,NS17,BPS18} for works on the relation between complex Langevin and Lefschetz thimbles), this behavior is washed out upon including an imaginary noise component, in which case the simulations sample a much larger region of the complex plane in all directions. Note, however, that there are periodic batches of higher probability remaining along $\imag\,z\approx0.5$, reflecting the thimble structure. The decay behavior of the distributions towards infinity is investigated in \cref{fig:hubbard_histogram_1D}, where, as in the previous subsections, projections onto the real and imaginary axes are shown.

For $N_I=0.0$, $\imag\,z$ is confined to a very narrow strip, while the distribution of $\real\,z$ shows oscillatory decay behavior. The decay towards infinity is faster than polynomial, which is also true for $N_I=11.641$, for which both $\real\,z$ and $\imag\,z$ show a fast decay. Note, however, that the probability distribution in the latter case is certainly nonzero in the vicinity of the drift term poles. The same can be said in the case of real noise: while the distribution of $\real\,z$ has local minima corresponding to the poles, it is still larger than zero there, even though a proper interpretation is somewhat less straightforward. In any case, histograms of $z$ correctly predict wrong convergence in the present setup.

\begin{figure}[t]
	\centering
	\includegraphics[scale=0.6]{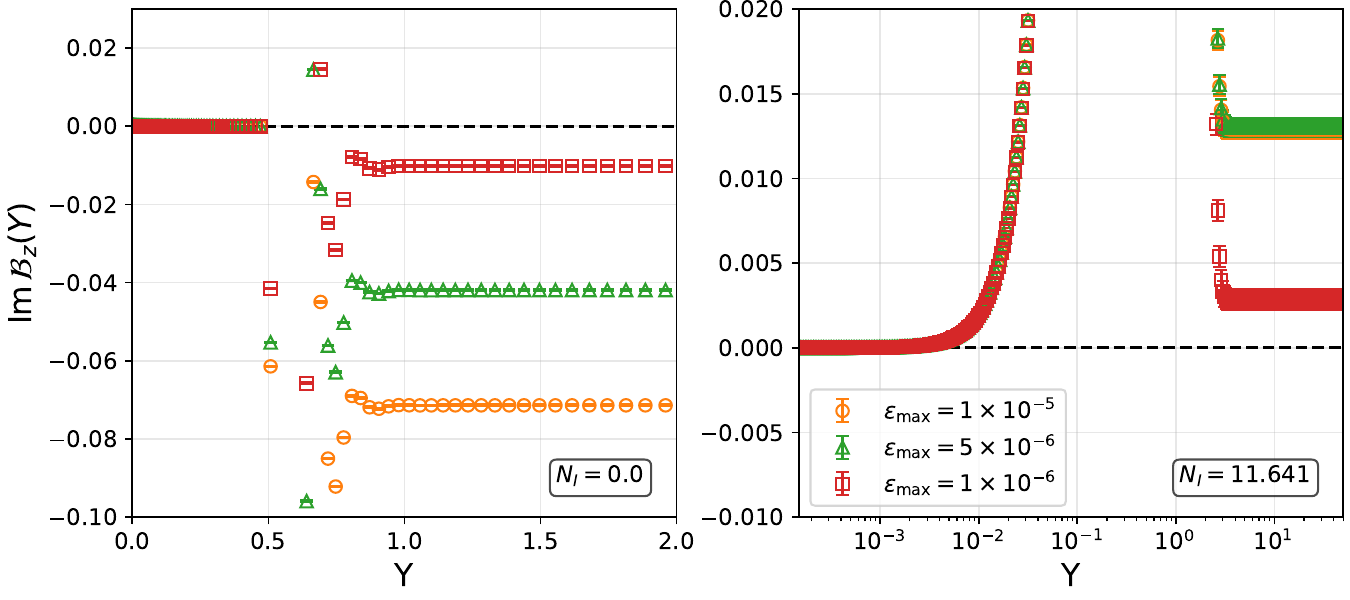}
	\caption{Imaginary part of the boundary term observable \eqref{eq:hubbard_boundary_terms} for $n=1$ in complex Langevin simulations of the model \eqref{eq:hubbard} with $U=1$, $\beta=30$, $\mu=0$ and three different maximum Langevin-time step-sizes $\eps_{\max}$, using the update \eqref{eq:imaginary_noise}, as a function of the cutoff $Y$. The real part of $\mathcal{B}_z(Y)$ is not shown, but is consistent with zero on the plateau in all cases considered. \emph{(left)}: $N_I=0.0$. \emph{(right)}: $N_I=11.641$, logarithmic horizontal axis.}
	\label{fig:hubbard_boundary_terms}
\end{figure}

\paragraph{Boundary terms}
The boundary-term observables \eqref{eq:boundary_terms} in the model \eqref{eq:hubbard} for obervables $\obs(z)=z^n$ take the rather complicated form
\begin{equation}\label{eq:hubbard_boundary_terms}
	\mathcal{B}_{z^n}(Y) = \left\langle\Theta(Y-\vert z\vert)\left(n(n-1)z^{n-2} + \frac{2\ii\beta nz^{n-1}}{1+e^{-\beta(\ii z+\mu+\frac{U}{2})}} - \frac{\beta n z^n}{U}\right)\right\rangle\;.
\end{equation}
As mentioned before, what is considered an advantage of the boundary-term criterion is its potential sensitivity to different observables. Thus, it would not be inconceivable that for $N_I=11.641$ the boundary terms of $\obs(z)=z$ vanish while those of all other observables are nonzero. This possibility is examined in \cref{fig:hubbard_boundary_terms}, where $\imag\,\mathcal{B}_z(Y)$ is shown for both $N_I=0.0$ and $N_I=11.641$. Having observed finite-step-size effects affecting the boundary terms in \cref{sec:one_pole}, here three different maximum Langevin-time step-sizes $\eps_{\max}$ are employed.

In all cases considered, $\mathcal{B}_z(Y)$ exhibits a plateau, with its real part being consistent with zero and its  imaginary part assuming a nonzero value. This would be a clear indication for the presence of boundary terms at infinity and, thus, for incorrect convergence. For $N_I=0.0$, this is expected from \cref{fig:hubbard_imag_noise}. For $N_I=11.641$, on the other hand, this comes as more of a surprise since, due to the tuning of the imaginary noise, correct results could be obtained for $\langle z\rangle$. Curiously, however, for both values of $N_I$ the boundary term observable on the plateau moves closer to zero upon decreasing the Langevin-time step-size, similar to what was observed in \cref{fig:one_pole_boundary_terms_continuum}. It should be mentioned that for $N_I=11.641$ the value of $\mathcal{B}_z(Y)$ at the plateau does not change significantly when going from $\eps_{\max}=10^{-5}$ to $\eps_{\max}=5\times10^{-6}$, but then jumps to a lower value for $\eps_{\max}=10^{-6}$. Due to the employed adaptive-step-size algorithm the actual step-size $\eps$ can vary over the course of a simulation and $\eps_{\max}$ is only its upper bound. Hence, such a behavior is not too surprising, even though it again highlights subtleties concerning the step-size-dependence of the boundary-term criterion.

For other monomial observables, the behavior of the boundary terms is similar, with plateaus at nonzero values that, generally, become smaller upon decreasing the maximum step-size. These findings are thus in conflict with the assertion that boundary terms can give information about correct results for individual observables. Indeed, for $\obs(z)=z$, complex Langevin reproduces the correct solution for $N_I=11.641$ but not for $N_I=0.0$ even though the results of \cref{fig:hubbard_boundary_terms} (assuming they extrapolate) suggest that boundary terms at infinity vanish in both cases. Finally, boundary terms at the poles of the drift term are once again not considered here.

\paragraph{Convergence conditions}
As can be seen from the limit $Y\to\infty$ of \cref{fig:hubbard_boundary_terms}, the convergence conditions \eqref{eq:convergence_conditions}, when taken at face value, are violated in the model \eqref{eq:hubbard} for both choices of $N_I$. This is a signal for incorrect convergence, or, rather, non-convergence. However, it is in conflict with the (naive) observation of \cref{tab:hubbard_dse} that the Dyson--Schwinger equations \eqref{eq:hubbard_dse} are fulfilled, since the validity of the Dyson--Schwinger equations implies the validity of the convergence conditions \cite{AJS11}: 
\begin{equation}
	\langle L_c z^n\rangle = n\langle Az^{n-1}\rangle\;.
\end{equation}
Since \cref{tab:hubbard_dse} was computed on the restricted ensemble, whereas the full ensemble was used for \cref{fig:hubbard_boundary_terms}, it can be concluded (without explicitly increasing the statistics) that, indeed, the Dyson--Schwinger equations for both $N_I=0.0$ and $N_I=11.641$ should actually be violated in the model \eqref{eq:hubbard} for the maximum step-sizes employed in \cref{fig:hubbard_boundary_terms}. In turn, this would imply that the Dyson--Schwinger equations can also detect incorrect convergence in \eqref{eq:hubbard}, assuming sufficient statistics. 

Note once more, however, that if the plateaus in \cref{fig:hubbard_boundary_terms} really become consistent with zero in the continuum limit, then the convergence conditions (and, thus, likely the Dyson--Schwinger equations as well) would be fulfilled for both $N_I=0.0$ and $N_I=11.641$. A possible interpretation of this would be that the complex Langevin evolution is affected by unwanted integration cycles. This possibility deserves a more detailed study in future work.

\begin{figure}[t]
	\centering
	\includegraphics[scale=0.5]{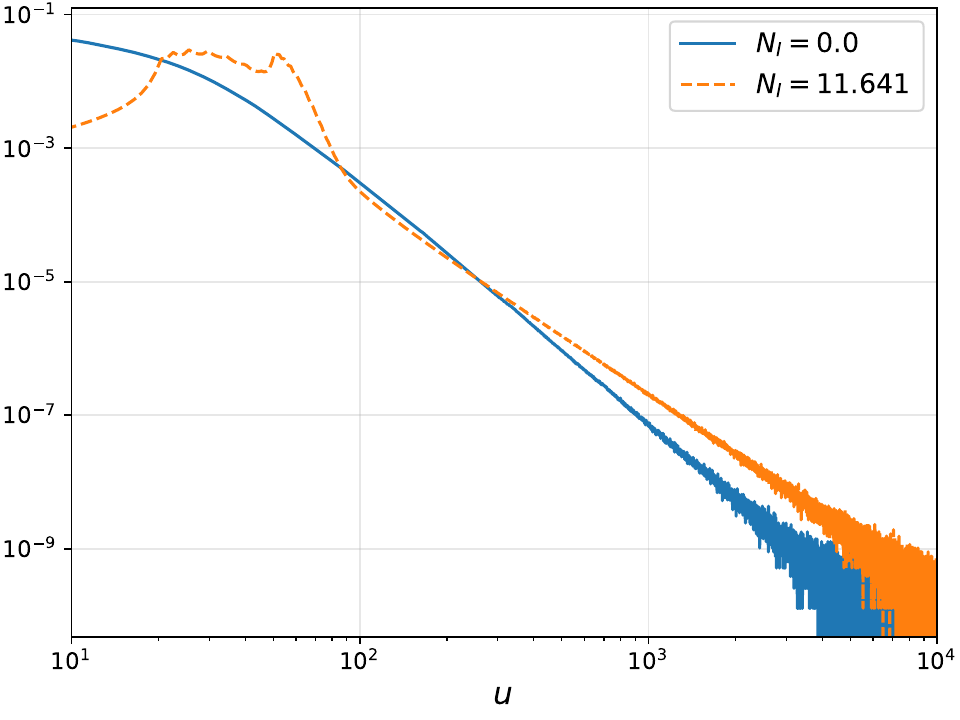}
	\caption{Histograms of the magnitude $u=\vert D\vert$ of the drift term \eqref{eq:hubbard_drift} of the model \eqref{eq:hubbard} with $U=1$, $\beta=30$, $\mu=0$ in complex Langevin simulations using the update prescription \eqref{eq:imaginary_noise} with $N_I=0.0$ and $N_I=11.641$. Both axes are logarithmic.}
	\label{fig:hubbard_drift}
\end{figure}

\paragraph{Drift criterion}
Histograms of the magnitude $u(z)=\vert D(z)\vert$ of the drift term of \eqref{eq:hubbard},
\begin{equation}\label{eq:hubbard_drift}
	D(z) =\frac{2\ii\beta}{1+e^{-\beta(\ii z+\mu+\frac{U}{2})}}-\frac{\beta z}{U}\;,
\end{equation}
with and without imaginary noise are shown in \cref{fig:hubbard_drift}. The observed decay is clearly not faster than exponential, from which one concludes incorrect convergence. This is in accordance with the expectation if one recalls that the drift criterion cannot be expected to make any statements about individual observables such as $\langle n\rangle$, but only about overall (non)convergence.

\begin{figure}
	\centering
	\includegraphics[scale=0.5]{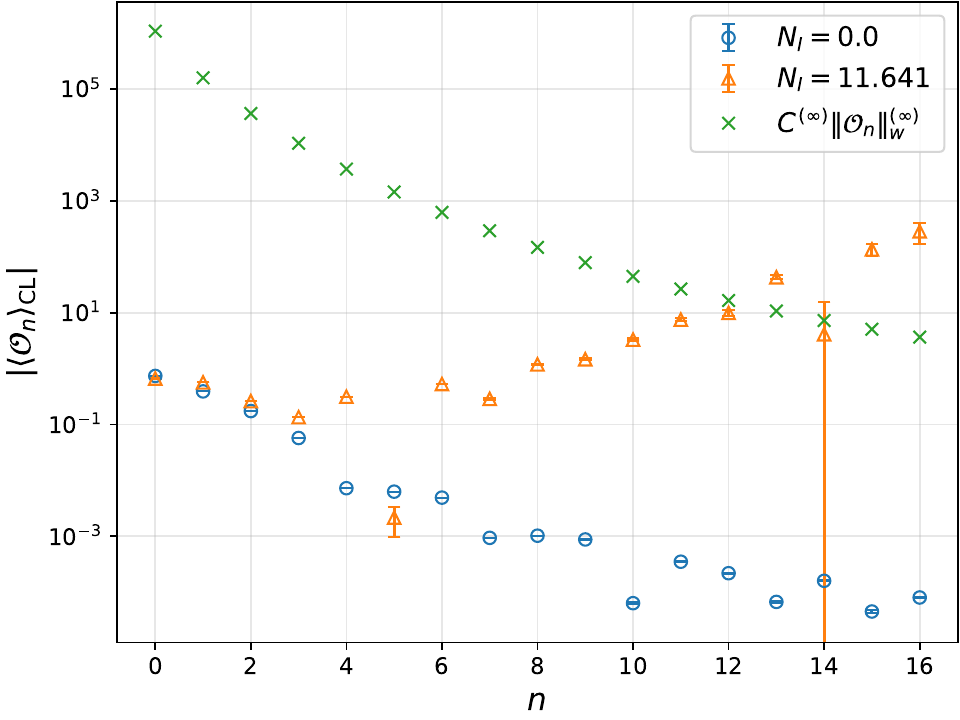}
	\caption{Comparison between the left- and right-hand sides of \eqref{eq:observable_bounds} for observables of the type \eqref{eq:hubbard_control_obseravble}, computed in complex-Langevin simulations of \eqref{eq:hubbard} for two values of $N_I$ in \eqref{eq:imaginary_noise}, as a function of $n$. The bounds $C^{(p)}\Vert\obs_n\Vert_w^{(p)}$ are computed with $b=1.01$ in \eqref{eq:hubbard_rho_decomposition} and $p=\infty$ in \eqref{eq:observable_bounds}. The vertical axis is logarithmic.}
	\label{fig:hubbard_bounds_vs_n}
\end{figure}

\paragraph{Observable bounds}
According to the discussion above, the Dyson--Schwinger equations are (naively) violated in the model \eqref{eq:hubbard} for both $N_I=0.0$ and $N_I=11.641$, i.e., the first of the two requirements of the observable-bound criterion is not met. In such a case it would be of little use to study a possible violation of the bounds \eqref{eq:observable_bounds}. However, for the sake of completeness, and due to the aforementioned possibility of the Dyson--Schwinger equations actually being valid in the limit $\eps_{\max}\to0$, a few remarks are given here nonetheless. First of all, one may decompose the density $\rho$ in \eqref{eq:hubbard} as
\begin{equation}\label{eq:hubbard_rho_decomposition}
	w(z) = \left(1+e^{\beta\left(\ii z+\mu+\frac{U}{2}\right)}\right)^2e^{-\frac{\beta}{2Ub}z^2}\;, \quad
	\rho_r(z) = \frac{1}{Z}e^{-\frac{\beta(b-1)}{2Ub}}\;,
\end{equation}
with a tunable parameter $b>1$. Experiments with different values of $p$ in \eqref{eq:observable_bounds} and $b$ again tend to produce the lowest bounds for small $b$ (e.g., $b=1.01$) and $p=\infty$. For polynomial control observables $\obs\in\mathfrak{A}$, the bounds become very large such that their violation seems implausible. Once again, however, one may consider control observables within the extended space $\mathcal{H}$. Simple choices are given by
\begin{equation}\label{eq:hubbard_control_obseravble}
	\obs_n(z) = \frac{z^n}{1+e^{\beta\left(\ii z + \mu + \frac{U}{2}\right)}}\;,
\end{equation}
with integer $n\geq0$. As can be seen in \cref{fig:hubbard_bounds_vs_n}, the bounds for $\obs_n$ with $p=\infty$ and $b=1.01$ decrease monotonically with $n$. For $N_I=0.0$, no violation of \eqref{eq:observable_bounds} is found, whereas for $N_I=11.6414$ the bounds are violated for sufficiently large $n$, signalling incorrect convergence. While from these results on their own one cannot yet conclude wrong convergence for $N_I=0.0$, they might nonetheless be useful for future studies of \eqref{eq:hubbard} using complex Langevin dynamics.

\begin{figure}[t]
	\centering
	\includegraphics[scale=0.5]{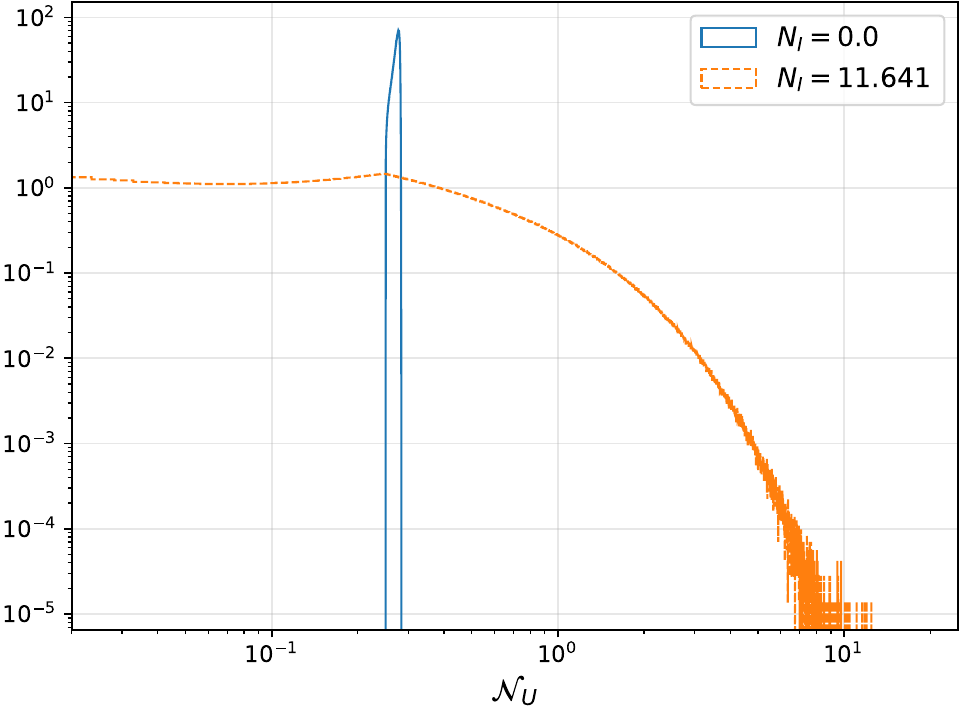}
	\caption{Histograms of $\mathcal{N}_U=(\imag\,z)^2$ in complex Langevin simulations of the model \eqref{eq:hubbard} with $U=1$, $\beta=30$, $\mu=0$, using the update \eqref{eq:imaginary_noise} with $N_I=0.0$ and $N_I=11.641$. Both axes are logarithmic.}
	\label{fig:hubbard_unitarity_norm}
\end{figure}

\paragraph{Unitarity norm}
The penultimate correctness criterion to be investigated in the context of \eqref{eq:hubbard} is the one based on the unitarity norm \eqref{eq:unitarity_norm}. As before, what is studied are histograms of this quantity on the restricted ensemble, which are shown for both $N_I=0.0$ and $N_I=11.641$ in \cref{fig:hubbard_unitarity_norm}. Given  \cref{fig:hubbard_histogram_2D,fig:hubbard_histogram_1D}, it does not come as a surprise that in the case of purely real noise the unitarity norm is confined to a very narrow range. The values that $\mathcal{N}_U$ is restricted to in this case are not very large, such that one might conclude correct convergence in this case. This, however, is known not to be true from \cref{tab:hubbard}. For $N_I=11.641$, on the other hand, the unitarity norm shows a more familiar behavior, its distribution seemingly decaying faster than polynomially but reaching rather large values. In that case, it is more natural to conclude incorrect convergence from the behavior of the unitarity norm. Note, however, that such a statement would still be very much heuristic in nature.

\paragraph{Configurational temperature}
Lastly, the inverse configurational temperature observable \eqref{eq:configurational_temperature} in the model \eqref{eq:hubbard} reads
\begin{equation}
	\tilde{\beta} = \frac{U}{\beta}\left\langle\frac{1+\gamma^2+2\gamma(1+U\beta)}{(\gamma(2U+\ii z)+\ii z)^2}\right\rangle\;, 
	\quad \textnormal{with} \quad
	\gamma = e^{\beta\left(\ii z + \mu + \frac{U}{2}\right)}\;.
\end{equation}
Complex Langevin simulations of \eqref{eq:hubbard} using \eqref{eq:imaginary_noise} give 
\begin{equation}
	\tilde{\beta}_\CL = 0.6(2)+1(1)\ii 
	\quad \textnormal{for} \quad 
	N_I=0.0\;,
\end{equation}
and 
\begin{equation}
	\tilde{\beta}_\CL = -0.3(2)-0.3(3)\ii
	\quad \textnormal{for} \quad
	N_I=11.641\;,
\end{equation}
respectively. Clearly, both of these results, which were evaluated on the restricted ensemble, deviate significantly from $\tilde{\beta}=1$, from which one concludes that the configurational-temperature criterion correctly predicts wrong convergence.

\paragraph{Summary}
To summarize the insights gained by the application of different correctness criteria to the one-site Hubbard model \eqref{eq:hubbard} with and without the use of complex noise in complex Langevin dynamics, it can be said that the histogram, drift and configurational-temperature criteria detect the incorrect convergence rather easily. For the boundary terms and convergence conditions, on the other hand, the situation is more complicated. For each individual maximum Langevin-time step-size considered in this work they both predict incorrect convergence (for all observables), but this behavior appears to change in the limit $\eps_{\max}\to0$. Similarly, the interpretation of the Dyson--Schwinger equations is not entirely clear either. While for larger step-sizes (and sufficient statistics) they are violated, upon decreasing $\eps_{\max}$ they are found to hold (in the restricted ensemble) again. Thus, in the limit of vanishing step size the Dyson--Schwinger equations, boundary terms, and convergence conditions all suggest correct convergence, which contradicts the comparison with exact solutions in \cref{tab:hubbard}. In any case, the findings of this subsection likely exclude the possibility of obtaining a correct result for a given observable in practice by tuning the imaginary noise parameter until the boundary terms for that observable vanish

Furthermore, the criterion based on the unitarity norm is again not fully conclusive. Finally, the observable-bounds criterion hinges on the validity of the Dyson--Schwinger equations. If they turn out to be violated, then the bounds provide no further insights. If the Dyson--Schwinger equations hold, on the other hand, one can find a violation of the bounds in the case of complex noise rather easily, while for real noise a similar violation could not be found with the considered parametrization and control observables.

\subsection{Quartic model on a complex time-contour}\label{sec:quartic_4D}
The final theory considered in this work is the only one describing more than one variable. Indeed, it is a simple discretization of the one-dimensional quantum field theory defined by the Lagrangian
\begin{equation}\label{eq:quartic_4D_continuum}
	\mathcal{L}[\phi] = \frac{1}{2}(\partial_t\phi(t))^2 - \frac{m^2}{2}\phi^2(t) - \frac{\lambda}{4}\phi^4(t)\;,
\end{equation}
describing a real scalar field $\phi$ with mass $m$ and quartic self-coupling $\lambda$. In contrast to the previous models, here, the Euclidean notation shall be abandoned and a Minkowski formulation is employed instead, such that the path integral density $\rho$ is defined as
\begin{equation}
	\rho[\phi] = e^{\ii S[\phi]}\;, \quad 
	S[\phi] = \int dt \mathcal{L}[\phi]\;.
\end{equation}
As with the previous models, \eqref{eq:quartic_4D_continuum} (and its generalization to higher-dimensional spacetimes)  has been studied using complex Langevin dynamics before \cite{BS05,BBS07,ALR21,ALR23,LS23,ARS24,Lar25}. In order to describe the real-time evolution of this theory, in practice one discretizes \eqref{eq:quartic_4D_continuum} not on a purely real time-contour, but resorts to regularized, Schwinger--Keldysch-type contours for reasons of numerical stability and removes the regulator by extrapolation. In such a scenario, the limiting factor appears to be the corresponding real-time extent $t_{\max}$; if it becomes too large, complex Langevin dynamics produce incorrect results \cite{LS23}, the origin of which as a function of $m$, $\lambda$, $t_{\max}$ and the regulator is not fully understood. The goal of this subsection is, on the one hand, to help contribute to this understanding and, on the other hand, to test the usefulness of the correctness criteria of \cref{sec:criteria} in the present setup.

\begin{figure}[t]
	\centering		
	\includegraphics[scale=1.]{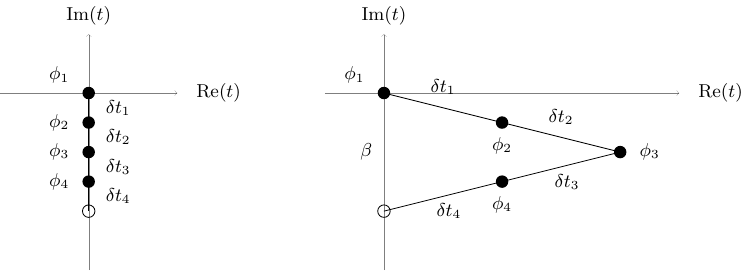}
	\caption{Lattice discretization of the considered time contours. The $\phi_i$ denote the four degrees of freedom and the $\delta t_i$ are the (generally) complex time differences between the lattice points. The open circles indicate the variable $\phi_{5}$ equivalent to $\phi_1$ via periodicity in the imaginary direction. \emph{(left)}: all $\delta t_i$ are equal and purely imaginary, with $\imag\,\delta t_i<0$. \emph{(right)}: $\delta t_1=\delta t_2\neq\delta t_3=\delta t_4$, with $\real\,\delta t_i\neq0$ and $\imag\,\delta t_i<0$.}
	\label{fig:quartic_4D_contour}
\end{figure}

In the context of this work, \eqref{eq:quartic_4D_continuum} is discretized as follows: First, the time-contour in the complex plane, on which periodic boundary conditions for the field $\phi$ are imposed, is divided into $N_t$ segments. Correspondingly, $N_t$ is also equal to the number of lattice points $t_i$ ($i=1,\dots,N_t$). With the definition $\phi_i:=\phi(t_i)$, the derivative of $\phi(t)$ in \eqref{eq:quartic_4D_continuum} is discretized using the forward-difference method as
\begin{equation}
	\partial_t\phi(t)\vert_{t=t_i} \approx \frac{\phi_{i+1}-\phi_i}{\delta t_i}\;, \quad \textnormal{with} \quad
	\delta t_{i} = t_{i+1} - t_i\;.
\end{equation}
Note that $\delta t_{i}$ will be different for different $i$ in general. With this, the discretization of \eqref{eq:quartic_4D_continuum} leads to the lattice action
\begin{equation}\label{eq:quartic_4D}
	S[\phi] = \sum_{i=1}^{N_t} \left(\frac{(\phi_{i+1}-\phi_i)^2}{2\delta t_i}-\delta t_i\frac{m^2}{2}\phi_i^2 - \delta t_i\frac{\lambda}{4}\phi_i^4\right)\;,
\end{equation}
with the periodic boundary condition $\phi_{N_t+1}=\phi_1$. Here, only coarse lattices with $N_t=4$ are considered, mainly for illustrative purposes. The lattice points are assumed to be of the form
\begin{align}\label{eq:contour}
	t_1 = 0\;, \quad 
	t_2 = \frac{t_{\max}}{2} - \ii \frac{\beta}{4}\;, \quad 
	t_3 = t_{\max} - \ii \frac{\beta}{2}\;, \quad 
	t_4 = \frac{t_{\max}}{2} - 3\ii\frac{\beta}{4}\;,
\end{align}
where $t_{\max}$ denotes the maximal extent of the time-contour in the real direction. Similarly, $\beta$ is the extent in the imaginary direction and plays the role of an inverse temperature as usual. It should also be noted that under periodic boundary conditions, one may define $t_5=-\ii\beta$. The setup is illustrated in \cref{fig:quartic_4D_contour}. 

With $t_{\max}=0$, one arrives back at the Euclidean formulation of the theory, which, for real $m$ and $\lambda$, does not have a sign problem since $S[\phi]\in\mathbb{R}$. This, in turn, suggests that on such a contour complex Langevin dynamics will give correct results. While this is still expected to be true for small nonzero $t_{\max}>0$, it is no longer the case once $t_{\max}$ becomes sufficiently large. It should be mentioned that in this context the notion of correct results does not correspond to the continuum formulation of the theory, but to expectation values obtained by solving the lattice model \eqref{eq:quartic_4D}.

To investigate the aforementioned behavior, three different real-time extents are considered: $t_{\max}=0.0$, $0.4$, and $0.8$, while the remaining parameters are fixed to $m=1$, $\lambda=4$, and $\beta=1$, respectively. The simulations of \eqref{eq:quartic_4D} use the improved update scheme \eqref{eq:improved_scheme} and a maximum Langevin-time step size of $\eps_{\max} = 10^{-5}$. While no exact results are available for the model \eqref{eq:quartic_4D}, for $N_t=4$ the system is still sufficiently simple to allow for straightforward numerical integration schemes to be applicable. The so-obtained ``exact'' results are compared with complex-Langevin simulation results below.

\begin{figure}[t]
	\centering
	\includegraphics[scale=0.5]{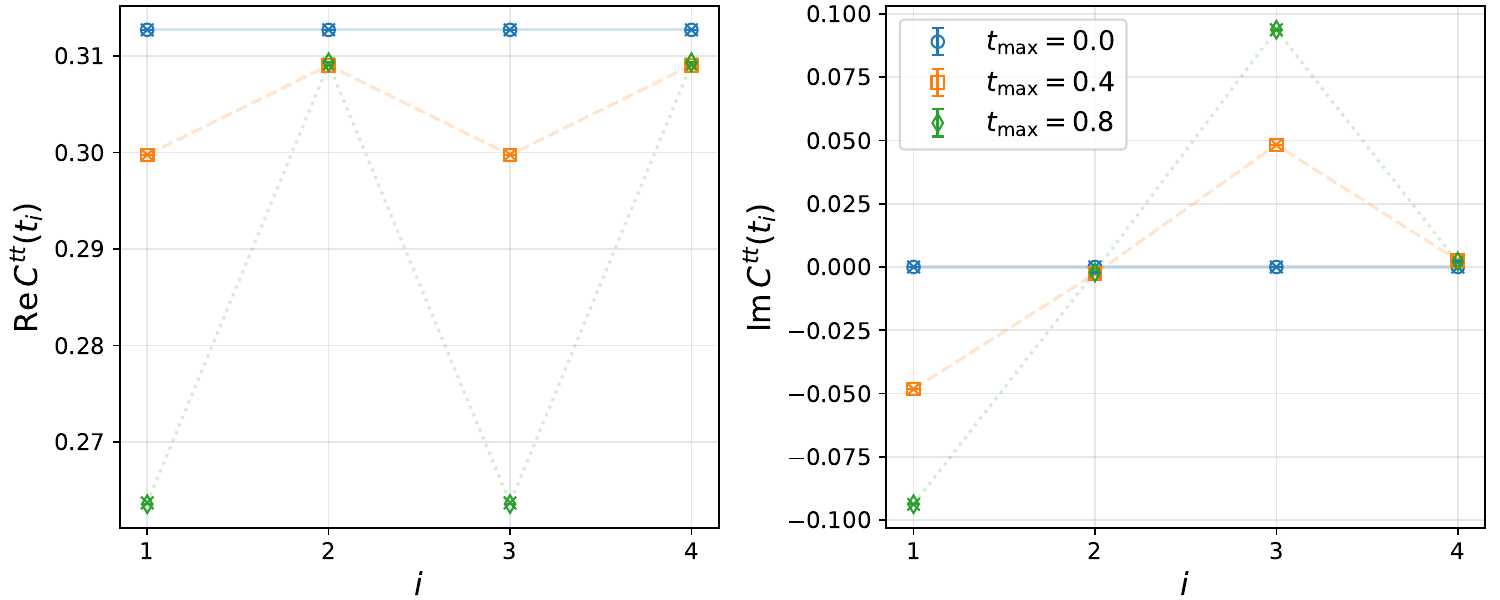}
	\caption{Equal-time correlation function $C^{tt}$ in \eqref{eq:equal_time_correlator} on the discrete time-contour \eqref{eq:contour} in complex Langevin simulations of the model \eqref{eq:quartic_4D} with $\beta=1$, $m=1$, $\lambda=4$, and different real-time extents $t_{\max}$. Exact results are shown as crosses connected by faint straight lines to guide the eye. \emph{(left)}: $\real\,C^{tt}(t_i)$. \emph{(right)}: $\imag\,C^{tt}(t_i)$.}
	\label{fig:equal_time_correlator}
\end{figure}

\begin{figure}[t]
	\centering
	\includegraphics[scale=0.5]{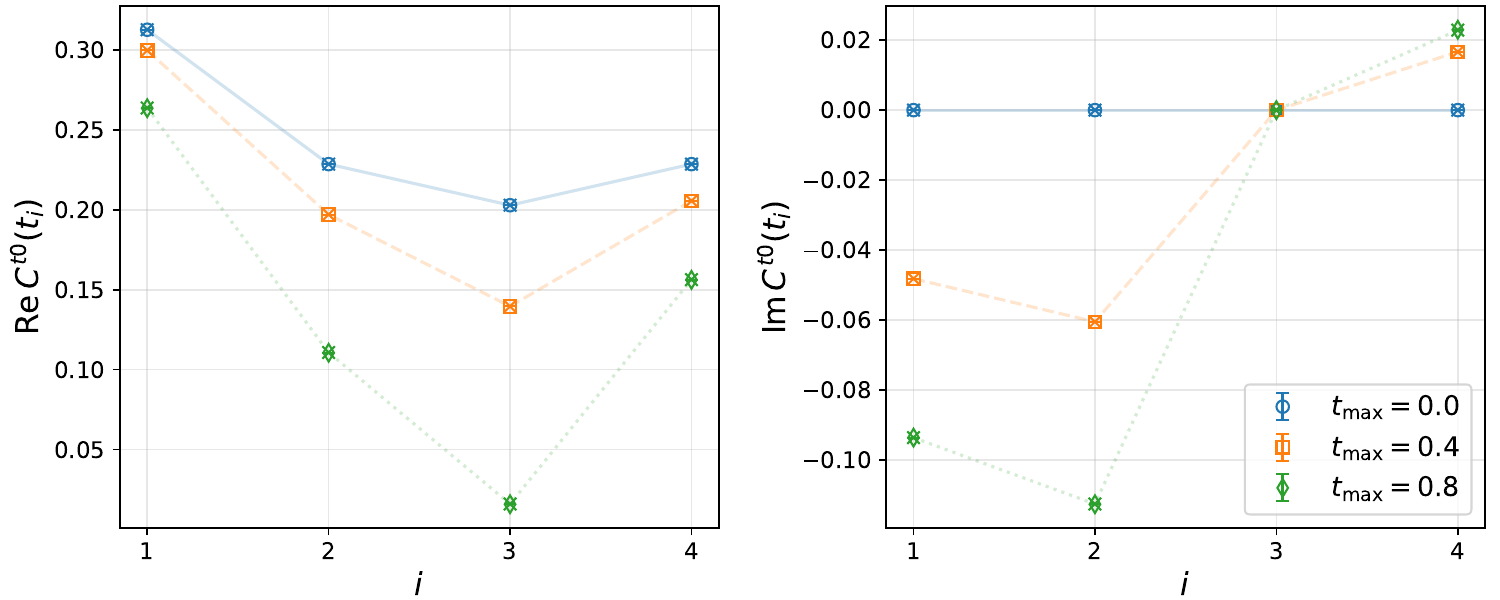}
	\caption{Similar to \cref{fig:equal_time_correlator}, but for the unequal-time correlation function $C^{t0}$ in \eqref{eq:unequal_time_correlator}.}
	\label{fig:unequal_time_correlator}
\end{figure}

Particularly interesting observables in the theory \eqref{eq:quartic_4D} are the equal-time correlation function
\begin{equation}\label{eq:equal_time_correlator}
	C^{tt}(t) = \langle\phi(t)\phi(t)\rangle\;,
\end{equation}
as well as the unequal-time correlator
\begin{equation}\label{eq:unequal_time_correlator}
	C^{t0}(t) = \langle\phi(t)\phi(0)\rangle\;,
\end{equation}
evaluated on the discretized contour, i.e., $t=t_i$. Results for $C^{tt}$ and $C^{t0}$ for different $t_{\max}$, including a comparison with exact results, are shown in \cref{fig:equal_time_correlator,fig:unequal_time_correlator}, respectively. In these plots, discretization effects become apparent, since in the continuum limit the equal-time correlator should be constant on the real branch of the contour, which is clearly not the case for the lattice employed in this work. As was mentioned above, however, this fact is of no concern here. Rather, what shall be interesting to study is the deviations of the complex Langevin results from the exact ones. While no such deviations can be seen on the scales shown in \cref{fig:equal_time_correlator,fig:unequal_time_correlator}, upon closer inspection a discrepancy indeed begins to emerge for $t_{\max}>0$, which becomes larger for larger real-time extents (keeping all other parameters fixed).

To see this, however, it shall be more convenient in the following to investigate monomial observables of the form $\obs[\phi]=\phi_1^{n_1}\phi_2^{n_2}\phi_3^{n_3}\phi_4^{n_4}$, with integer $n_i\geq0$ and $\sum_{i=1}^4n_i=2$, from which the correlation functions \eqref{eq:equal_time_correlator} and \eqref{eq:unequal_time_correlator} for $t=t_i$ can be trivially reconstructed. Comparisons between complex Langevin and exact results for observables of this type and different $t_{\max}$ are shown in \cref{tab:quartic_4D}. As can be seen, for the two smaller real-time extents the exact solutions are reproduced by the complex Langevin results, whereas a statistically significant (albeit still small) discrepancy can be observed for $t_{\max}=0.8$. This deviation is only expected to grow for even larger $t_{\max}$. Increasing the real-time extent while keeping $\beta$, $m$, $\lambda$, and $N_t$ fixed means that the real parts of the lattice spacings $\delta t_i$ and, thus, the ratios $r_i=\frac{\real\,\delta t_i}{\imag\,\delta t_i}$ increase (in magnitude) as well. In the present setup, $\vert r_i\vert=2t_{\max}$. A straightforward interpretation of these findings would be that complex Langevin produces correct results as long as $\vert r_i\vert$ is sufficiently small; for instance, $\vert r_i\vert\leq1$. This makes clear the problems one faces when attempting to remove the regulator, which would lead to $\imag\,\delta t_i=0$. 

The remainder of this subsection is concerned with the question which (if any) correctness criteria of \cref{sec:criteria} are capable of detecting the incorrect convergence for $t_{\max}=0.8$ observed in \cref{tab:quartic_4D}. As in that table, the considered observables are $\obs[\phi]=\phi_i\phi_j$ with $i, j=1,2,3,4$.

\begin{table*}[t]
    \centering
    \renewcommand{\arraystretch}{1.2}
	\resizebox{\textwidth}{!}{
    \begin{tabular}{|c|cc|cc|cc|}
    \hline
    \multirow{2}{*}{$\obs[\phi]$} & \multicolumn{2}{c|}{$t_{\max}=0.0$} & \multicolumn{2}{c|}{$t_{\max}=0.4$} & \multicolumn{2}{c|}{$t_{\max}=0.8$} \\
    & $\langle\obs(z)\rangle_\CL$ & correct & $\langle\obs(z)\rangle_\CL$ & correct & $\langle\obs(z)\rangle_\CL$ & correct \\
    \hline
	$\phi_1^2$ & 
				 $0.312678(9)$ & 
				 $0.312680$ & 
				 $0.299743(9)-0.048219(3)\ii$ & 
				 $0.299743-0.048219\ii$ & 
				 $0.263590(6)-0.093670(5)\ii$ & 
				 $0.263723-0.093680\ii$ \\
	$\phi_2^2$ & 
				 $0.312667(9)$ & 
				 $0.312680$ & 
				 $0.30899(1)-0.002642(3)\ii$ & 
				 $0.309002-0.002643\ii$ & 
				 $0.309317(9)-0.002166(6)\ii$ & 
				 $0.309134-0.001826\ii$ \\
	$\phi_3^2$ & 
				 $0.312675(9)$ & 
				 $0.312680$ & 
				 $0.299733(9)+0.048213(3)\ii$ & 
				 $0.299743+0.048219\ii$ & 
				 $0.263606(7)+0.093665(5)\ii$ & 
				 $0.263723+0.093680\ii$ \\
	$\phi_4^2$ & 
				 $0.312671(9)$ & 
				 $0.312680$ & 
				 $0.30899(1)+0.002640(3)\ii$ & 
				 $0.309002+0.002643\ii$ & 
				 $0.309316(9)+0.002161(5)\ii$ & 
				 $0.309134+0.001826\ii$ \\
	$\phi_1\phi_2$ & 
					 $0.228578(9)$ & 
					 $0.228586$ & 
					 $0.19703(1)-0.060547(3)\ii$ & 
					 $0.197038-0.060545\ii$ & 
					 $0.110552(7)-0.112736(4)\ii$ & 
					 $0.110684-0.112668\ii$ \\
	$\phi_2\phi_3$ & 
					 $0.228576(9)$ & 
					 $0.228586$ & 
					 $0.20560(1)-0.016680(2)\ii$ & 
					 $0.205617-0.016677\ii$ & 
					 $0.156176(7)-0.023000(4)\ii$ & 
					 $0.156289-0.022883\ii$ \\
	$\phi_3\phi_4$ & 
					 $0.228578(9)$ & 
					 $0.228586$ & 
					 $0.197023(9)+0.060542(3)\ii$ & 
					 $0.197038+0.060545\ii$ & 
					 $0.110556(6)+0.112733(5)\ii$ & 
					 $0.110684+0.112668\ii$ \\
	$\phi_4\phi_1$ & 
					 $0.228579(9)$ & 
					 $0.228586$ & 
					 $0.205609(9)+0.016676(2)\ii$ & 
					 $0.205617+0.016677\ii$ & 
					 $0.156162(8)+0.022999(4)\ii$ & 
					 $0.156289+0.022883\ii$ \\
	$\phi_1\phi_3$ & 
					 $0.202868(9)$ & 
					 $0.202875$ & 
					 $0.139592(9)-2(2)\times10^{-6}\ii$ & 
					 $0.139605$ & 
					 $0.015751(7)-4(4)\times10^{-6}\ii$ & 
					 $0.015914$ \\
	$\phi_2\phi_3$ & 
					 $0.202866(9)$ & 
					 $0.202875$ & 
					 $0.19172(1)-1(2)\times10^{-6}\ii$ & 
					 $0.191731$ & 
					 $0.161702(8)-3(3)\times10^{-6}\ii$ &
					 $0.161700$ \\
    \hline
    \end{tabular}}
    \caption{Comparison of expectation values $\langle\phi_1^{n_1}\phi_2^{n_2}\phi_3^{n_3}\phi_4^{n_4}\rangle $ between complex-Langevin simulations of the model \eqref{eq:quartic_4D} with $m=1$, $\lambda=4$, $\beta=1$, and different real-time extents $t_{\max}$ with `exact' solutions, computed via numerical integration. The parentheses indicate the statistical uncertainties rounded to their respective first signficant digits. For $t_{\max}=0.0$, the imaginary parts of all observables are negligibly small.}
    \label{tab:quartic_4D}
\end{table*}

\paragraph{Dyson--Schwinger equations}
The Dyson--Schwinger equations \eqref{eq:dse} of the model \eqref{eq:quartic_4D} read
\begin{equation}\label{eq:quartic_4D_dse}
	\langle A_i\obs[\phi]\rangle = 0\;, \quad \textnormal{with} \quad
	A_i\obs[\phi] = \left(\frac{\partial}{\partial\phi_i}+D_i[\phi]\right)\obs[\phi]\;,
\end{equation}
where the $i$-th component of the drift term is given by the adaption of \eqref{eq:drift} to the Minkowskian metric (without a kernel), i.e, $D_i=\ii\frac{\partial S[\phi]}{\partial\phi_i}$. In the model \eqref{eq:quartic_4D}, it can be expressed as
\begin{equation}\label{eq:quartic_4D_drift}
	-\ii D_i[\phi] = \frac{\phi_i-\phi_{i+1}}{\delta t_i} + \frac{\phi_i-\phi_{i-1}}{\delta t_{i-1}} - \delta t_i m^2\phi_i - \delta t_i \lambda \phi_i^3\;.
\end{equation}
The validity of \eqref{eq:quartic_4D_dse} is investigated in \cref{tab:quartic_4D_dse}, considering $i=1$ exemplarily; the results for the other components are similar. Since the number of monomial observables one needs to measure in order to compute \eqref{eq:quartic_4D_dse} becomes quite large, such a measurement was not performed online and the restricted ensemble was used instead. As can be seen, no significant deviation of $\langle A_1\obs[\phi]\rangle_{\CL}$ from zero is observed for any observable $\obs[\phi]$ and real-time extent $t_{\max}$. This would be a sign for correct convergence across the board, implying that the Dyson--Schwinger equations are incapable of detecting the incorrect convergence for $t_{\max}=0.8$. However, it might again be attributable simply to insufficient statistics in the restricted ensemble. In fact, this appears to be the most likely scenario.

\begin{table*}[t]
    \centering
    \renewcommand{\arraystretch}{1.1}
	\resizebox{\textwidth}{!}{
    \begin{tabular}{|c|ccc|}
    \hline
    $\obs[\phi]$ & $\langle A_1\obs(z)\rangle_\CL$ ($t_{\max}=0.0$) & $\langle A_1\obs(z)\rangle_\CL$ ($t_{\max}=0.4$) & $\langle A_1\obs(z)\rangle_\CL$ ($t_{\max}=0.8$) \\
    \hline
	$\phi_1^2$ & 
				 $-4(6)\times10^{-4}$ & 
				 $3(6)\times10^{-4}-1(2)\times10^{-4}\ii$ & 
				 $1(1)\times10^{-3}-1(1)\times10^{-3}\ii$ \\
	$\phi_2^2$ & 
				 $3(4)\times10^{-4}$ & 
				 $7(5)\times10^{-4}-0(2)\times10^{-4}\ii$ & 
				 $-6(5)\times10^{-4}-8(6)\times10^{-4}\ii$ \\
	$\phi_3^2$ & 
				 $-1(5)\times10^{-4}$ & 
				 $5(5)\times10^{-4}+1(1)\times10^{-4}\ii$ & 
				 $5(4)\times10^{-4}-0(3)\times10^{-4}\ii$ \\
	$\phi_4^2$ & 
				 $4(5)\times10^{-4}$ & 
				 $-2(5)\times10^{-4}+4(2)\times10^{-4}\ii$ & 
				 $-2(5)\times10^{-4}+1(3)\times10^{-4}\ii$ \\
	$\phi_1\phi_2$ & 
		 		 	 $2(4)\times10^{-4}$ & 
		 		 	 $9(5)\times10^{-4}+0(1)\times10^{-4}\ii$ & 
		 		 	 $-0(5)\times10^{-4}-3(3)\times10^{-4}\ii$ \\
	$\phi_2\phi_3$ & 
					 $-2(4)\times10^{-4}$ & 
					 $4(4)\times10^{-4}+0(1)\times10^{-4}\ii$ & 
					 $2(4)\times10^{-4}+1(2)\times10^{-4}\ii$ \\
	$\phi_3\phi_4$ & 
					 $-0(4)\times10^{-4}$ & 
					 $-0(4)\times10^{-4}+3(1)\times10^{-4}\ii$ & 
					 $1(3)\times10^{-4}+1(2)\times10^{-4}\ii$ \\
	$\phi_4\phi_1$ & 
					 $4(4)\times10^{-4}$ & 
					 $1(5)\times10^{-4}+3(1)\times10^{-4}\ii$ & 
					 $-3(4)\times10^{-4}+2(3)\times10^{-4}\ii$ \\
	$\phi_1\phi_3$ & 
					 $-2(4)\times10^{-4}$ & 
					 $5(4)\times10^{-4}+1.1(9)\times10^{-4}\ii$ & 
					 $1(3)\times10^{-4}+0(4)\times10^{-4}\ii$ \\
	$\phi_2\phi_3$ & 
					 $2(4)\times10^{-4}$ & 
					 $-1(4)\times10^{-4}+2(1)\times10^{-4}\ii$ & 
					 $0(4)\times10^{-4}+1(2)\times10^{-4}\ii$ \\
    \hline
    \end{tabular}}
   	\caption{Validity of the Dyson--Schwinger equations \eqref{eq:quartic_4D_dse} for $i=1$ in complex-Langevin simulations of the model \eqref{eq:quartic_4D} with $m=1$, $\lambda=4$, $\beta=1$ and various real-time extents $t_{\max}$, computed in the restricted ensemble. The parentheses indicate the statistical uncertainties rounded to their first respective significant digits. For $t_{\max}=0.0$, the imaginary parts are negligibly small. Results for $\langle A_i\obs(z)\rangle_\CL$ with $i>1$ are similar.}
    \label{tab:quartic_4D_dse}
\end{table*}

\begin{figure}[t]
	\centering
	\includegraphics[scale=0.5]{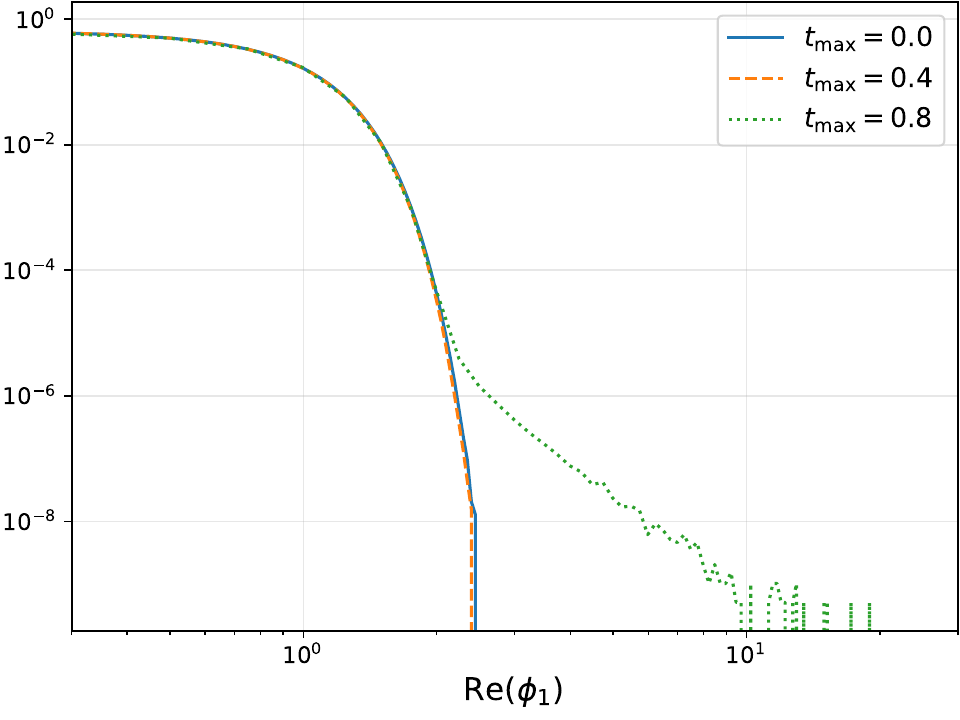}
	\caption{Histograms of the real part of $\phi_1$ resulting from complex Langevin simulations of the model \eqref{eq:quartic_4D} with $\beta=1$, $m=1$, $\lambda=4$, and different real-time extents $t_{\max}$, arising from projecting histograms in the complex plane onto the real axis via integrating over the imaginary axis. Histograms for $\real\,\phi_{i\neq1}$ show the same qualitative behavior. The same is true for $\imag\,\phi_i$, except that for $t_{\max}=0.0$ this quantity vanishes. Both axes are logarithmic.}
	\label{fig:quartic_4D_histogram_1D}
\end{figure}

\paragraph{Histograms}
Once again, one may investigate histograms of the dynamical variables in an attempt to gain some insights in the convergence properties. For the model \eqref{eq:quartic_4D}, however, there are now four degrees of freedom, the respective distributions of which are, in general, different. With the present choice of parameters, it was verified that there is no qualitative difference between the respective histograms, such that one may focus on, say, $\phi_1$ without loss of generality. Since the drift term of \eqref{eq:quartic_4D} has no pole and the decay properties are anyways harder to read off from histograms in the complex plane, the latter are not shown here. Instead, one may again consider their projections onto the real (or imaginary) axis. This is shown in \cref{fig:quartic_4D_histogram_1D}.

There is a (perhaps surprisingly) clear trend in this figure. Namely, for $t_{\max}=0.0$ and $0.4$, the decay of the distributions towards infinity is faster than polynomial, while there is a pronounced algebraic tail for $t_{\max}=0.8$. From this, one would conclude that only the latter case gives rise to incorrect convergence, which is true according to \cref{tab:quartic_4D}. 

\begin{figure}[t]
	\centering
	\includegraphics[scale=0.6]{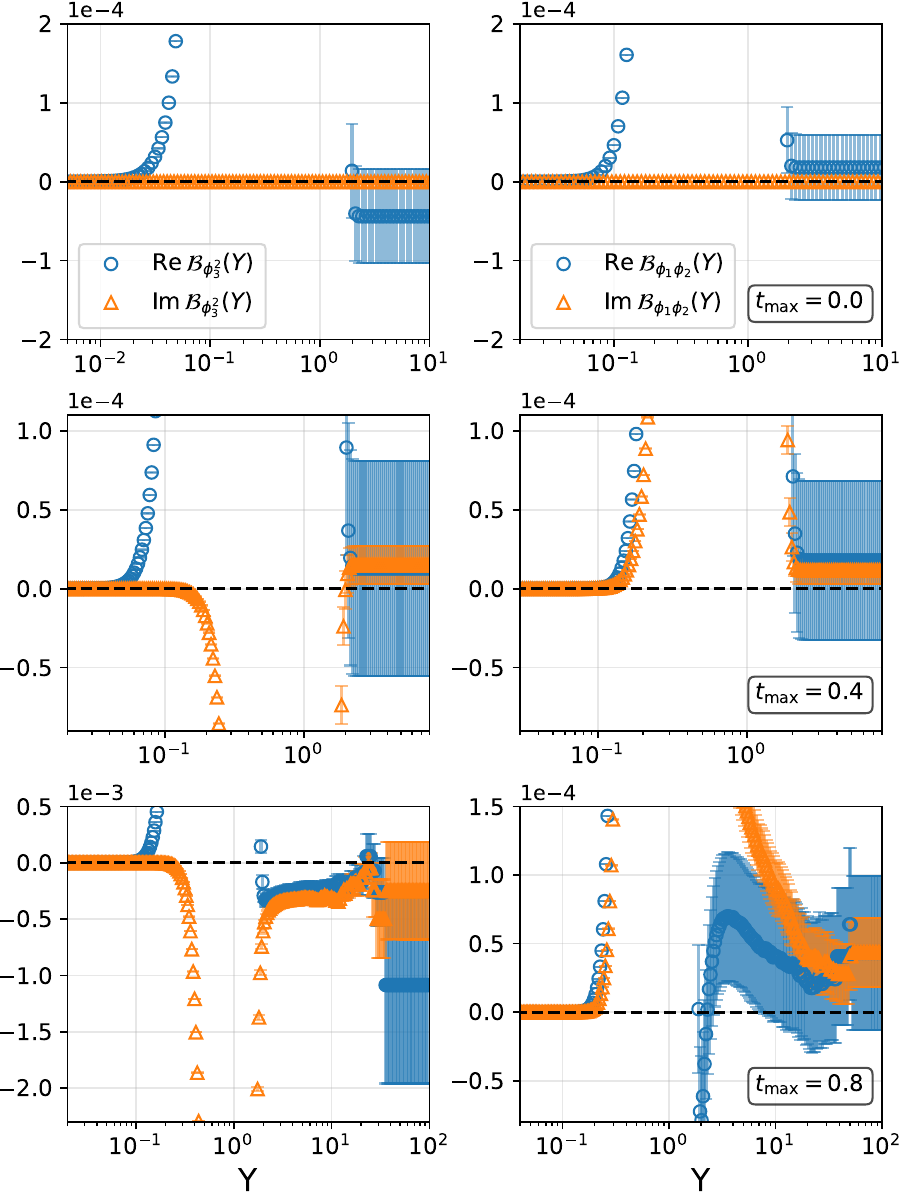}
	\caption{Boundary-term observables \eqref{eq:quartic_4D_boundary_terms} in complex Langevin simulations of the model \eqref{eq:quartic_4D} with $N_t=4$, $\beta=1$, $m=1$, $\lambda=4$ as a function of the cutoff $Y$. The upper, middle, and lower rows show results for $t_{\max}=0.0$, $0.4$, and $0.8$, respectively, while the left and right columns show the boundary-term observables for $\obs[\phi]=\phi_3^2$ and $\obs[\phi]=\phi_1\phi_2$, respectively. The horizontal axes are logarithmic, the data points for $\imag\,\mathcal{B}_{\obs[\phi]}(Y)$ are slightly offset horizontally for better visibility, and the dashed horizontal lines mark zero.}
	\label{fig:quartic_4D_boundary_terms}
\end{figure}

\paragraph{Boundary terms}
In the model \eqref{eq:quartic_4D} with a Minkowskian signature,  \eqref{eq:boundary_terms} can be written as
\begin{equation}\label{eq:quartic_4D_boundary_terms}
	\mathcal{B}_{\obs[\phi]}(Y) = \left\langle\Theta\left(Y-\max_{j}\vert\phi_j\vert\right)\left(\frac{\partial}{\partial\phi_i}+D_i[\phi]\right)\frac{\partial}{\partial\phi_i}\obs[\phi]\right\rangle\;,
\end{equation}
with the drift term $D_i$ defined in \eqref{eq:quartic_4D_drift}. Exemplary plots of the boundary-term observables as a function of the cutoff $Y$ for $\obs[\phi]=\phi_3^2$ and $\obs[\phi]=\phi_1\phi_2$ are shown in \cref{fig:quartic_4D_boundary_terms}.

For $t_{\max}=0.0$, both observables have boundary terms consistent with zero, hinting at correct convergence, which agrees with the results of \cref{tab:quartic_4D}. The same is true for $t_{\max}=0.4$. For the largest real-time extent, $t_{\max}=0.8$, on the other hand, the behavior changes rather drastically. Indeed, for $\obs[\phi]=\phi_3^2$, one finds a plateau-like region at a value different from zero for small $Y$, whereas upon increasing $Y$ another, more pronounced plateau emerges, at a value consistent with zero within less than twice the statistical uncertainty. Such behavior, while not uncommon in the literature (see, e.g., \cite{SSS20}) has not occurred in any other model considered in this work. One may thus conjecture that it is more prone to happen when multiple degrees of freedom are involved. The usual interpretation of such a result is that it indicates nonvanishing boundary terms and, thus, incorrect results due to the first plateau. This is indeed consistent with the findings of \cref{tab:quartic_4D}. 

For the other observable, $\obs[\phi]=\phi_1\phi_2$, the situation is again different. A clear signal for a plateau is only observed for rather large values of the cutoff $Y$, at a value of $\imag\,\mathcal{B}_{\phi_1\phi_2}$ consistent with zero within around twice the statistical uncertainty, the real part agreeing within one standard deviation. For this observable, complex Langevin dynamics produce a slightly incorrect result according to \cref{tab:quartic_4D}. The boundary term observable, however, is not entirely conclusive. Taken at face value and interpreted in a similar way as in previous subsections, one would (erroneously) conclude correct convergence. However, upon increasing the statistics, it is plausible that signs of nonvanishing boundary terms might solidify also for this observable. It should be mentioned that due to the improved update prescription \eqref{eq:improved_scheme} used for the simulations of \eqref{eq:quartic_4D}, finite-step-size effects for the boundary-term observables are believed to be under control in the present case. Finally, since the deviation observed in \cref{tab:quartic_4D} is still rather small numerically, the corresponding boundary terms are in any case expected to be small, and possibly hard to distinguish from zero, as well.

\paragraph{Convergence conditions}
The model \eqref{eq:quartic_4D} provides a first example of the correctness criterion based on boundary terms and the convergence conditions producing different conclusions. Indeed, according to \cref{fig:quartic_4D_boundary_terms}, upon taking the $Y\to\infty$ limit of $\mathcal{B}_{\phi_3^2}(Y)$ for $t_{\max}=0.8$, one finds that the convergence conditions are valid, while, as per the discussion above, boundary terms predict incorrect convergence. Hence, this is a situation in which the boundary-term obseravbles are more conclusive than the convergence conditions. The same cannot be said for the observable $\obs[\phi]=\phi_1\phi_2$ for the same real-time extent, however.

\begin{figure}[t]
	\centering
	\includegraphics[scale=0.5]{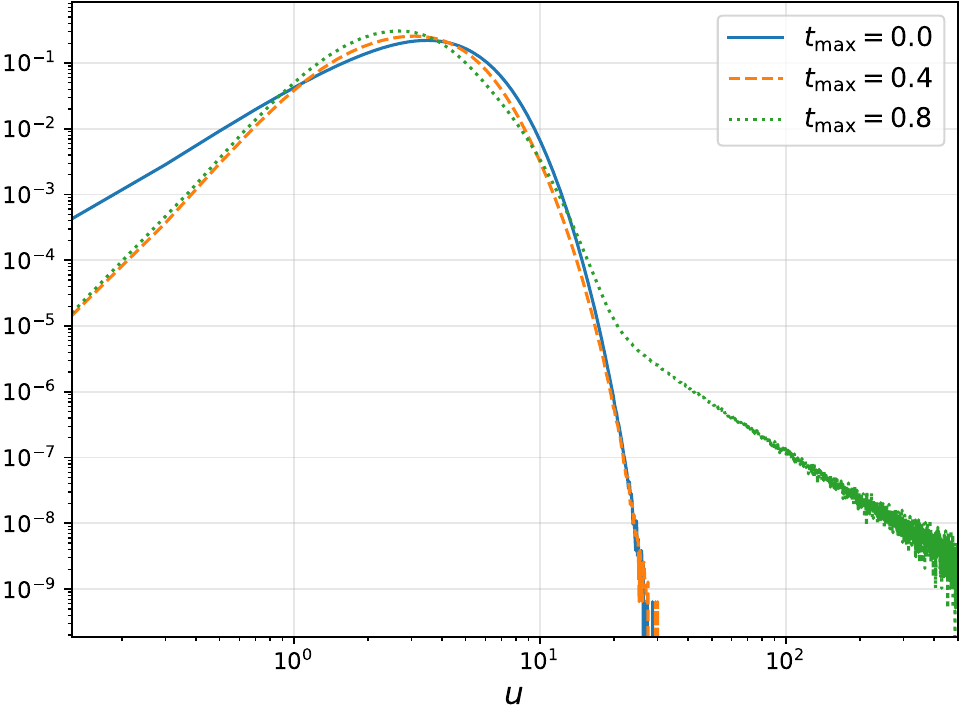}
	\caption{Histograms of the maximum drift-term magnitude $u=\max_i\vert D_i\vert$ (with $D_i$ given in \eqref{eq:quartic_4D_drift}) in complex Langevin simulations of the model \eqref{eq:quartic_4D} with $\beta=1$, $m=1$, $\lambda=4$ and different real-time extents $t_{\max}$. Both axes are logarithmic.}
	\label{fig:quartic_4D_drift}
\end{figure}

\paragraph{Drift criterion}
In models with multiple degrees of freedom like \eqref{eq:quartic_4D}, the drift criterion involves taking the maximum of the magnitudes of all drift term components according to \eqref{eq:max_drift}. For the three different real-time extents, such histograms are shown in \cref{fig:quartic_4D_drift}. Similarly to \cref{fig:quartic_4D_histogram_1D}, one finds an algebraic decay for $t_{\max}=0.8$ and an exponential (or faster) decay for the smaller real-time extents. Thus, the drift criterion properly predicts correct and incorrect convergence also in this case.

\paragraph{Observable bounds}
Given that the deviations between complex Langevin and exact results for $t_{\max}=0.8$ are only very small according to \cref{tab:quartic_4D} in spite of their statistical significance, finding a suitable polynomial control variable violating the bounds \eqref{eq:observable_bounds} is likely a highly nontrivial task. This was one of the conclusions of \cref{sec:one_pole} as well. Thus, while a few simple observables like $\phi_i^2$ were checked explicitly and found not to violate the bounds, a more systematic study of \eqref{eq:observable_bounds} for the model \eqref{eq:quartic_4D} is not performed in this work. In fact, if the Dyson--Schwinger equations indeed turn out to be violated for $t_{\max}=0.8$ upon increasing the statistics, which is the expected outcome, such an investigation would be obsolete anyways, since the first requirement of the observable-bound criterion would already be violated.

\begin{figure}[t]
	\centering
	\includegraphics[scale=0.5]{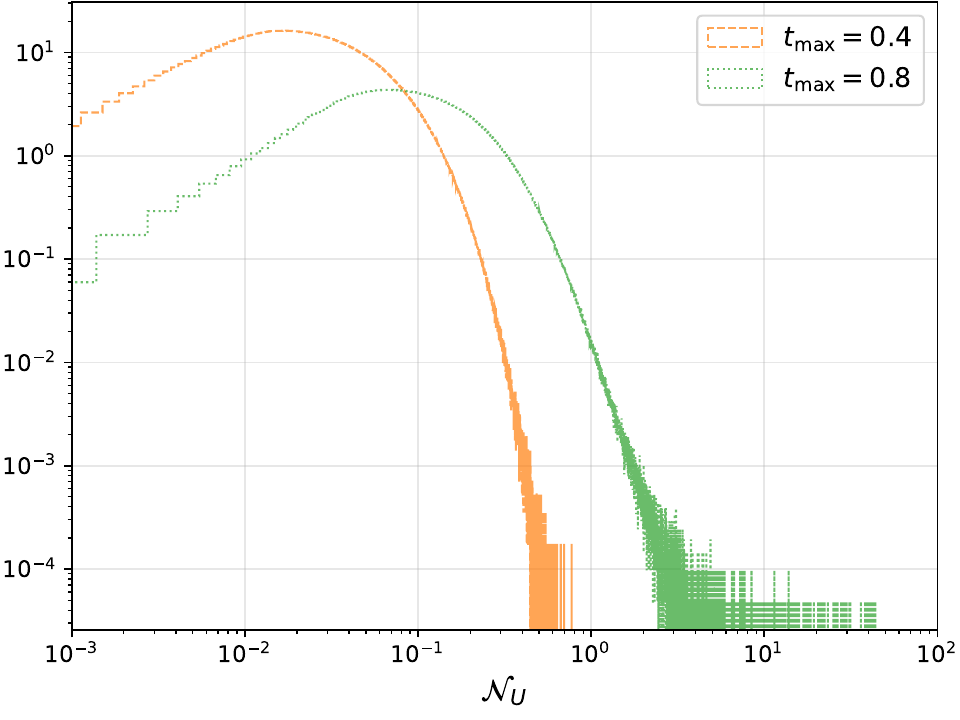}
	\caption{Histograms of $\mathcal{N}_U=\sum_{i}\imag(\phi_i)^2$ in complex Langevin simulations of the model \eqref{eq:quartic_4D} with $\beta=1$, $m=1$, $\lambda=4$ and different real-time extents $t_{\max}$, computed on the restricted ensemble. For $t_{\max}=0.0$ the unitarity norm is zero and thus not shown. Both axes are logarithmic.}
	\label{fig:quartic_4D_unitarity_norm}
\end{figure}

\paragraph{Unitarity norm}
Histograms of the unitarity norm \eqref{eq:unitarity_norm} in the restricted ensemble are shown in \cref{fig:quartic_4D_unitarity_norm}. Since for $t_{\max}=0.0$ the simulation trajectories converge to the respective real axes and remain there indefinitely, the unitarity norm is zero in that case. For the nonzero real-time extents, the unitarity-norm distribution becomes broader with increasing $t_{\max}$. While, again, there is no clear-cut threshold for $\mathcal{N}_U$ beyond which results should be interpreted as incorrect, one once again observes a correlation between incorrect results and a tendency of the unitarity norm to become large and its distribution to decay slowly. 

\paragraph{Configurational temperature}
Finally, the configurational-temperature observable $\tilde{\beta}$ in the model \eqref{eq:quartic_4D} is defined as 
\begin{equation}
	\tilde{\beta} = -\left\langle\frac{\partial}{\partial\phi_i}\frac{D_i[\phi]}{D_j[\phi]D_j[\phi]}\right\rangle\;,
\end{equation}
with the drift term $D_i[\phi]$ in \eqref{eq:quartic_4D_drift}. Evaluating this quantity on the restricted ensemble for different real-time extents $t_{\max}$ gives
\begin{align}
	\begin{aligned}
		\tilde{\beta}_{CL} = 
			\begin{cases}
				1.004(3) \ &\textnormal{for} \quad t_{\max}=0.0\\
				0.6(2) -0.1(1)\ii \ &\textnormal{for} \quad t_{\max}=0.4\\
				0.3(4) - 0.8(6)\ii \ &\textnormal{for} \quad t_{\max}=0.8
			\end{cases}\;,
	\end{aligned}
\end{align}
where $\imag\,\tilde{\beta}_\CL$ is negligible for $t_{\max}=0.0$. This result suggests correct convergence for $t_{\max}=0.0$ and incorrect convergence for the other real-time extents. For $t_{\max}=0.0$ and $0.8$, these findings are consistent with \cref{tab:quartic_4D}, but for $t_{\max}=0.4$, this is not the case. It is unclear whether this is a deficiency of the configurational-temperature criterion itself or merely a consequence of the theory \eqref{eq:quartic_4D} with four degrees of freedom not being close enough to the thermodynamic limit, where the criterion is assumed to be valid. Another possibility is that the criterion is indeed valid, which would, in turn, imply that the results for $t_{\max}=0.4$ in \cref{tab:quartic_4D} should turn out to be incorrect upon increasing the statistics. This, however, appears to be unlikely, as it would contradict all other criteria.

\paragraph{Summary}
While conceptionally less simple than the previous models due to the larger number of variables, distinguishing correct from incorrect convergence appears to be relatively straightforward in the lattice model \eqref{eq:quartic_4D} when employing the histogram or drift criteria. A boundary-term analysis allows for the same conclusions, albeit with some subtleties arising for different observables. The convergence conditions fail in this case. Regarding the Dyson--Schwinger equations, the generated statistics in the restricted ensemble likely was insufficient in the present case, but one would generally expect them to detect incorrect convergence in \eqref{eq:quartic_4D} as well. The unitarity-norm criterion once again allows only for heuristic conclusions, while the configurational temperature either fails or is not well defined for the model with four variables. Finally, observable bounds were not investigated in detail for \eqref{eq:quartic_4D}, but it again appears to be very difficult to find a suitable control observable.
		
	\section{Summary \& conclusions}
	\label{sec:conclusions}Complex Langevin simulations sometimes produce incorrect results for expectation values of observables and various diagnostic tools exist to assess the (in)correctness of simulation results in practice. This work has examined eight different such correctness criteria: Dyson--Schwinger equations, histograms, boundary terms, the convergence conditions, the drift criterion, observable bounds, the unitarity norm, and the configurational temperature. They were applied to three models with a single degree of freedom (a quartic model, one whose drift term has a single pole, and one with multiple poles) as well as to a quartic model with four degrees of freedom arising from a simple lattice discretization of a one-dimensional quantum field theory on a complex time-contour. The aim of this investigation was to examine in detail the advantages and disadvantages of each correctness criterion and to provide a practical guideline for their usage. The results of this work can be summarized as follows.

First of all, it comes as no surprise that the convergence conditions of \cref{sec:convergence_conditions} do not constitute a reliable correctness criterion. After all, what they really measure is equilibration and not correctness. While in most of the cases studied they allow for the same conclusions as a boundary-term analysis, a counterexample was found as well. In more realistic scenarios, a situation in which the convergence conditions hold but boundary terms are nonzero appears to be the norm rather than the exception. In particular, such a scenario seems more likely to occur when multiple degrees of freedom are involved.

Secondly, from the results of this work one would conclude that the configurational temperature defined in \cref{sec:configurational_temperature} is not a satisfactory criterion either, at least not in the models considered here. In fact, it was observed to produce false positives as well as false negatives and the inverse configurational-temperature observable was even found to diverge for most kernels within the model \eqref{eq:quartic_1D}. This, however, might simply be a consequence of the assumption of a thermodynamic limit being violated by the low numbers of degrees of freedom investigated. Clearly, the configurational-temperature observable will suffer from large fluctuations if the complex-Langevin equilibrium distribution is nonzero close to its poles.

While the observable bounds of \cref{sec:observable_bounds} are rigorous and failsafe in principle, the present investigation has also shed some more light on their obvious weakness, which is their practical applicability. While the violation of the bounds \eqref{eq:observable_bounds} unambigously indicates incorrect convergence, there is (to this date) no way of using them to actually prove correctness. What is more, finding appropriate control variables violating the bounds in the case of incorrect convergence is a highly nontrivial endeavor that could not be accomplished in some of the cases investigated. This is especially complicated when the deviation between complex-Langevin and exact results is small. An important takeaway from the present investigation is, however, that it can be advisable to make use of the freedom of studying control observables within the extended space $\mathcal{H}$ if possible.

There is an interesting correlation between incorrect results and the appearance of large unitarity norms (see \cref{sec:unitarity_norm}) in complex Langevin simulations. In particular, results with a larger deviation from exact solutions tend to be associated with unitarity norms with broader histograms and a slower decay towards infinity. In order to turn this observation into a practical correctness criterion, one would have to define some sort of heuristic threshold beyond which results are to be considered incorrect. Thus, at least for the cases studied, the unitarity norm serves as a reasonable guideline but not a reliable correctness criterion. Also note that a counterexample to this behavior was found as well in \cref{sec:hubbard}

The validity of the Dyson--Schwinger equations discussed in \cref{sec:dse} is a necessary condition for correctness, which means that their violation automatically implies incorrectness. Once again, however, the converse does not hold, as the complex Langevin evolution might receive contributions from unwanted integration cycles spoiling the results. This is particularly apparent in the models \eqref{eq:quartic_1D} and \eqref{eq:one_pole}, in which such contributions are known to be important. It is still unclear what the role of unwanted integration cycles is in more realistic theories (with or without a kernel). If one can find an argument preventing them from contributing to the complex-Langevin dynamics, then the Dyson--Schwinger equations are certainly a viable correctness criterion. Note, however, that in a few of the cases studied, the statistics (in the restricted ensemble) were insufficient for revealing their violation. In general, evaluating the Dyson--Schwinger equations is more cumbersome than many other correctness criteria.

The decay of the distributions of the dynamical degrees of freedom towards infinity as well as near poles of the drift term is studied when employing the histogram criterion of \cref{sec:histograms}. Indeed, a slow decay of such histograms is associated with incorrect convergence, which was confirmed by the findings of this work. The converse, however, is not necessarily true as a few cases with fast-decaying distributions but incorrect results were encountered in the model \eqref{eq:quartic_1D}. Nonetheless, whenever no unwanted integration cycles contribute, the histogram criterion appears to be very reliable.

The situation of the boundary-term observables of \cref{sec:boundary_terms} is a rather peculiar one. While in principle they should be capable of indicating (in)correct convergence on the level of individual observables, the reality appears to be more subtle. First of all, it is not always straightforward to find plateaus in the boundary-term observables that one could extrapolate to the limit of infinite cutoff in the first place. Second, in the (admittedly pathological) example involving a complex noise term of \cref{sec:hubbard}, the boundary terms for a given observable, whose complex-Langevin expectation value was correct, were nonetheless nonzero 
for all maximum Langevin-time step-sizes considered. However, boundary terms are quite susceptible to finite-step-size effects. In this work, this manifested itself in the form of boundary terms being nonzero for larger step-sizes but decreasing in magnitude as the maximum allowed step-size was decreased. Finally, boundary terms also seem to be insensitive to contributions from unwanted integration cycles, similar to the Dyson--Schwinger equations. Note again  that only boundary terms at infinity were studied in this work, since boundary terms at poles are expected to vanish in the equilibrium limit \cite{Sei20}. 

Lastly, the drift criterion of \cref{sec:drift_histograms} is among the most powerful criteria when it comes to detecting overall (in)correctness. While it cannot distinguish between individual observables, this should not necessarily be considered a downside. After all, a situation in which some observables come out correct while others do not seems only realistic with finite statistics. In other words, given unlimited statistics either all observables (perhaps apart from pathological ones) should be correct or all of them should be incorrect and the drift criterion appears to be able to distinguish between those cases. An additional advantage of the drift criterion is that it is relatively cheap to evaluate since the drift term must be computed for each update anyways. However, the criterion does not come without its own downsides: On the one hand, as was argued in \cref{app:kernel_1D}, in the model \eqref{eq:quartic_1D} the criterion cannot detect contributions from unwanted integration cycles. This, however, seems to be an exception, as in the model \eqref{eq:one_pole} unwanted cycles play a role as well but the incorrectness they cause is nonetheless detected by the drift criterion. Thus, \eqref{eq:quartic_1D} might be special in this regard, owing to its simplicity. On the other hand, as discussed in \cref{app:drift_criterion}, it should also be stressed that the drift criterion sometimes indicates incorrect convergence even in situations in which the complex Langevin dynamics reproduce the correct solutions within the statistical uncertainties, in particular in compact models.

More generally, besides the drift criterion histogram analyses, boundary terms and Dyson--Schwinger equations are all found to be viable, but the latter two in general require the computation of additional observables and/or some post-processing and are sometimes plagued by nontrivial subtleties. In practice, it makes sense to evaluate multiple correctness criteria and compare their predictions if possible. The proper choice of criterion might also depend on which model one intends to investigate. Finally, a comment regarding more realistic theories such as lattice models with more than four degrees of freedom or ones with compact variables such as gauge theories is in order. While the application of the various correctness criteria might be much more cumbersome (and even prohibitive for some) in those theories, many if not most of the arising complications have already been covered in this work by studying models with poles and multiple degrees of freedom. Thus, it is expected that most of the conclusions drawn in this work can be transferred to realistic theories without complication.
		
	\acknowledgments
	The author is indebted to Erhard Seiler and D\'enes Sexty for useful comments on this manuscript. Discussions with Enno Carstensen, Anosh Joseph, Arpith Kumar, Erhard Seiler, D\'enes Sexty, Ion-Olimpiu Stamatescu, and Georg Wieland are gratefully acknowledged. This research was funded in whole, or in part, by the Austrian Science Fund (FWF) [\href{https://doi.org/10.55776/P36875}{10.55776/P36875}]. The data analysis performed in this work is based on the \verb|Python| ecosystem for scientific computing \cite{python}, in particular via the packages \cite{numpy,matplotlib,pandas,scipy,cmcrameri}, but the creation and maintenance of their dependencies is acknowledged as well. Additional numerical computations were done with the help of the \verb|Wolfram Mathematica| software system \cite{mathematica}.

	\paragraph{Open Access Statement}
For the purpose of open access, the author has applied a CC BY public copyright licence to any Author Accepted Manuscript version arising from this submission.

	\paragraph{Data Availability Statement}
The data set underlying this work \cite{data} as well as the employed simulation code \cite{code} and analysis scripts \cite{scripts} are available online.

	\appendix
	\section{Constant kernel in the one-dimensional quartic model}
	\label{app:kernel_1D}This appendix summarizes the use of a constant kernel $H$ in the one-dimensional model \eqref{eq:quartic_1D}. For this model, the complex-Langevin equation reads
\begin{equation}
	\frac{\partial z}{\partial\tau} = -H^2\lambda z^3 + H\eta(\tau)\;.
\end{equation}
As shown in \cite{OOS89}, one may introduce a different variable $r$ as $z=Hr$, resulting in (since $H\neq0$)
\begin{equation}
	\frac{\partial r}{\partial\tau} = -H^4\lambda r^3 + \eta(\tau)\;.
\end{equation}
This equation, however, is nothing but the complex Langevin equation for a theory with the action
\begin{equation}\label{eq:alternative_lambda}
	\bar{S}(z) = \frac{\bar{\lambda}}{4}z^4\;, \quad 
	\textnormal{with} \quad
	\bar{\lambda} = H^4\lambda\;.
\end{equation}
This means that by an appropriate choice of kernel one may transform between theories with different couplings $\lambda$. In particular, one can always transform onto a theory with a real action, whose real axis, however, differs from the original one by a complex rescaling with $H$. Since for theories with real actions correct convergence of Langevin dynamics is essentially guaranteed, a kernel, which can be constructed for every choice of $\lambda\neq0$, completely solves the sign problem in such a case. 

Things become more complicated upon adding more terms to the action. Indeed, as was demonstrated in \cite{OSZ91}, in the model (see also \cite{AGS13})
\begin{equation}
	S(z) = \frac{\lambda}{4}z^4 + \frac{\sigma}{2}z^2\;,
\end{equation}
a constant kernel solves the sign problem only for certain combinations of $\lambda$ and $\sigma$, while generally one has to resort to $z$-dependent kernels. 

The coupling $\lambda$ and kernel $H$ used in this work can be connected to those employed in previous works \cite{OOS89,HMS25,MSS25} (labelled as $\tilde\lambda$ and $\tilde{H}$, respectively) in a straightforward way. Writing
\begin{equation}
	\lambda = e^{\frac{5\ii\pi}{12}}\;, \quad H=e^{-\frac{m\ii\pi}{48}}\;, \quad \tilde{\lambda} = e^{\frac{5\ii\pi}{6}}\;, \quad \tilde{H}=e^{-\frac{\tilde{m}\ii\pi}{48}}
\end{equation}
and requiring $H^4\lambda=\tilde{H}^4\tilde{\lambda}$, one obtains $m=\tilde{m}-5$. In particular, this means that the results of \cite{HMS25} could be reproduced in the current setup by shifting the values of $m$ used in \cite{HMS25} by $-5$. For instance, it implies that in this work correct convergence is expected around $m=5$, whereas the complex Langevin evolution would sample the imaginary integration cycle around $m=29$ and a nontrivial linear combination of the real and imaginary cycles close to $m=17$. Moreover, for $m=0$ or $m=1$ one expects results to be incorrect. This is indeed the behavior observed in \cref{sec:quartic_1D}.

Note that due to the fourth power of $H$ in \eqref{eq:alternative_lambda}, there are multiple choices of $H$ that lead to the same $\tilde{\lambda}$. All of these choices give rise to identical complex Langevin dynamics on the respective rotated real axes. More precisely, for every $H$, $-H$ and $\pm\ii H$ result in the same $\tilde{\lambda}$. However, since the respective rotations for $H$ and $\ii H$, and, thus, the corresponding equilibrium distributions, differ by $90$ degrees, the resulting complex Langevin results will also be different in general ($-H$ and $-\ii H$, on the other hand, are equivalent to $H$ and $\ii H$, respectively). This observation has profound consequences. Indeed, while some kernel $H$ might lead to correct results, the kernel $H'=\ii H$ will not, even though the underlying dynamics are exactly the same. Identical dynamics implies that the decay behavior of the drift term magnitudes must also be identical in both cases. This demonstrates that the drift criterion in \cref{sec:drift_histograms} cannot distinguish between these two kernels. An important conclusion is that the drift criterion is not (or at least not always) sensitive to unwanted integration cycles. 
	\section{Resolution of discrepancies in the one-dimensional quartic model}
	\label{app:resolution}\begin{figure}[t]
	\centering
	\includegraphics[scale=0.5]{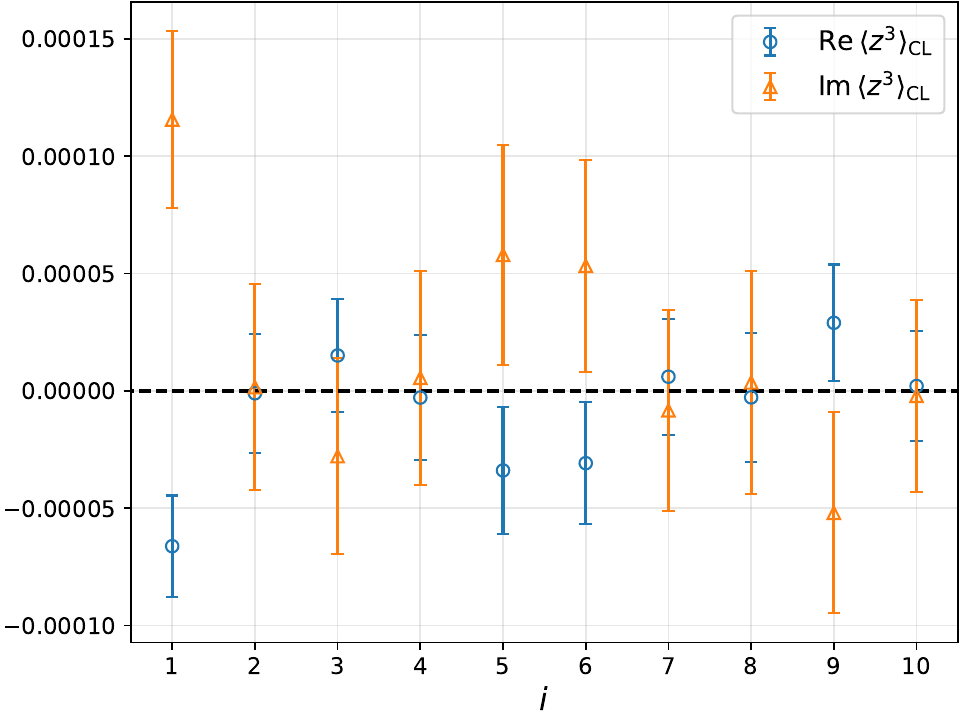}
	\caption{Estimators for the observable $\obs(z)=z^3$ in complex Langevin simulations of the model \eqref{eq:quartic_1D} using a kernel of the form \eqref{eq:quartic_1D_kernel} with $\phi=\frac{m}{48}\pi$ and $m=4$, resulting from independent data sets, each involving $N_{\mathrm{runs}}=100$ runs as described in \cref{sec:setup}. The integer $i$ enumerates these data sets.}
	\label{fig:quartic_1D_statistics}
\end{figure}

Here, the discrepancy between complex Langevin and exact results in the model \eqref{eq:quartic_1D}, observed for $m=4$ and small odd $n$ in \cref{tab:quartic_1D_low}, is investigated. As it turns out, this deviation is a statistical anomaly. As outlined in \cref{sec:setup}, the results in \cref{tab:quartic_1D_low} were computed using $N_{\mathrm{runs}}=100$ independent data points, each being an average first over parallel simulations and then over Langevin time. For the present analysis, nine more such data sets (each thus consisting of $100$ uncorrelated data points) were generated in an identical setup apart from the random number seeds. The average for $\obs(z)=z^3$ computed in each data set is shown in \cref{fig:quartic_1D_statistics}, where $i$ labels the sets and $i=1$ corresponds to the original one that went into \cref{tab:quartic_1D_low}.

One observes that the expectation values fluctuate around the correct result $\langle z^3\rangle_{\gamma_1}=0$, as they should. As a matter of coincidence, the deviation happens to be strongest for the first data set $i=1$, which was used for \cref{tab:quartic_1D_low}. The same is true for $\obs(z)=z$, which also shows a deviation from zero according to \cref{tab:quartic_1D_low}. Curiously, observables $z^n$ with even $n$ do not seem to exhibit this behavior.

	\section{False negatives in the drift criterion}
	\label{app:drift_criterion}The drift criterion discussed in \cref{sec:drift_histograms} sometimes indicates incorrect convergence of complex Langevin dynamics even if the latter reproduce the correct expectation values. To demonstrate this, consider the following one-dimensional model\footnote{The author is indebted to Erhard Seiler for bringing the model \eqref{eq:one_link} and the corresponding violation of the drift criterion to his attention.}, discussed, for instance in \cite{ASS17}:
\begin{equation}\label{eq:one_link}
	\rho(z) = \frac{1}{Z}[1+\kappa\cos(z-\ii\mu)]^{n_p}e^{\beta\cos(z)}\;, \quad
	Z = \int_{-\pi}^\pi dz [1+\kappa\cos(z-\ii\mu)]^{n_p}e^{\beta\cos(z)}\;, 
\end{equation}
with real parameters $\kappa$, $\mu$, and $\beta$, and a nonnegative integer $n_p$. Here, these parameters are chosen as $\beta=0.3$, $\mu=1$, $\kappa=1$, and $n_p=1$ for definiteness. Contrary to the other models studied in this work, \eqref{eq:one_link} describes a compact variable $z$.

Expectation values of observables of the form $\obs(z)=e^{\pm\ii kz}$ with integer $k>0$, obtained in complex-Langevin simulations of \eqref{eq:one_link} using the update scheme \eqref{eq:euler_maruyama} and a maximum Langevin-time step-size $\eps_{\max}=10^{-5}$, are shown and compared with exact solutions in \cref{tab:one_link_U1}. Indeed, it can be seen that the complex Langevin evolution reproduces the correct results within the statistical uncertainties. It should be mentioned that the statistical ensemble generated for the study of \eqref{eq:one_link} is smaller than the ensembles for the models considered in the main text by a factor of ten.

\begin{table*}[t]
    \centering
    \renewcommand{\arraystretch}{1.2}
    \begin{tabular}{|c|cc|cc|}
    \hline
    $k$ & $\langle e^{-\ii kz}\rangle_\CL$ & correct & $\langle e^{\ii kz}\rangle_\CL$ & correct \\
    \hline
    $1$ & $1.2283(1)$ & $1.228351$ & $0.28263(4)$ & $0.282645$ \\
    $2$ & $0.1730(4)$ & $0.173161$ & $0.03182(2)$ & $0.031834$ \\
    $3$ & $0.0125(5)$ & $0.012712$ & $0.002129(5)$ & $0.002132$ \\
    $4$ & $0.0007(9)$ & $0.000629$ & $0.000100(2)$ & $0.000100$ \\
    \hline
    \end{tabular}
    \caption{Comparison of expectation values $\langle e^{\pm\ii kz}\rangle$ between complex-Langevin simulations of the model \eqref{eq:one_link} with $\beta=0.3$, $\mu=1$, $\kappa=1$, and $n_p=1$, and correct results, computed via simple numerical integration, for different integers $k>0$. The parentheses indicate the statistical uncertainties rounded to their respective first significant digits. The imaginary parts of all observables are found to be consistent with zero and are thus not shown.}
    \label{tab:one_link_U1}
\end{table*}

\begin{figure}[t]
	\centering
	\includegraphics[scale=0.5]{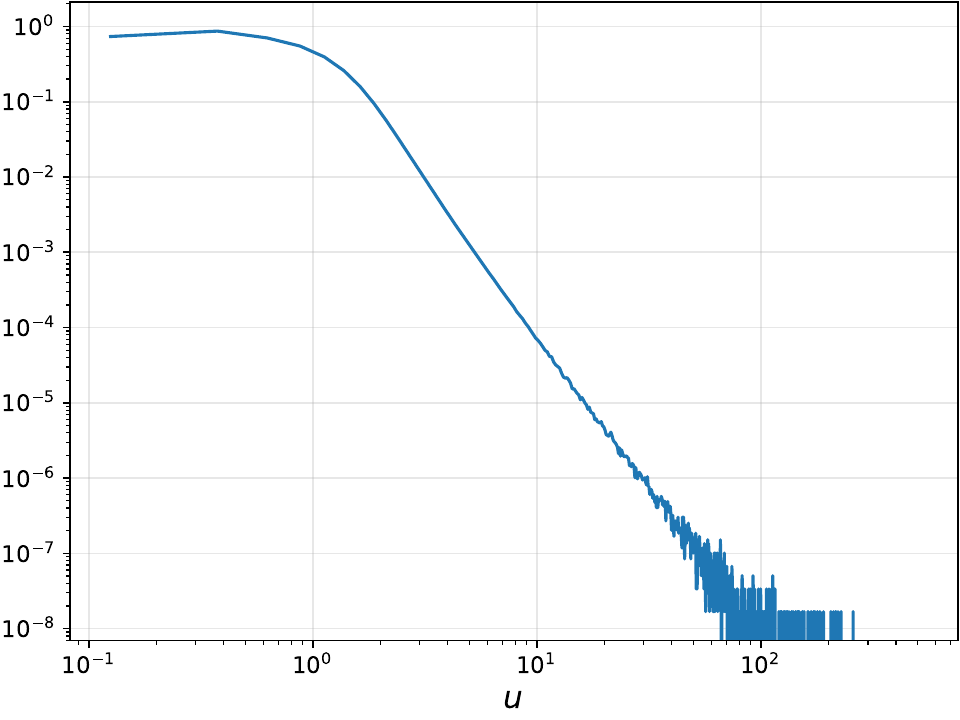}
	\caption{Histograms of the drift-term magnitude $u=\vert D\vert$ (with $D$ given in \eqref{eq:one_link_U1_drift}) in complex Langevin simulations of the model \eqref{eq:one_link} with $\beta=0.3$, $\mu=1$, $\kappa=1$, and $n_p=1$. Both axes are logarithmic.}
	\label{fig:one_link_U1_drift}
\end{figure}

Now, consider the drift term of \eqref{eq:one_link},
\begin{equation}\label{eq:one_link_U1_drift}
	D(z) = -\beta\sin(z) - \frac{n_p\kappa\sin(z-\ii\mu)}{1+\kappa\cos(z-\ii\mu)}\;.
\end{equation}
A histogram of $\vert D(z)\vert$ resulting from complex-Langevin simulations of \eqref{eq:one_link} is shown in \cref{fig:one_link_U1_drift}. As can be seen, the drift distribution's decay is clearly not exponential or faster, which implies that the drift criterion suggests incorrect convergence in the present case. This, however, is in conflict with \cref{tab:one_link_U1}, confirming that the drift criterion can motivate false negative conclusions. It is possible that the failure of the drift criterion in the context of the model \eqref{eq:one_link} is related to fact that the latter describes a compact degree of freedom, but it is not clear in which way exactly. Indeed, it could be argued that compact models like \eqref{eq:one_link} deserve a similar study like the present work on their own. 

\bibliographystyle{JHEP}
\bibliography{bibliography}

\end{document}